\documentclass[12pt]{article}

\usepackage{amsmath}
\usepackage{amsfonts}
\usepackage{amssymb}
\usepackage{graphics,graphicx,tikz}
\usetikzlibrary{patterns}
\usepackage[linktocpage]{hyperref}
\numberwithin{equation}{section} 
\usepackage{caption,subcaption}
\usepackage{cancel}
\def\ii{i}

\newcommand{\RT}{\mathrm{III}}
\newcommand{\RN}{\mathrm{IX}}
\newcommand{\Ht}{\mathrm{H}}
\newcommand{\cRT}{\mathrm{cIII}}
\newcommand{\cRN}{\mathrm{cIX}}
\newcommand{\cRNbar}{\overline{\mathrm{cIX}}}
\newcommand{\cHt}{\mathrm{cH}}
\newcommand{\beq}{\begin{equation}}
\newcommand{\eeq}{\end{equation}}
\newcommand{\bea}{\begin{align}}
\newcommand{\eea}{\end{align}}

\usepackage{color}

\begin{document}
\begin{titlepage}
\hfill \hbox{CERN-TH-2021-046}
\vskip 0.1cm
\hfill \hbox{NORDITA 2021-028}
\vskip 0.1cm
\hfill \hbox{QMUL-PH-21-17}
\vskip 0.4cm
\begin{flushright}
\end{flushright}
\begin{center}
{\Large \bf The Eikonal Approach to  Gravitational Scattering and Radiation at $\mathcal O(G^3)$} 
\vskip 1.0cm {\large  Paolo Di Vecchia$^{a, b}$, Carlo Heissenberg$^{b,c}$,
Rodolfo Russo$^{d}$, \\ Gabriele Veneziano$^{e, f}$ } \\[0.6cm]

{\it $^a$ The Niels Bohr Institute, University of Copenhagen, Blegdamsvej 17, \\DK-2100 Copenhagen \O, Denmark}\\
{\it $^b$ NORDITA, KTH Royal Institute of Technology and Stockholm University, \\
 Hannes Alfv\'ens väg 12, SE-
 11419
  Stockholm, Sweden}\\
{\it $^c$ Department of Physics and Astronomy, Uppsala University,\\ L\"agerhyddsv\"agen 1,  Box 516, SE-
	75237 Uppsala, Sweden}\\
{\it $^d$ Centre for Research in String Theory, School of Physics and Astronomy \\ Queen Mary University of London, Mile End Road, E1 4NS London, United Kingdom}\\
{\it $^e$ Theory Department, CERN, CH-1211 Geneva 23, Switzerland}\\
{\it $^f$Coll\`ege de France, 11 place M. Berthelot, 75005 Paris, France}
\end{center}
\begin{abstract}
  Using $\mathcal N=8$ supergravity as a theoretical laboratory, we extract the 3PM gravitational eikonal for two colliding massive scalars from the classical limit of the corresponding elastic two-loop amplitude. We employ the eikonal phase to obtain the physical deflection angle and to show how its non-relativistic (NR) and ultra-relativistic (UR) regimes are smoothly connected. Such a smooth interpolation rests on keeping contributions to the loop integrals originating from the full soft region,  rather than  restricting it to its potential sub-region. This task is efficiently carried out by using the method of differential equations with  complete near-static boundary conditions. 
  In contrast to the potential-region result, the physical deflection angle includes radiation-reaction contributions that are essential for recovering the finite and universal UR limit implied by general analyticity and crossing arguments.
  We finally discuss the real emission of massless states, which accounts for the imaginary part of the 3PM eikonal and for the dissipation of energy-momentum. Adopting a direct approach based on unitarity and on the classical limit of the inelastic tree-level amplitude, we are able to treat $\mathcal N=8$ and General Relativity on the same footing, and to complete the conservative 3PM eikonal in Einstein's gravity by the addition of the radiation-reaction contribution. We also show how this approach can be used to compute waveforms, as well as  the differential and  integrated spectra, for the different radiated massless fields.
\end{abstract}

\end{titlepage}

\tableofcontents

\section{Introduction}
\label{intro}

Effective field theories of Quantum Gravity are amenable to perturbative quantisation, so it is conceptually straightforward to calculate scattering amplitudes in the standard weak-coupling expansion, where the Newton constant $G$ is formally regarded as small. Powerful connections between gauge and gravity theories, such as Kawai--Lewellen--Tye relations  and colour-kinematics duality, render such calculations quite efficient as far as integrand construction is concerned, especially in the presence of supersymmetry, while systematic methods based on differential equations and integration-by-parts (IBP) identities automatise to a great extent the evaluation of the Feynman integrals.

The question then arises as to how classical gravity theories, in particular General Relativity (GR), emerge from such effective quantum theories in the classical limit, and in particular how classical observables can be recovered from perturbative scattering amplitudes, exporting as much as possible of the above technical tools.
This question raises an immediate challenge: considering for simplicity four-dimensional $2\to2$ scattering in the centre-of mass-frame, where the total energy is given by $\sqrt s$, the classical regime lies, for dimensional reasons, at trans-Planckian energies $G s \gg \hbar$, a manifestly non-perturbative kinematic region. 
The problem becomes tractable in the post-Minkowskian (PM) approximation, where one expands in powers of the dimensionless ratio $G \sqrt s/b\ll1$, with $b$ the impact parameter: the classical PM regime is captured by the Lorentz-invariant on-shell amplitudes when enforcing the hierarchy of length scales
\begin{equation}\label{hie1}
	\frac{\hbar}{\sqrt s} \ll G \sqrt s \ll  b\,,
\end{equation}
where $\frac{\hbar}{\sqrt s}$ and $G\sqrt s$ can be regarded as the effective Compton wavelength and Schwarzschild radius, respectively. Notice that the first inequality in \eqref{hie1}, which is equivalent to $G\sqrt s\gg \ell_P$ in terms of the Planck length $\ell_P=\sqrt{G\hbar}$, is incompatible with the perturbative expansion of quantum field theory.

Luckily enough, gravitational scattering at high energy and large impact parameter was extensively studied in the late eighties~\cite{ Amati:1987wq, tHooft:1987vrq, Amati:1987uf,Muzinich:1987in,Sundborg:1988tb} and the fundamental progress achieved then indeed sheds light on the problem outlined above.
The first process studied in the regime~\eqref{hie1} was the elastic gravitational scattering of two massless states which, to lowest order in the perturbative expansion, is given by the one-graviton-exchange diagram. The graviton couples to the energy itself, and the tree-level scattering amplitude diverges strongly at high energies, violating partial-wave unitarity bounds. But it was soon understood that unitarity is restored by the observation that the lowest-order tree diagram, when looked at in impact-parameter space, is just the first term of an exponential series which corresponds to summing over the exchange of arbitrarily many gravitons. This non-perturbative resummation yields a phase factor $\exp(2i \delta_0)$, called the leading eikonal, which can be used to compute \emph{classical} observables such as the deflection angle or the Shapiro time delay. It turns out that, at this lowest order, the eikonal phase is nothing but the phase-shift obtained for the scattering of one massless particle in the Aichelburg-Sexl shock-wave metric \cite{Aichelburg:1970dh} produced by the other particle~\cite{tHooft:1987vrq}. Let us mention as an aside that the original motivations for studying  trans-Planckian-energy collisions of light particles (or strings) were essentially theoretical, in the spirit of the old gedanken experiments of quantum mechanics. They had to do, in particular, with solving Hawking's celebrated information paradox~\cite{Hawking:1974sw} by trying to construct a unitary S-matrix even in the regime $G \sqrt{s} < b$ in which one expects the collision to lead to black-hole formation. In spite of some interesting progress \cite{Amati:2007ak,Ciafaloni:2011de,Ciafaloni:2014esa}, this has remained, so far, an unfinished program. 

For present-day purposes, it is important to stress that the gravitational eikonal exponentiation is not restricted to the case of massless scattering, as already pointed out in~\cite{Kabat:1992tb}. On the contrary, it has become clear over the last few years that it consistently provides the necessary non-perturbative resummation to extract the proper classical limit of the amplitude and to calculate relevant classical observables in many different setups, including the scattering of objects with masses $m_1$ and $m_2$, in the whole range $m_1+m_2\le \sqrt s <\infty$, i.e.~for \emph{generic} relative velocities between the two bodies
\cite{KoemansCollado:2019ggb,Parra-Martinez:2020dzs,AccettulliHuber:2020oou,DiVecchia:2020ymx,Bern:2020uwk}. For these reasons, since the first direct detection of gravitational waves, the emphasis in the study of the eikonal exponentiation has naturally shifted towards its application to the classical relativistic two body problem, for which various GR techniques, like the PM expansion mentioned above, the post-Newtonian (PN) expansion, and the effective-one-body (EOB) approach \cite{Buonanno:1998gg} had been developed in the past to describe the inspiral phase of two merging black holes. 
In particular, as emphasized by Damour \cite{Damour:2017zjx}, quantum scattering amplitudes can provide useful inputs to the EOB description of black-hole binaries by fixing some of the parameters appearing in its effective potential, and there has been considerable interest in the use of amplitudes-based techniques for extracting the conservative part of the EOB potential~\cite{Goldberger:2004jt,Goldberger:2016iau,Luna:2017dtq,Cheung:2018wkq,Kosower:2018adc,Bjerrum-Bohr:2018xdl,Bjerrum-Bohr:2019kec,Kalin:2019rwq,Kalin:2019inp,Cristofoli:2020uzm,Mogull:2020sak,Huber:2020xny}.

The techniques developed for the massless eikonal must be adapted to make room for the massive scattering problem at arbitrary relative velocities, while keeping suitable constraints to enforce the classical regime and ensure the validity of the post-Minkowskian approximation. 
To this effect, we shall incorporate the non-trivial mass scales $m_{1,2}$ in the hierarchy \eqref{hie1} according to
\begin{equation}\label{hie2}
	 \frac{\hbar}{\sqrt s} \lesssim \frac{\hbar}{m_{1,2}} \ll G m_{1,2} \lesssim G \sqrt s \ll b\,,
\end{equation}
or more precisely, since dimensional regularization is needed to deal with the standard infrared divergences that arise along the way, using its $(4-2\epsilon)$-dimensional counterpart \eqref{hierarchy}. In this setup, the Schwarzschild radii $G m_{1,2}$ are taken to be much larger than the two types of quantum scales corresponding to the Planck length $\ell_P \sim \sqrt{G \hbar}$ and to the Compton wavelengths $\frac{\hbar}{m_{1,2}}$, so that the leading eikonal $2\delta_0$ effectively scales as $G m_1 m_2/\hbar\gg1$ and captures the 1PM two-body dynamics. Higher-order PM corrections are instead parametrised by the quantity $G m_{1,2}/b\ll1$, which is indeed the typical size of Einstein's deflection angle, for arbitrary values of $\frac{s}{m_1 m_2}$. In this way, $2\delta_1$ is suppressed by an extra factor of $G m_{1,2}/b$ compared to the leading eikonal $2\delta_0$  and captures the 2PM two-body dynamics. In the massive case, this 2PM eikonal $2\delta_1$ is non-trivial and its $D$-dimensional GR expression was discussed in~\cite{KoemansCollado:2019ggb,Cristofoli:2020uzm}.  In the massless case, the analogous 2PM contribution vanishes for kinematic reasons, and the same happens in the case of maximally supersymmetry gravity in $D=4$ even when the external states are given a susy-preserving mass (for instance by assuming that they have a non-zero Kaluza-Klein momentum)~\cite{Caron-Huot:2018ape}. Thus in the latter cases, the first non-trivial correction to the eikonal, $2\delta_2$, comes at two loop (or 3PM) order.

In the massless, non-supersymmetric case, the 3PM eikonal was first derived more than thirty years ago~\cite{Amati:1990xe}. Unlike its lower-PM counterparts, $2\delta_2$ has both a real and an imaginary part: the latter was computed in \cite{Amati:1990xe} from the three-particle cut involving the two massless energetic particles and a graviton, and $\operatorname{Re}2\delta_2$ was determined from $\operatorname{Im}2\delta_2$ through arguments based on analyticity and crossing. The analysis of \cite{Bellini:1992eb} suggested that the same value of $ \operatorname{Re}2\delta_2$ should hold also for supersymmetric theories and this was confirmed in \cite{DiVecchia:2019kta} for ${\cal{N}}=8$ supergravity. This universality was finally extended to massless theories with different amounts of supersymmetry including GR \cite{Bern:2020gjj}. Universality of the real part depends crucially on a universal $\log(s)$-enhanced term in the $\operatorname{Im}2\delta_2$, while the rest of the imaginary part is  not universal and depends on the spectrum of  massless states that contribute to the three-particle cut.

In the massive case, the first GR results at 3PM order~\cite{Bern:2019nnu,Bern:2019crd} used the matching with an effective theory of interacting scalars to extract a conservative Hamiltonian relevant for the classical dynamics~\cite{Cheung:2018wkq}, which could be directly exported from the scattering setup to the case of bound trajectories. These results were later confirmed in \cite{Cheung:2020gyp,Kalin:2020fhe}, while the 3PM analysis in the maximally supersymmetric theory is more recent \cite{Parra-Martinez:2020dzs}. 

Before commenting further on these results, an important practical observation is in order. 
Of course, the program of extracting classical physics from scattering amplitude would hardly simplify matters compared to the PM expansion of classical gravity if one had to \emph{first} evaluate the exact quantum amplitude and \emph{then} take its classical limit/resummation. 
The goal is therefore to drop purely quantum contributions, such as short-range interactions localized near $b=0$, as early as possible in the calculation of the amplitude. This can be achieved systematically by performing the near forward expansion of the amplitude, i.e. the expansion for small $q\sim\frac{\hbar}{b}$, via the method of regions \cite{Beneke:1997zp,Smirnov:2002pj}. This tool simplifies the integration by reducing the expansion of a given Feynman integral to the expansion of its integrand in suitable scaling limits, called regions, which leads, in  turn, to the evaluation of much simpler integrals. In this language, neglecting analytic-in-$q^2$ terms in the amplitude simply amounts to dropping the so-called hard region, defined by the scaling $\ell\sim\mathcal O(m_{1,2})$ for the momenta $\ell$ of the gravitons exchanged, and focus on the soft region, $\ell\sim \mathcal O(q)$, which contributes to the long-range effects.

All the investigations of massive 3PM scattering mentioned above focused instead on the contributions arising from the potential region, a sub-region of the soft region defined for small velocities $v$ by the non-relativistic scaling $\ell^0\sim \mathcal O(vq)$, $\vec{\ell}\sim\mathcal O(q)$, hence breaking manifest Lorentz invariance. The rationale behind this further expansion was that the potential region, tailored to capturing conservative effects, would directly provide the conservative EOB data. Moreover, the restriction to the potential region allows one to consistently drop certain bubble-like contributions already at the integrand level, thus further simplifying the task at hand. However, the resulting conservative 3PM deflection angle exhibited a $\log(s)$-enhancement in the high-energy limit $\sqrt s\gg m_{1,2}$. This feature, which is tantamount to a $\log(s)$-enhancement in $\operatorname{Re}2\delta_2$, was also claimed to be universal \cite{Parra-Martinez:2020dzs}, in tension with the already mentioned and supposedly universal ultrarelativistic massless result \cite{Amati:1990xe}, which did not show any sign of such an enhancement. This apparent discontinuity between the high-energy limit of massive scattering and massless scattering prompted a lot of discussion and some attempt, such as the one in \cite{Damour:2019lcq}, was made to solve it, but was eventually contradicted by subsequent checks at 6PN level~\cite{Blumlein:2020znm,Bini:2020nsb}.

This contradiction became even sharper with Ref.~\cite{DiVecchia:2020ymx}, where we first extended the analyticity-plus-crossing analysis of \cite{Amati:1990xe} to the ultrarelativistic \emph{massive} case arriving at exactly the same conclusions. However in Ref.~\cite{DiVecchia:2020ymx} we also presented for the first time the solution of the puzzle. Following \cite{Parra-Martinez:2020dzs}, but performing the calculation of the various loop diagrams taking into account the full soft region --rather its potential part-- we computed the $2\to2$ scattering amplitude and the associated eikonal $\operatorname{Re}2\delta_2$ in massive ${\cal{N}}=8$ supergravity at 3PM level and we found that the resulting deflection angle is free of $\log(s)$-divergences, thus restoring agreement with \cite{Amati:1990xe} and the universality in the ultra-relativistic limit. The physical interpretation of the new contributions emerging in this calculation was immediately suggested by Damour: the potential region provides only the conservative part of the deflection angle, while the soft region provides also the effects of dissipation due to the emitted gravitational waves, i.e. radiation reaction effects. While they are crucial to restore universality in the ultra-relativistic limit, these radiation-reaction terms can be more easily identified in the non-relativistic regime as those associated with odd powers of the velocity, or equivalently with half-integer PN order. In conclusion, while the unphysical ``conservative deflection angle" has a spurious high-energy divergence, the physical (in principle measurable) deflection angle has a smooth limit.

By adding to the conservative result obtained in \cite{Bern:2019nnu,Bern:2019crd}  a linear-response contribution~\cite{Bini:2012ji} due to an ${\cal O}(G^2)$ radiative loss of angular momentum\footnote{It is somewhat puzzling to have an ${\cal O}(G^2)$ loss of angular momentum with an energy loss ${\cal O}(G^3)$. Actually, in order for the (unambiguous) ADM  angular momentum to agree with the one at the infinite past ($ u \to - \infty$) of future null infinity, the latter should be computed in a shear-free Bondi frame (see \cite{Ashtekar:1979} and references therein), and this implies, in the problem at hand, a loss of angular momentum ${\cal O}(G^3)$. Nevertheless, the final result of~\cite{Damour:2020tta} is sound because the linear response formula of \cite{Bini:2012ji} involves a ``mechanical angular momentum" corresponding to a different, suitably chosen Bondi frame~\cite{VV}.}, Damour~\cite{Damour:2020tta} managed to extend to GR the result of the physical deflection angle presented in \cite{DiVecchia:2020ymx} for $\mathcal N=8$, obtaining the cancellation of the $\log(s)$-divergence also in Einstein's gravity. His result has been confirmed by a completely different, amplitude-based approach in \cite{DiVecchia:2021ndb}. This alternative method consists in computing the infrared-divergent part of the  three-particle unitarity cut due to the emission of a soft graviton. As in the massless case, this determines the infrared-divergent part of $\operatorname{Im} 2 \delta_2$ (which can be equivalently derived from the exponentiation of infrared divergences in momentum space \cite{Heissenberg:2021tzo}) and also a specific logarithmic term in the finite part of $\operatorname{Im} 2 \delta_2$. This in its turn uniquely fixes the radiation-reaction contributions to $\operatorname{Re} 2 \delta_2$ via analyticity, and therefore the needed correction terms to retrieve the physical deflection angle from the conservative one.

In this paper we provide a fully self-contained discussion of the 3PM eikonal in massive $\mathcal N=8$ supergravity. In particular, we give the technical tools necessary to verify the $\mathcal{N}=8$ result for $\operatorname{Re} 2 \delta_2$ presented in~\cite{DiVecchia:2020ymx} and also derive the corresponding $\operatorname{Im} 2 \delta_2$. We extend the analysis of the imaginary part also to the non-supersymmetric case and, by building on previous results~\cite{Bern:2019nnu,Bern:2019crd,Damour:2020tta,DiVecchia:2021ndb}, present the full 3PM eikonal for the scattering of massive scalars in Einstein's gravity. As already mentioned, this requires evaluating the classical limit of the relevant two-loop Feynman integrals in the full soft region, and we achieve this goal by extending the method of ordinary differential equations (ODE) developed in \cite{Parra-Martinez:2020dzs}, combining ODE and IBP-reduction techniques with the complete near-static boundary conditions\footnote{This analysis has been performed independently in~\cite{Herrmann:2021toa} whose results were used in~\cite{Herrmann:2021lqe}.}. This allows us to find from first principles the full 3PM eikonal in ${\cal N}=8$ and to elucidate the analytic structure of the classical result by writing it explicitly in a real-analytic and crossing symmetric way.  We also extend the analysis of the 3-particle unitarity cut of~\cite{DiVecchia:2020ymx,DiVecchia:2021ndb} to generic centre-of-mass energies and beyond the Weinberg limit for the intermediate massless states, to include all classical contributions relevant to the scattering of massive states. This allows us to obtain the complete expression of $\operatorname{Im} 2 \delta_2$ in GR, thus generalising the result of~\cite{Amati:1990xe}.
As a simple extension of this analysis, we are also able to calculate  the total energy emitted at 3PM in the scattering process, both in $\mathcal N=8$ and in GR, finding perfect agreement with the results of~\cite{Herrmann:2021lqe}.
We finally describe how our approach can  also be applied to obtain waveforms, as well as  full differential emission spectra, for the various massless fields as functions of the frequency/retarded time and  the angles. For GR  these results date back to \cite{Kovacs:1978eu}, while the complete spectrum has been recently computed in~\cite{Jakobsen:2021smu,Mougiakakos:2021ckm} by using the world-line formalism and diagrammatic techniques (see also \cite{Gruzinov:2014moa, Ciafaloni:2018uwe} for previous results in the massless case and~\cite{DiVecchia:2021ndb} for the zero-frequency limit in the massive case). 

The detailed outline of the paper is as follows.
In Section~\ref{sec:preliminaries} we describe the kinematics of the process under study and then give a quick reminder of the eikonal approach to connect quantum scattering amplitudes to classical gravitational observables. 
In Section~\ref{sec:toolkit} we give details on the tools used to perform loop calculations, both for the elastic $2\to2$ amplitude and for the 3-particle unitarity cut thereof. These are based on IBP techniques and ODE tailored to the evaluation of Feynman integrals in the soft region. 
In Section~\ref{sec:eikonal3PM} we present the actual results for the massive $\mathcal{N}=8$ case up to two loops in momentum space. Integration covers the full soft region and is manifestly relativistic invariant. We then compute the corresponding eikonal phase and physical deflection angle.
In Section~\ref{analytic} the same results are recast in an explicitly analytic and crossing symmetric form. This allows us to connect some contributions to the real part of the amplitude to corresponding imaginary parts endowed with a simple physical interpretation (e.g. elastic vs. inelastic contributions). One such connection is the shortcut used in \cite{DiVecchia:2021ndb} in order to obtain the radiation-reaction term of the deflection angle.
In Section~\ref{sec:directunitarity} we study the 3-particle unitarity cut of the elastic $2\to2$ amplitude, which we obtain directly from the appropriate ``square'' of the inelastic $2\to3$ amplitude at tree level. We show how the corresponding integrations can be carried out in momentum space, borrowing tools from the reverse unitarity method \cite{Anastasiou:2002yz,Anastasiou:2002qz,Anastasiou:2003yy,Anastasiou:2015yha,Herrmann:2021lqe}. These techniques can be used to calculate $\operatorname{Im}2\delta_2$,  which allows us to reconstruct the full GR eikonal up to 3PM, and the total radiated energy-momentum, for which we find agreement with Ref.~\cite{Herrmann:2021lqe}. We also indicate (and show in an example) how our method can be adapted to obtain the wave-forms in impact parameter space and to calculate the full differential spectrum of the emitted massless field waves from a completely amplitude-based approach.
Section~\ref{sec:outlook} summarizes our results and gives a brief outlook. The two appendices contain additional technical material. Appendix~\ref{app:SoftBC} shows how to obtain the complete near-static boundary conditions that are needed for the differential equations in Section~\ref{sec:toolkit} in order to encompass the contributions of the full soft region. In Appendix~\ref{PN} we give the PN expansion of our results illustrating how the radiation-reaction terms correspond to odd powers of the velocity in the PN expansion of the real phase while containing logarithms of $v$ in the imaginary parts.

\section{Eikonal Approach for $2\to2$ Scattering}
\label{sec:preliminaries}

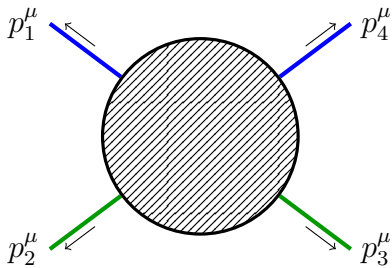
\begin{figure}[h!]
	\centering
	\begin{tikzpicture}
		\draw[<-] (-1.8,0)--(-1.4,.3);
		\draw[<-] (-1.8,3)--(-1.4,2.7);
		\draw[<-] (1.8,3)--(1.4,2.7);
		\draw[<-] (1.8,0)--(1.4,.3);
		\draw[color=green!60!black,ultra thick] (-2,0)--(0,1.5);
		\draw[color=blue,ultra thick] (0,1.5)--(2,3);
		\draw[color=blue,ultra thick] (-2,3)--(0,1.5);
		\draw[color=green!60!black,ultra thick] (0,1.5)--(2,0);
		\filldraw[color=white, fill=white, very thick](0,1.5) circle (1.3);
		\filldraw[pattern=north east lines, very thick](0,1.5) circle (1.3);
		\node at (-2,3)[left]{$p_1^{\mu}$};
		\node at (-2,0)[left]{$p_2^{\mu}$};
		\node at (2,0)[right]{$p_3^{\mu}$};
		\node at (2,3)[right]{$p_4^{\mu}$};
	\end{tikzpicture}
	\caption{Schematic representation of a $2\to2$ scattering process involving two objects with masses $m_1$ (blue) and $m_2$ (green).}
	\label{2-to-2}
\end{figure}

We are interested in the classical scattering of two spinless massive objects with masses $m_1$ and $m_2$. Our approach is to start from the elastic $2\to2$ amplitude depicted in Fig.~\ref{2-to-2}, which is evaluated in the usual perturbative expansion in powers of the Newton constant $G$, and then apply the eikonal resummation to extract its proper classical asymptotics.
In this section, we first spell out our conventions concerning the process under scrutiny and then provide a short summary of the eikonal strategy for the calculation of the classical deflection angle.

\subsection{The process under study}
\label{ssec:conventions}

With reference to Fig.~\ref{2-to-2}, we consider formally outgoing external momenta $p_1^{\mu}$, $p_2^{\mu}$, $p_3^{\mu}$ and $p_4^{\mu}$ subject to the conservation and mass-shell conditions
\begin{equation}\label{}
	p_1^{\mu}+p_2^{\mu}+p_3^{\mu}+p_4^{\mu}=0\,,\qquad
	-p_1^2=m_1^2=-p_4^2\,,\qquad
	-p_2^2=m_2^2=-p_3^2\,.
\end{equation}
The relative speed $ v$ of the asymptotic states is sized by the relativistic factor
\begin{equation}\label{sigma}
	\sigma=-\frac{p_1 \cdot p_2}{m_1 m_2} = \frac{1}{\sqrt{1- v^2}}
\end{equation}
which tends to 1 in the static limit, $ v\to 0$, and to $+\infty$ in the ultrarelativistic regime, $ v\to 1$.
We work in the centre-of-mass frame and, following \cite{Parra-Martinez:2020dzs}, we introduce the variables $\bar p_1^{\,\mu}$, $\bar p_2^{\,\mu}$ and the perturbative transferred momentum $q^{\mu}$ according to
\begin{align}\label{covBreit}
	&p_1^\mu=-\bar{p}_1^{\,\mu}+q^\mu/2 \,,\qquad  p_4^\mu = \bar{p}_1^{\,\mu} + q^\mu/2 \\
	&p_2^\mu=-\bar{p}_2^{\,\mu}-q^\mu/2 \,,\qquad  p_3^\mu = \bar{p}_2^{\,\mu} - q^\mu/2 \,,
\end{align}
together with
\begin{equation}
	u_1^\mu = \frac {\bar{p}_1^{\,\mu}} {\bar m_1}, \qquad u_2^\mu = \frac {\bar{p}_2^{\,\mu}} {\bar m_2},
	\label{eq:u1u2def}
\end{equation}
where 
\begin{equation}
	\bar m_1^2 = -\bar{p}_1^2 = m_1^2 + \frac {q^2} 4, \qquad \bar m_2^2 = - \bar{p}_2^2 = m_2^2 + \frac {q^2} 4\,. \label{eq:mbar}
\end{equation}
The usefulness of the new variables lies in the orthogonality relations
\begin{align}
	p_1^2 - p_4^2 = -2 \, \bar{p}_1 \cdot q = 0\,,\qquad
	p_2^2 - p_3^2 = 2 \, \bar{p}_2\cdot q = 0\,,
\end{align}
which imply that $q^\mu$ is constrained to a $(D-2)$-plane, although we still regard it as a $D$-vector.
We also introduce the parameter
\begin{equation}
	y=-u_1\cdot u_2
\end{equation}
which, for small $q$, is related to $\sigma$ of Eq.~\eqref{sigma} by
\begin{equation}\label{}
	y = \sigma - \frac{2m_1m_2+\sigma(m_1^2+m_2^2)}{8 m_1^2 m_2^2}\,q^2+\mathcal O\left(q^4\right)\,.
\end{equation}
Here and in the following, we also employ $q$ as a shorthand notation for $\sqrt{q^2}$.

In order to rationalize certain square roots that naturally emerge, it is also convenient to further define
\begin{align}
	z = \sigma - \sqrt{\sigma^2-1}\,,\qquad
	x = y - \sqrt{y^2-1}\,, \label{xdef}
\end{align}
so that
\begin{equation}\label{}
	\log{\frac{1}{z}}=\operatorname{arccosh}\sigma
\end{equation}
and similarly for $\log{\frac{1}{x}}$.
Comparing with Eq.~\eqref{sigma}, we see in particular that $\log \frac{1}{z}$ plays the role of a relative rapidity in this context, since
\begin{equation}\label{}
	z = \sqrt{\frac{1- v}{1+ v}}\,,\qquad
	\log \frac{1}{z} = \operatorname{arctanh}  v \,.
\end{equation} 
Finally, we will also employ the  shorthand definition
\begin{equation}\label{taudef}
	\tau = \sqrt{y^2-1}= \frac{1-x^2}{2x}\,, 
\end{equation}
especially in the discussion of differential equations and their boundary conditions.

In terms of the variables introduced above, the Mandelstam invariants
\begin{equation}\label{eq:mandelstam}
	s=-(p_1+p_2)^2\,,\qquad t=-(p_1+p_4)^2\,,\qquad u=-(p_1+p_3)^2
\end{equation} 
take the form
\begin{equation}\label{}
	s=m_1^2 + 2 m_1 m_2 \sigma + m_2^2\,,\qquad t=-q^2\,,\qquad u =m_1^2 - 2 m_1 m_2 \sigma + m_2^2+q^2\,.
\end{equation}
Let us also record here some useful relations that link $p$, the momentum of either particle in the centre-of-mass frame, to the  centre-of-mass energy $E=\sqrt{s}$:
\begin{eqnarray}
	4pE= 4m_1m_2 \sqrt{ \left( \frac{s-m_1^2-m_2^2}{2m_1m_2}\right)^2 -1}= 4m_1m_2 \sqrt{\sigma^2 -1}= \frac{2m_1 m_2 (1-z^2)}{z} \,\,.
	\label{T4}
\end{eqnarray}

\subsection{A reminder of the eikonal approach}
\label{ssec:eikonalgen}
As usual, we decompose the $S$-matrix according to \begin{equation}\label{}
	S=1+iT\,,\qquad 
	T = (2 \pi)^D \delta^{(D)}\left(\sum p_i\right) A\,,
\end{equation} 
so that the elastic $2\to2$ amplitude $\mathcal A(s,q^2)$ is defined by the matrix element
\begin{equation}\label{}
	\left\langle p_4,p_3|T|p_1,p_2\right\rangle =
	(2 \pi)^D \delta^{(D)}\left(p_1+p_2+p_3+p_4\right) \mathcal A(s,q^2)\,.
\end{equation}
Its perturbative expansion in powers of the coupling $G$ takes the form
\begin{equation}\label{genampl}
	\mathcal A(s,q^2) = \mathcal A_0(s,q^2) + \mathcal A_1(s,q^2) + \mathcal A_2(s,q^2) + \cdots \,,
\end{equation} 
with $\mathcal A_j(s,q^2)\sim \mathcal O(G^{j+1})$ the $j$-loop contribution. For brevity the dependence of $\mathcal A$ on the masses $m_1$, $m_2$ and on the dimensional regulator $\epsilon=\frac{4-D}{2}$ is left implicit. 
The classical weak-coupling regime obtains considering the following hierarchy of length scales,
\begin{equation}\label{hierarchy}
	\frac{\hbar}{m_{1,2}}\ll \left(G m_{1,2}\right) \left(\frac{\hbar}{q}\right)^{2\epsilon}\lesssim \left(G E\right) \left(\frac{\hbar}{q}\right)^{2\epsilon}\ll \frac{\hbar}{q}\,,
\end{equation}
where the two quantities in the middle are constructed in analogy to the Schwarzschild radius. The first inequality ensures that quantum effects be negligible, i.e. suppressed by the ratio between the Compton wavelength and the size of the objects under consideration, while the last inequality allows us to focus on a regime of weak gravitational interactions, i.e.~of near-forward scattering.

In this near-forward limit, namely for small momentum transfer $q^2\to 0$, each of the above contributions to the amplitude in Eq.~\eqref{genampl} can be schematically expanded as follows
\begin{subequations}\label{Aexpqspace}
	\begin{align}
		\mathcal A_0(s,q^2) &= \frac{a_0(s)}{q^2}+\cdots\,,\\
		\mathcal A_1(s,q^2) &= \frac{a_1^\mathrm{scl}(s)}{(q^2)^{1+\epsilon}}+\frac{a_1^\mathrm{cl}(s)}{(q^2)^{\frac{1}{2}+\epsilon}}+\frac{a_1^\mathrm{q}(s)}{(q^2)^{\epsilon}}+\cdots\,,\\
		\mathcal A_2(s,q^2) &= \frac{a_2^\mathrm{sscl}(s)}{(q^2)^{1+2\epsilon}}+\frac{a_2^\mathrm{scl}(s)}{(q^2)^{\frac{1}{2}+2\epsilon}}+\frac{a_2^\mathrm{cl}(s)}{(q^2)^{2\epsilon}}+\cdots\,,
	\end{align}
\end{subequations}
up to terms that are either proportional to non-negative integer powers of $q^2$ or subleading for small $q^2$. The superscripts refer to the $\hbar$ power counting, which marks the appearance of both classical $\mathcal O(\hbar^{-1})$, quantum $\mathcal O(\hbar^{0},\hbar^1,\ldots)$ and super-classical $\mathcal O(\hbar^{-2},\hbar^{-3},\ldots)$ terms.
The most natural interpretation of the singular, super-classical terms is that such contributions arise from the formal small-$G$ expansion of an exponential of the type $e^{2i\delta}$ that oscillates infinitely rapidly in the classical limit, where its phase $\operatorname{Re}2\delta$ becomes large. 

To see that this is indeed the case, however, one needs first to rewrite the amplitude in terms of the $(D-2)$-dimensional impact-parameter $b^\mu$ according to
\begin{equation}\label{impparsp}
	\widetilde{\mathcal A}(s,b) = \int\frac{d^{D-2} q}{(2\pi)^{D-2}}\, \frac{\mathcal A(s,  q^2)}{4p E}\, e^{-i  b \cdot  q} \,.
\end{equation}
In impact-parameter space, the
following exponentiation becomes manifest,
\begin{equation}\label{eikonalgen}
	1+i\widetilde{\mathcal A}(s,b) = \left(1+2i\Delta(s,b)\right)\,e^{2i\delta(s,b)}\,,
\end{equation}
where $\delta(s,b)$ is the classical eikonal, while $\Delta(s,b)$ is a quantum remainder.
Employing term by term the formula
\begin{equation}\label{FTD-2}
	\int\frac{d^{D-2}q}{(2\pi)^{D-2}}\, \frac{e^{-i b \cdot  q}}{( q^2)^\alpha} = \frac{1}{\pi^{1-\epsilon}4^\alpha \Gamma(\alpha)}\,\frac{\Gamma(1-\epsilon-\alpha)}{(b^2)^{1-\epsilon-\alpha}}\,,
\end{equation}
the Fourier transform of Eq.~\eqref{Aexpqspace} to impact parameter grants
\begin{align}
	\widetilde{\mathcal A_0}(s,b) &= \frac{f_0(s)}{(b^2)^{-\epsilon}}\,,\\
	\widetilde{\mathcal A_1}(s,b) &= \frac{f_1^\mathrm{scl}(s)}{(b^2)^{-2\epsilon}}+\frac{f_1^\mathrm{cl}(s)}{(b^2)^{\frac{1}{2}-2\epsilon}}+\frac{f_1^\mathrm{q}(s)}{(b^2)^{1-2\epsilon}}+\cdots\,,\\
	\widetilde{\mathcal A_2}(s,b) &= \frac{f_2^\mathrm{sscl}(s)}{(b^2)^{-3\epsilon}}+\frac{f_2^\mathrm{scl}(s)}{(b^2)^{\frac{1}{2}-3\epsilon}}+\frac{f_2^\mathrm{cl}(s)}{(b^2)^{1-3\epsilon}}+\cdots\,,
\end{align}
where for instance
\begin{equation}\label{}
	f_0(s)= a_0(s)\, \frac{\Gamma(-\epsilon)}{4\pi^{1-\epsilon}}\,,\qquad
	f_1^\mathrm{scl}(s)
	=
	a_1^\mathrm{scl}(s)\,
	\frac{\Gamma(-2\epsilon)}{\pi^{1-\epsilon}4^{1+\epsilon} \Gamma(1+\epsilon)}
\end{equation}
and analogous relations link the $f$'s to the $a$'s for higher-order terms.
Expanding $\delta$ and $\Delta$ appearing in Eq.~\eqref{eikonalgen} in powers of $G$,
\begin{equation}\label{deltaexp}
	\delta = \delta_0+\delta_1+\delta_2+\cdots\,,\qquad
	\Delta = \Delta_1+\cdots
\end{equation} 
the eikonal resummation amounts to identifying
\begin{equation}
	2\delta_0 = \frac{f_0}{(b^2)^{-\epsilon}}\,,\qquad 
	2\delta_1=\frac{f_1^\mathrm{cl}}{(b^2)^{\frac{1}{2}-2\epsilon}}\,,\qquad 
	2\Delta_1 = \frac{f_1^\mathrm q}{(b^2)^{1-2\epsilon}}\,,\qquad 
	2\delta_2=\frac{f_2^\mathrm{cl}-i f_0 f_1^\mathrm q}{(b^2)^{1-3\epsilon}}\,,
\end{equation}
and checking that
\begin{equation}\label{}
	f_1^\mathrm{scl}=\frac{i}{2!}\,f_0^2\,,\qquad
	f_2^\mathrm{sscl}=-\frac{1}{3!}\,f_0^3\,,\qquad
	f_2^\mathrm{scl}=i\,f_0 f_1^\mathrm{cl}\,.
\end{equation}
The eikonal data is therefore determined by the classical and quantum terms of the amplitude, while the superclassical terms should be  simply regarded as redundant objects.

Once the eikonal phase is properly retrieved, the inverse Fourier transform
\begin{equation}
\label{eik}
	i\, \frac{{\mathcal A}(s, Q^2)}{4 p E} = \int d^{D-2} b\left(e^{2i\delta(s,b)}-1\right)\, e^{i b \cdot  Q} 
\end{equation}
leads one to identify the total classical momentum transfer as
\begin{equation}\label{}
	 Q^\mu = - \frac{\partial\operatorname{Re} 2\delta(s,b)}{\partial  b^\mu}\,,
\end{equation}
since the integral is dominated by the stationary point of the (large) phase 
\begin{equation}\label{}
	\operatorname{Re}2\delta(s,b)+b\cdot Q\,.
\end{equation}
Distinguishing between the real and imaginary part is crucial because, starting from two-loop order, the eikonal
acquires an imaginary part connected to the suppression of the probability of the purely elastic process compared to the inelastic process including  Bremsstrahlung.
The deflection angle $\chi$ can be obtained by noting that
\begin{equation}\label{}
	|Q| = 2p\sin\frac{\chi}{2}\,.
\end{equation}
Note that the impact parameter $ b^\mu$ used in all previous equations is by construction oriented along the direction of  the momentum transfer $Q^\mu$.
It is sometime more informative to recast the final result as a function of the impact parameter $b_J^\mu$ that is orthogonal to the asymptotic momentum in the centre-of-mass frame, so that the angular momentum $J$ is given by $p b_J = J$. To this effect, it is sufficient to note that
\begin{equation}\label{}
 b_J = b \cos\frac{\chi}{2}\,,
\end{equation}
so that 
\begin{equation}\label{}
	\tan\frac{\chi}{2} = -\frac{1}{2p} \frac{\partial \operatorname{Re} 2\delta}{\partial b_J}\,.
\end{equation}
This is the basic relation that we are going to employ in the following in order to retrieve the scattering angle $\chi$ from the eikonal.

\subsection{Unitarity in $b$-space}
\label{sec:unib}

In this paper we shall make use of (perturbative) unitarity constraints in impact parameter space up to the two-loop level. In momentum space the standard unitarity equation reads
\begin{equation}
\label{unitarity}
	-i (T- T^{\dagger}) =T T^{\dagger} \,. 
\end{equation}
We will consider \eqref{unitarity} between our initial and final two-particle states, using the loop expansion \eqref{genampl}, but, of course, enforcing the full unitarity constraint involves elastic as well as inelastic amplitudes. Below we shall briefly review the implications of these constraints on the elastic  impact-parameter amplitude as parametrized in \eqref{eikonalgen}.
 
 At tree level \eqref{eikonalgen} simply defines the leading eikonal phase $2 \delta_0$ which is real. If one could neglect all inelastic intermediate states, \eqref{eikonalgen} with 
 $\delta = \delta_0$ and $\Delta =0$ would be the exact solution of the unitarity constraint, modulo a subtle point to be mentioned shortly.
 
Up to  two loops, the possible intermediate states entering the right-hand side of the unitarity equation \eqref{unitarity} consist of either two or three particles.  Given that there is conservation for each heavy-particle species, the possible two-particle intermediate states are either the same as the initial/final ones or their (fermionic or KK) partners. The possible 3-particle intermediate states, occurring first at two-loop order,  contain in addition a  massless particle.

At one loop, the unitarity constraint \eqref{unitarity} reads:
 \begin{equation}
\label{unitarity2l}
	2 \operatorname{Im} \mathcal{A}_1 =  \mathcal{A}_0 \otimes \mathcal{A}_0  + \sum_{in} \mathcal{A}_0^{(in)} \otimes \mathcal{A}_0^{(in)} \,, 
\end{equation}
where we indicated by $\otimes$ the standard on shell convolution with the appropriate phase space factors (see e.g.~Section~\ref{sec:directunitarity})  and by $\mathcal{A}_0^{(in)}$  a generic $2 \rightarrow 2$ inelastic tree-level amplitude.
On the other hand, going to impact parameter space and using \eqref{eikonalgen} gives:
\begin{equation}
\label{unitarity1l}
	 \operatorname{Im} \tilde{\mathcal{A}}_1 = \frac12 ( 2 \delta_0)^2 + 2  \operatorname{Im} \delta_1 + 2  \operatorname{Im} \Delta_1 = \frac12 (\tilde{\mathcal{A}}_0)^2  + \dots + FT[ \mathcal{A}_0^{(in)} \otimes \mathcal{A}_0^{(in)}] \,, 
\end{equation}
where we  indicated by $FT[\,\cdot\,]$ the Fourier transform according to \eqref{impparsp} and
the dots correspond to a possible lack of exact factorization when one converts the convolution in momentum space to a product in impact-parameter space. It turns out that in our frame (in which $q$ is orthogonal to both $p_2-p_3$ and $p_1-p_4$), 
$FT[\mathcal{A}_0 \otimes \mathcal{A}_0] =  ( 2 \delta_0)^2$, 
so that $b$-space factorization is exact in $D=4$~\footnote{We refer to Appendix B of \cite{Amati:1990xe} for details on this latter point.}.  
As a consequence, 
$2 \operatorname{Im} \delta_1 =0$
while $2 \operatorname{Im} \Delta_1$ is just related to the inelastic  two-body channels mentioned above\footnote{For instance, it vanishes in pure gravity while it gets a contribution from $n$ species  of two-gravitino intermediate states in ${\cal N} = n$ supergravity.}.

 In general, $\operatorname{Re}\tilde{\mathcal{A}}_1$ has a non-vanishing classical real part of ${\cal O}(\hbar^{-1})$, denoted by $2 \operatorname{Re} \delta_1$ in \eqref{deltaexp}, 
as well as a quantum contribution (${\cal O}(\hbar^{0})$) denoted there by $2 \operatorname{Re} \Delta_1$. As discussed later in the paper, $\Delta_1$ should be kept up to ${\cal O}(\epsilon)$ since it intervenes in the determination of $\delta_2$. It will be ignored here since it does not play an important role in the present discussion.

Consider now \eqref{unitarity} at two-loop order. In momentum space it  takes the form:
\begin{equation}
\label{unitarity2lx}
	2 \operatorname{Im} \mathcal{A}_2 = 2 \mathcal{A}_0 \otimes \operatorname{Re} \mathcal{A}_1  + \mathcal{A}_0^{(2 \rightarrow 3)} \otimes \mathcal{A}_0^{(3 \rightarrow 2)} \,, 
\end{equation}
where we have used the fact that the tree-level amplitudes $\mathcal{A}_0$ and $\mathcal{A}_0^{(2 \rightarrow 3)}$ are real. On the other hand, assuming eikonal exponentiation as in \eqref{eikonalgen}:
\begin{equation}
\label{exp2l}
	\tilde{\mathcal{A}}_2 = - \frac{ (2 \delta_0)^3}{6} +i (2 \delta_0)(2 \delta_1 + 2 \Delta_1 ) + 2 \delta_2  \,, 
\end{equation}
that can be separated into an equation for its real and one for its imaginary part. For the latter we get:
\begin{equation}
\label{exp2lim}
	\operatorname{Im} \tilde{\mathcal{A}}_2 =  (2 \delta_0)(2 \operatorname{Re} \delta_1 +2 \operatorname{Re} \Delta_1) + 2\operatorname{Im} \delta_2  \,, 
\end{equation}
Using again that the two-particle convolution in momentum space factorizes in $b$ space, we find:
\begin{equation}
\label{exp2lim2}
	 FT[\mathcal{A}_0 \otimes {\rm Re}\mathcal{A}_1] =  (2 \delta_0)(2 \operatorname{Re} \delta_1 +2 \operatorname{Re} \Delta_1)   \, ,
\end{equation}
 and inserting this in the Fourier transform of \eqref{unitarity2l} gives
\begin{equation}
\label{exp2lim3}
	4 \operatorname{Im} \delta_2   = FT[\mathcal{A}_0^{(2 \rightarrow 3)} \otimes \mathcal{A}_0^{(3 \rightarrow 2)}]    \,.
\end{equation}
This is the sought-for connection between $\operatorname{Im} \delta_2$ and the three-particle cut of the elastic amplitude  to be used later in this paper. We will show there how the r.h.s. of \eqref{exp2lim3} can be factorized through a suitable definition of the Fourier transform of the $2 \rightarrow 3$ amplitude. We also point out that $2 \rightarrow 3$ processes in which the final state does not consist of the initial two particles and a massless quantum do not contribute to the classical eikonal $\delta_2$, but only to its quantum counterpart $\Delta_2$ in \eqref{deltaexp} which is not of interest at the 3PM level.
The above arguments  can be generalized to higher loops but we shall not need them in the context of this paper.

\section{IBP+ODE Toolkit}
\label{sec:toolkit}

In this section we apply the method of differential equations \cite{Parra-Martinez:2020dzs} to the calculation of the integrals that are needed to evaluate the $2\to2$ amplitude in the limit of small momentum transfer up to two-loop order. In general, the flowchart of the method can be summarized as follows:
\begin{enumerate}
	\item Express the $2\to2$ amplitude in terms of elementary (possibly scalar) integrals $I_T$, each corresponding to a certain topology $T$ (which, by following the notation of~\cite{Parra-Martinez:2020dzs}, we dub $\mathrm{II}$, $\RT$, $\RN$, $\Ht$ and for their crossed counterparts $\overline{\mathrm{II}}$, $\overline{\RT}$, $\overline{\RN}$, $\overline{\Ht}$);
	\item Consider the limit of small momentum transfer $q \to 0$ and expand each integrand in the soft region, defined by the scaling $\ell_i\sim \mathcal O(q)$ of the loop momenta $\ell_i$;
	\item Perform Integration By Parts (IBP) reduction to identify a basis of master integrals and find the differential equations that link them to one another;
	\item Solve these equations order by order in $\epsilon$ using as boundary conditions the values of the integrals for small velocities.
\end{enumerate}   
In step number 2 one neglects contributions arising from the hard region, $\ell_i\sim\mathcal O(q^0)$, which amounts to disregarding terms proportional to non-negative integer powers of $q^2$ in the amplitude. This is fully justified, for the present purposes, because it amounts to neglecting  contributions to the eikonal whose support is localized at $b=0$ in impact-parameter space.
Step number 3 is automated using \texttt{LiteRed} \cite{Lee:2012cn,Lee:2013mka} and \texttt{FIRE6} \cite{Smirnov:2008iw}.
For $\mathcal N=8$ supergravity, steps 1--3 were already performed in \cite{Parra-Martinez:2020dzs} (except for the odd-$q$ master integrals associated to the $\Ht$ family), where however step 4 was simplified by further restricting the evaluation of the integrals to the potential region, in the static limit. 

In this section,we perform step 1-to-4 for the \emph{full} soft region, evaluating the static limit of the master integrals without any extra ingredient (in particular, without restricting to the potential region). We first discuss ordinary (planar and non-planar) topologies arising at one and two loops, and then turn to the analysis of the integrals subject to the three-particle cut.

\subsection{Box and crossed box}
\label{ssec:Box}

\begin{figure}[h!]
	\centering
	\begin{tikzpicture}
		\draw[color=green!60!black,ultra thick] (-2,0)--(2,0);
		\draw[color=blue,ultra thick] (-2,3)--(2,3);
		\draw (-1,0)--(-1,3);
		\draw (1,0)--(1,3);
		\node at (0,3)[above]{$1$};
		\node at (0,0)[below]{$2$};
		\node at (-1,1.5)[left]{$3$};
		\node at (1,1.5)[right]{$4$};
		\node at (-2,3)[left]{$p_1$};
		\node at (-2,0)[left]{$p_2$};
		\node at (2,0)[right]{$p_3$};
		\node at (2,3)[right]{$p_4$};
	\end{tikzpicture}
	\caption{The box diagram.}
	\label{Box}
\end{figure}
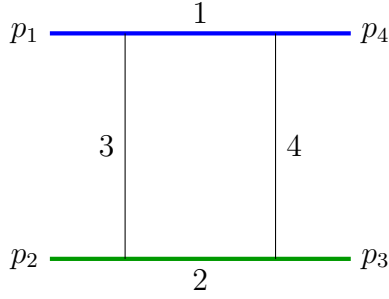

The relevant integrals at one loop are
\begin{equation}\label{}
	G_{i_{1}, i_{2}, i_{3}, i_{4}}= \int_\ell \frac{1}{\rho_{1}^{i_{1}} \rho_{2}^{i_{2}} \rho_{3}^{i_{3}} \rho_{4}^{i_{4}}}\,,
\end{equation}
where
\begin{equation}\label{}
	\int_\ell=\int \frac{\mathrm{d}^{D} \ell\, e^{\gamma_{\mathrm{E}} \epsilon}}{i \pi^{D / 2}}\,,
\end{equation}
\begin{equation}\label{}
	\rho_{1}=2 u_{1} \cdot \ell-i0, \quad \rho_{2}=-2 u_{2} \cdot \ell-i0, \quad \rho_{3}=\ell^{2}-i0, \quad \rho_{4}=(\ell-q)^{2}-i0
\end{equation}
and $i_k$ are integers. The association between the labels $i_1$, $i_2$, $i_3$, $i_4$ and the internal lines is depicted in Fig.~\ref{Box}.

The master integrals can be taken as follows
\begin{align}\label{BoxMasters}
	f_{1}=\epsilon\, q^{2} G_{0,0,2,1}, 
	\quad f_{2}=-\epsilon\, q\, G_{1,0,1,1}, \quad f_{3}=\epsilon^{2} \tau \, q^{2} G_{1,1,1,1}\,.
\end{align}
The scalar box integral\footnote{To restore the proper loop-integral normalization, $\frac{d^D\ell}{(2\pi)^{D}}$, we need to eventually multiply by a factor $\frac{i}{(4\pi)^2}\left(4\pi\,e^{-\gamma_E}\right)^{\epsilon}$ for each loop.}
\begin{equation}\label{fullbox}
	I_\mathrm{II}=\int_\ell \frac{1}{[\bar m_1 \rho_1+(\ell^2-q\cdot \ell)][\bar m_2 \rho_2+(\ell^2-q\cdot \ell)]\rho_3\rho_4}
\end{equation}
can be then expanded in the soft region and IBP-reduced to obtain
\begin{subequations}
\begin{align}
	\label{box1}
	I_{\mathrm{II}}&=\frac{1}{\epsilon^{2} \bar{m}_{1} \bar{m}_{2} \sqrt{y^2-1}} \frac{1}{q^{2}} f_{3}\\
	\label{box2}
	&+\frac{(y+1)\left(\bar{m}_{1}+\bar{m}_{2}\right)}{\bar{m}_{1}^{2} \bar{m}_{2}^{2}(y^2-1)} \frac{1}{q} f_{2} \\
	\label{box3}
	&-\frac{(1+2 \epsilon)\left(2 \bar{m}_{2} \bar{m}_{1} y+\bar{m}_{1}^{2}+\bar{m}_{2}^{2}\right)}{8 \epsilon^{2} \bar{m}_{1}^{3} \bar{m}_{2}^{3}\left(y^{2}-1\right)^{3 / 2}} f_{3}-\frac{(1+2 \epsilon)\left[\left(\bar{m}_{1}^{2}+\bar{m}_{2}^{2}\right) y+2 \bar{m}_{1} \bar{m}_{2}\right]}{8 \epsilon \bar{m}_{1}^{3} \bar{m}_{2}^{3}\left(y^{2}-1\right)} f_{1}\,,
\end{align}
\end{subequations}
where \eqref{box1} is superclassical, \eqref{box2} is classical, \eqref{box3} is the first quantum correction and we neglect higher-order quantum contributions.

The differential equations for the master integrals are conveniently expressed in terms of the parameter $x$ defined in Eq.~\eqref{xdef} and read
\begin{equation}\label{diffBox}
	\frac{\mathrm{d} \vec{f}}{\operatorname{dlog} x}=\epsilon {A} \vec{f}
	\,,\qquad
	A=\left(\begin{matrix}
		0 & 0 & 0 \\
		0 & 0 & 0 \\
		1 & 0 & 0
	\end{matrix}\right)\,.
\end{equation}
Their solutions
\begin{equation}\label{mysol}
	f_1=c_1\,,\qquad
	f_2=c_2\,,\qquad
	f_3=\epsilon\, c_1 \operatorname{log}x +c_3
\end{equation}
are uniquely determined by fixing the integration constants $c_1$, $c_2$, $c_3$ using the boundary conditions at $x=1$.
For the first two master integrals, one obtains by standard methods
\begin{equation}\label{f1Box}
	f_1=-\frac{e^{\gamma_E\epsilon}}{(q^2)^{\epsilon}}\,\frac{\Gamma(1+\epsilon)\Gamma(1-\epsilon)^2}{\Gamma(1-2\epsilon)}\,.
\end{equation}
and
\begin{equation}\label{f2Box}
	f_2=\frac{e^{\gamma_E\epsilon}}{(q^2)^{\epsilon}}\,\frac{\sqrt\pi\,\Gamma\left(\tfrac{1}{2}+\epsilon\right)\Gamma\left(\tfrac{1}{2}-\epsilon\right)^2}{4\Gamma(-2\epsilon)}\,.
\end{equation}
In this simple case, $f_3$ can be also evaluated by standard methods for any $x$, obtaining 
\begin{equation}\label{f3Box}
	f_3=\frac{e^{\gamma_E\epsilon}}{(q^2)^{\epsilon}}\,\frac{\Gamma(1+\epsilon)\Gamma(1-\epsilon)^2}{2\Gamma(-2\epsilon)}(\log x + i\pi)
\end{equation}
consistently with the general solution \eqref{mysol}.
Substituting back into the decomposition given by Eqs.~\eqref{box1}, \eqref{box2} and \eqref{box3}, one then finds, up to subleading orders in $q$,
\begin{subequations}
\begin{align}\label{}
	I_\mathrm{II}&=
	-\frac{e^{\gamma_E  \epsilon } \Gamma (1-\epsilon )^2 \Gamma (\epsilon +1) \left(\log z+i \pi \right)}{m_1
		m_2 \sqrt{\sigma ^2-1} \epsilon  \Gamma (1-2 \epsilon ) (q^2)^{1+\epsilon}} \\
	&+\frac{e^{\gamma_E  \epsilon }\sqrt{\pi } \left(m_1+m_2\right)  \Gamma \left(\frac{1}{2}-\epsilon \right)^2 \Gamma \left(\epsilon
		+\frac{1}{2}\right)}{4 m_1^2 m_2^2 (\sigma -1) \Gamma (-2 \epsilon )(q^2)^{\frac{1}{2}+\epsilon}}\\
	&+\frac{e^{\gamma_E  \epsilon } \Gamma (1-\epsilon )^2 \Gamma (\epsilon +1) \left[\sqrt{\sigma ^2-1} \left(m_1^2\sigma+m_2^2\sigma +2 m_1 m_2\right)+s (\log z+i \pi  )\right]}{4 m_1^3 m_2^3 (\sigma^2 -1)^{3/2} \Gamma (1-2 \epsilon )(q^2)^{\epsilon}}\,,
\end{align}
\end{subequations}
where $s$ is the Mandelstam invariant $s=-(p_1+p_2)^2$.

The crossed box is given by replacing $p_1\leftrightarrow p_4$ in the box. It can be obtained by first expressing $I_\mathrm{II}$ in terms of $z$ and then implementing crossing symmetry according to
\begin{equation}\label{crossingxsigma}
	z \mapsto -\frac{1}{z}-i0+\frac{q^2}{m_1 m_2 (1-z^2)}+\mathcal O(q^4)\,.
\end{equation}
The result is
\begin{subequations}
	\begin{align}
		I_{\overline{\mathrm{II}}}&=
		\frac{e^{\gamma_E  \epsilon } \Gamma (1-\epsilon )^2 \Gamma (\epsilon +1) \log \left(z\right)}{m_1 m_2 \sqrt{\sigma
				^2-1} \epsilon  \Gamma (1-2 \epsilon)(q^2)^{1+\epsilon}} \\
		&-\frac{ e^{\gamma_E  \epsilon } \sqrt{\pi } \left(m_1+m_2\right) \Gamma \left(\frac{1}{2}-\epsilon \right)^2 \Gamma \left(\epsilon
			+\frac{1}{2}\right)}{4 m_1^2 m_2^2 (\sigma +1) \Gamma (-2 \epsilon)(q^2)^{\frac{1}{2}+\epsilon}}\\
		\begin{split}
			&-\frac{e^{\gamma_E  \epsilon } \Gamma (\epsilon +1) \Gamma (1-\epsilon )^2 \left((m_1^2 +m_2^2) \epsilon-2 (1+\epsilon)m_1 m_2 \sigma 
				\right) \log z}{4 m_1^3 m_2^3 (\sigma^2-1)^{3/2} \epsilon  \Gamma (1-2 \epsilon)(q^2)^\epsilon}\\
			&-\frac{e^{\gamma_E  \epsilon } \Gamma (\epsilon +1) \Gamma (1-\epsilon )^2 \left((m_1^2+m_2^2) \sigma 
				\epsilon -2 (1+\epsilon)m_1
				m_2 \right)}{4 m_1^3 m_2^3 (\sigma^2-1) \epsilon  \Gamma (1-2 \epsilon)(q^2)^\epsilon}\,,
		\end{split}
	\end{align}
\end{subequations}
up to subleading orders in $q$.

\subsection{Planar double box}
\label{ssec:Double box}

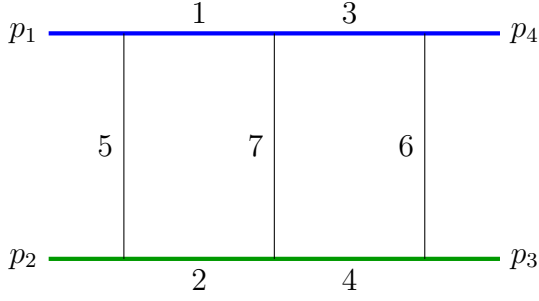
\begin{figure}[h!]
	\centering
	\begin{tikzpicture}
		\draw[color=green!60!black,ultra thick] (-3,0)--(3,0);
		\draw[color=blue,ultra thick] (-3,3)--(3,3);
		\draw (-2,0)--(-2,3);
		\draw (0,0)--(0,3);
		\draw (2,0)--(2,3);
		\node at (-1,3)[above]{$1$};
		\node at (-1,0)[below]{$2$};
		\node at (1,3)[above]{$3$};
		\node at (1,0)[below]{$4$};
		\node at (-2,1.5)[left]{$5$};
		\node at (0,1.5)[left]{$7$};
		\node at (2,1.5)[left]{$6$};
		\node at (-3,3)[left]{$p_1$};
		\node at (-3,0)[left]{$p_2$};
		\node at (3,0)[right]{$p_3$};
		\node at (3,3)[right]{$p_4$};
	\end{tikzpicture}
	\caption{The double box diagram.}
	\label{DoubleBox}
\end{figure}

In this case the relevant soft integrals are
\begin{equation}\label{bigG}
	G_{i_{1}, i_{2}, \ldots, i_{9}}=\int_{\ell_1} \int_{\ell_2} \frac{1}{\rho_{1}^{i_{1}} \rho_{2}^{i_{2}} \ldots \rho_{9}^{i_{9}}}\,.
\end{equation}
Here the first four entries are associated to  massive lines in the soft limit,
\begin{equation}\label{rho1234}
	\rho_{1}=2 \ell_{1} \cdot u_{1}, \qquad \rho_{2}=-2 \ell_{1} \cdot u_{2}, \qquad \rho_{3}=-2 \ell_{2} \cdot u_{1}, \qquad \rho_{4}=2 \ell_{2} \cdot u_{2},
\end{equation}
while $i_5$, $i_6$, $i_7$ correspond to massless lines
\begin{equation}\label{rho567}
	\rho_{5}=\ell_{1}^{2}\,,\qquad
	\rho_{6}=\ell_{2}^{2}\,, \qquad
	\rho_{7}=\left(\ell_{1}+\ell_{2}-q\right)^{2}
\end{equation}
as depicted in Fig.~\ref{DoubleBox}.
The last two entries
\begin{equation}\label{rho89}
	\rho_{8}=\left(\ell_{1}-q\right)^{2}\,,
	\qquad  \rho_{9}=\left(\ell_{2}-q\right)^{2}\,,
\end{equation}
are needed to complete the basis.
The Feynman prescription has been left implicit but it can be restored by adding $-i0$ on the right-hand sides of \eqref{rho1234}, \eqref{rho567} and \eqref{rho89}.
We consider the master integrals
\begin{subequations}
	\begin{align}
		f_{\RT,1}={}& -\epsilon ^2  q^2 G_{0,0,0,0,0,0,1,2,2}\,,\label{eq:fRT1}\\ 
		f_{\RT,2}={}&-\epsilon ^4 \tau\, G_{0,1,1,0,0,0,1,1,1}\,, \label{eq:fRT2}\\ 
		f_{\RT,3}={}&\epsilon ^3 q^2 \tau\, G_{0,1,1,0,0,0,2,1,1}\,,\\ 
		f_{\RT,4}={}&\epsilon ^2 q^2 G_{0,2,2,0,0,0,1,1,1} +\epsilon ^3 y \, q^2 G_{0,1,1,0,0,0,2,1,1}\,,\\ 
		f_{\RT,5}={}&\epsilon ^3 \tau\,  q^2 G_{1,1,0,0,1,1,2,0,0}\,,\\ 
		f_{\RT,6}={}&-\epsilon ^3 (1-6 \epsilon)\, G_{1,0,1,0,1,1,1,0,0}\,, \label{eq:fRT6} \\ 
		f_{\RT,7}={}&-\epsilon ^4 \tau^2 q^2 G_{1,1,1,1,1,1,1,0,0}\,, 
	\end{align}
\end{subequations}
which depend on even powers of $q$, and 
\begin{subequations}
	\begin{align}
		f_{\mathrm{\RT,8}}={}&-\epsilon^3 q\,
		G_{1,0,0,0,1,1,2,0,0}\,,\\ 
		f_{\mathrm{\RT,9}}={}&\epsilon^3 q\, G_{0,2,1,0,1,1,1,0,0}\,,\\ 
		f_{\mathrm{\RT,10}}={}&\epsilon^4\tau \, q\, G_{1,1,1,0,1,1,1,0,0}\,,\label{eq:fRT10}
	\end{align}
\end{subequations}
which depend on odd powers of $q$.
The scalar double box integral
\begin{equation}\label{}
	\begin{split}
		I_{\RT} &= \int_{\ell_1} \int_{\ell_2}
		\frac{1}{\left[\bar m_1 \rho_1 + (\ell_1^2-\ell_1\cdot q)\right]
			\left[\bar m_2 \rho_2 + (\ell_1^2-\ell_1\cdot q)\right]}\\
		&\frac{1}{\left[\bar m_1 \rho_3 + (\ell_2^2-\ell_2\cdot q)\right]\left[\bar m_2 \rho_4 + (\ell_2^2-\ell_2\cdot q)\right]\rho_5\rho_6\rho_7}
	\end{split}
\end{equation}
can be decomposed as follows in the soft limit,
\begin{equation}\label{softexpRT}
	I_{\RT}
	=
	\frac{ I_{\RT}^{(2)}}{q^2} + \frac{I_{\RT}^{(1)}}{q}+I_{\RT}^{(0)}+\cdots
\end{equation}
with
\begin{align}
	I_{\RT}^{(2)}=-\frac{f_{\RT,7}}{\epsilon^4\bar m_1^2 \bar m_2^2 (y^2-1)}\,,\qquad 
	I_{\RT}^{(1)}=-\frac{2(\bar m_1+\bar m_2)f_{\RT,10}}{\epsilon^3 \bar m_1^3\bar m_2^3(y-1)\sqrt{y^2-1}}
\end{align}
and
\begin{subequations}
	\begin{align}
		I_{\RT}^{(0)}
		&= f_{\RT,1}\,
		\frac{3 \epsilon  \left(y^2 \left(\bar{m}_1^2+\bar{m}_2^2\right)+4 y \bar{m}_1
			\bar{m}_2+\bar{m}_1^2+\bar{m}_2^2\right)+2 y \bar{m}_2 \bar{m}_1+\bar{m}_1^2+\bar{m}_2^2}{24
			\left(y^2-1\right)^2 \epsilon ^3 \bar{m}_1^4 \bar{m}_2^4}\\
		&+ f_{\RT,2}\,
		\frac{y \left(\bar{m}_1^2+\bar{m}_2^2\right)+2 \bar{m}_1 \bar{m}_2}{\left(y^2-1\right)^{3/2}
			\epsilon ^2 \bar{m}_1^4 \bar{m}_2^4}
		\\
		&+f_{\RT,3}\,
		\frac{y \left(\bar{m}_1^2+\bar{m}_2^2\right)+2 \bar{m}_1 \bar{m}_2}{8 \left(y^2-1\right)^{3/2}
			\epsilon ^3 \bar{m}_1^4 \bar{m}_2^4}
		\\
		&+f_{\RT,4}\,
		\frac{3 \epsilon  \left(y^2 \left(\bar{m}_1^2+\bar{m}_2^2\right)+4 y \bar{m}_1
			\bar{m}_2+\bar{m}_1^2+\bar{m}_2^2\right)+2 y \bar{m}_2 \bar{m}_1+\bar{m}_1^2+\bar{m}_2^2}{12
			\left(y^2-1\right)^2 \epsilon ^3 \bar{m}_1^4 \bar{m}_2^4}
		\\
		&-f_{\RT,5}\,
		\frac{(2 \epsilon +3) \left(y \left(\bar{m}_1^2+\bar{m}_2^2\right)+2 \bar{m}_1
			\bar{m}_2\right)}{12 \left(y^2-1\right)^{3/2} \epsilon ^3 \bar{m}_1^4 \bar{m}_2^4}
		\\
		&+f_{\RT,6}\,
		\frac{3 \epsilon  \left(y^2 \left(\bar{m}_1^2+\bar{m}_2^2\right)+4 y \bar{m}_1
			\bar{m}_2+\bar{m}_1^2+\bar{m}_2^2\right)+2 y \bar{m}_2 \bar{m}_1+\bar{m}_1^2+\bar{m}_2^2}{12
			\left(y^2-1\right)^2 \epsilon ^3 \bar{m}_1^4 \bar{m}_2^4}
		\\
		&+f_{\RT,7}\,
		\frac{(4 \epsilon +3) \left(2 y \bar{m}_2 \bar{m}_1+\bar{m}_1^2+\bar{m}_2^2\right)}{12
			\left(y^2-1\right)^2 \epsilon ^4 \bar{m}_1^4 \bar{m}_2^4}\,.
	\end{align}
\end{subequations}
The differential equations for the master integrals
\begin{equation}
	\mathrm{d} \vec {f}_{\RT}= \epsilon\left[A_{\RT,0}\,\operatorname{dlog}(x)+A_{\RT,+1}\,\operatorname{dlog}(x+1)+A_{\RT,-1}\,\operatorname{dlog}(x-1)\right]\vec{f}_{\RT}\label{eq:RTSoftDE}
\end{equation}
decompose into two disjoint sectors, depending on even and odd powers of $q$, 
\begin{align}
	A_{\RT,i}=\left(
	\begin{matrix}
		A_{\RT, i}^{(\mathrm{e})} & 0 \\
		0 & A_{\RT,i}^{(\mathrm{o})} \\
	\end{matrix}
	\right)\,,
\end{align}
where the matrices are given by
\begin{align}\label{eq:RTmateven}
	A_{\RT,0}^{(\mathrm{e})}={}&\left(
	\begin{matrix}
		0 & 0 & 0 & 0 & 0 & 0 & 0 \\
		-\frac{1}{2} & -6 & 0 & -1 & 0 & 0 & 0 \\[2pt]
		-\frac{3}{2} & 0 & 2 & -2 & 0 & 0 & 0 \\
		0 & 12 & 2 & 0 & 0 & 0 & 0 \\
		-\frac{3}{4} & 0 & 0 & 0 & 0 & 0 & 0 \\
		0 & 0 & 0 & 0 & 0 & 0 & 0 \\
		0 & 0 & 1 & 0 & -2 & 0 & 0 \\
	\end{matrix}
	\right),\quad 
	A_{\RT,\pm1}^{(\mathrm{e})} = \left(
	\begin{matrix}
		0 & 0 & 0 & 0 & 0 & 0 & 0 \\
		0 & 6 & 0 & 0 & 0 & 0 & 0 \\
		0 & 0 & -2 & 0 & 0 & 0 & 0 \\
		0 & 0 & 0 & 0 & 0 & 0 & 0 \\
		0 & 0 & 0 & 0 & 0 & 0 & 0 \\
		0 & 0 & 0 & 0 & 0 & 0 & 0 \\
		0 & 0 & 0 & 0 & 0 & 0 & 0 \\
	\end{matrix}
	\right),
\end{align}
\begin{align} \label{eq:RTmatodd}
	A_{\RT,0}^{(\mathrm{o})}={}\left(
	\begin{matrix}
		0 & 0 & 0 \\
		0 & -2 & 0 \\
		0 & 1 & 0 \\
	\end{matrix}
	\right),\quad
	A_{\RT,+1}^{(\mathrm{o})}={}\left(
	\begin{matrix}
		0 & 0 & 0 \\
		3 & 6 & 0 \\
		0 & 0 & 0 \\
	\end{matrix}
	\right),\quad
	A_{\RT,-1}^{(\mathrm{o})}={}\left(
	\begin{matrix}
		0 & 0 & 0 \\
		-3 & -2 & 0 \\
		0 & 0 & 0 \\
	\end{matrix}
	\right).
\end{align}

In order to solve the above differential equations, we need to first determine the static boundary conditions by studying $x$-dependence of the master integrals in the limit $x\to1$. The details of this calculation are presented in Appendix~\ref{app:SoftBC}.
For the even sector, we have
\begin{subequations}
	\begin{align}
		\label{fIII1}
		f_{\RT,1}&=-\frac{e^{2\gamma_E\epsilon}}{(q^2)^{2\epsilon}}
		\frac{\Gamma\left(1-\epsilon\right)^3\Gamma\left(1+2\epsilon\right)}{\Gamma\left(1-3\epsilon\right)}\,,\\
		\label{fIII2}
		f_{\RT,2}\big|_{x=1}&=0\,, \\
		\label{fIII3}
		f_{\RT,3}\big|_{x\to1}& \sim i\pi \bigg(\frac{e^{i\pi/2}e^{\gamma_E}}{(1-x)q^2}\bigg)^{2\epsilon} \frac{\epsilon \Gamma(1+2 \epsilon) \Gamma(1-2 \epsilon)^{2} \Gamma(1-\epsilon)\Gamma\left(\frac{1}{2}+\epsilon\right)}{2 \Gamma(1-4 \epsilon) \Gamma\left(\frac{1}{2}\right)}\,, \\
		\label{fIII4}
		f_{\RT,4}\big|_{x=1}&=\frac{e^{2 \gamma_E  \epsilon }}{(q^2)^{2\epsilon}}
		\,\Gamma(1+2\epsilon)\left(
		\frac{\Gamma(1-\epsilon)^3}{2\Gamma(1-3\epsilon)}
		-
		\frac{\pi\epsilon^2}{3}
		\frac{\Gamma\left(\frac{1}{2}-\epsilon\right)^3}{\Gamma\left(\frac{1}{2}-3\epsilon\right)}
		\right)\,,\\
		\label{fIII5}
		f_{\RT,5}&=-
		\frac{e^{2\gamma_E\epsilon}}{(q^2)^{2\epsilon}}
		\, \frac{\Gamma(1-\epsilon)^3\Gamma(1+2\epsilon)}{4\Gamma(-3\epsilon)}\,(\log x+i\pi )\,,\\
		\label{fIII6}
		f_{\RT,6}&=-\frac{e^{2\gamma_E\epsilon}}{(q^2)^{2\epsilon}}
		\frac{\epsilon^2\pi\Gamma(1+2\epsilon)\Gamma\left(\tfrac{1}{2}-\epsilon\right)^3}{6\Gamma\left(\tfrac{1}{2}-3\epsilon\right)}\,,\\
		\label{fIII7}
		f_{\RT,7}\big|_{x=1}&=	-\frac{e^{2\gamma_E\epsilon}}{(q^2)^{2\epsilon}}\frac{\epsilon\pi^2\Gamma(1+2\epsilon)\Gamma(1-\epsilon)^3}{6\Gamma(-3\epsilon)}\,.
	\end{align}
\end{subequations}
Eqs.~\eqref{fIII1}, \eqref{fIII5}, \eqref{fIII6} are actually valid for generic $x$, Eq.~\eqref{fIII3} refers to the leading asymptotic behaviour as $x\to1$, while eqs.~~\eqref{fIII2}, \eqref{fIII4} and  \eqref{fIII7} only apply for $x=1$.
For the odd sector, we have instead
\begin{subequations}
	\begin{align}
		\label{fIII8}
		f_{\RT,8}&=\frac{e^{2\gamma_E\epsilon}}{(q^2)^{2\epsilon}}
		\frac{}{}
		\frac{\sqrt\pi\,\epsilon^2\Gamma(1-\epsilon)^2\Gamma\left(\frac{1}{2}+2\epsilon\right)\Gamma\left(\frac{1}{2}-2\epsilon\right)\Gamma\left(\frac{1}{2}-\epsilon\right)}{2\Gamma(1-2\epsilon)\Gamma(1-3\epsilon)}\,,\\
		\begin{split}
\label{fIII9}
f_{\RT,9}\big|_{x\to1}
&\sim -\frac{e^{2\gamma_E\epsilon}}{(q^2)^{2\epsilon}}
\frac{3\cdot 2^{-2+2\epsilon} \pi \epsilon^2 \Gamma(1-\epsilon)\Gamma\left(\frac{1}{2}-2\epsilon\right)\Gamma\left(\frac{1}{2}+2\epsilon\right)}{\Gamma(1-3\epsilon)}\\
&+ \bigg(\frac{e^{i\pi/2}e^{\gamma_E}}{(1-x)q^2}\bigg)^{2\epsilon} \frac{\sqrt{\pi}\,\epsilon^2\Gamma(1+\epsilon)\Gamma(1-\epsilon)\Gamma\left(\frac{1}{2}+2\epsilon\right)\Gamma\left(\frac{1}{2}-2\epsilon\right)^2}{2\Gamma(1-4\epsilon)}\,,
		\end{split}
		\\
		\label{fIII10}
		f_{\RT,10}\big|_{x=1}&=-\frac{e^{2\gamma_E\epsilon}}{(q^2)^{2\epsilon}}
		\frac{i 2^{-2+2\epsilon}\pi^2\epsilon^3\Gamma\left(\frac{1}{2}-\epsilon\right)\Gamma\left(\frac{1}{2}-2\epsilon\right)\Gamma\left(\frac{1}{2}+2\epsilon\right)}{\Gamma\left(\frac{1}{2}-3\epsilon\right)}\,.
	\end{align}
\end{subequations}

The solutions of the differential equations can be then determined in a perturbative fashion order by order in $\epsilon$. Letting for $j=1,2,\ldots,10$,
\begin{equation}
	f_{\RT,j}=\frac{1}{(q^2)^{2\epsilon}}\sum_{k=0}^\infty f_{\RT,j}^{(k)}\epsilon^k\,,
\end{equation}
allows one to recast \eqref{eq:RTSoftDE} in terms of the recursion relations
\begin{equation}
	\mathrm{d} \vec {f}^{\,(k)}_{\RT}= \left[A_{\RT,0}\,\operatorname{dlog}(x)+A_{\RT,+1}\,\operatorname{dlog}(x+1)+A_{\RT,-1}\,\operatorname{dlog}(x-1)\right]\vec{f}^{\,(k-1)}_{\RT}\,,\label{eq:recRTSoftDE}
\end{equation}
which can be then solved for $k=0,1,2,\ldots$ taking into account the $\epsilon$-expansion of the boundary conditions at each step. 

We find, for the even sector,
\begin{equation}\label{}
	\begin{array}{l}
		f_{\RT,1}^{(0)}=-1\,,\\[4pt]
		f_{\RT,2}^{(0)}=0\,,\\[4pt]
		f_{\RT,3}^{(0)}=0\,,\\[4pt]
		f_{\RT,4}^{(0)}=\frac{1}{2}\,,\\[4pt]
		f_{\RT,5}^{(0)}=0\,,\\[4pt]
		f_{\RT,6}^{(0)}=0\,,\\[4pt]
		f_{\RT,7}^{(0)}=0\,,
	\end{array}\,
	\begin{array}{l}
		f_{\RT,1}^{(1)}=0\,,\\[4pt]
		f_{\RT,2}^{(1)}=0\,,\\[4pt]
		f_{\RT,3}^{(1)}=\frac{1}{2}\log(- x)\,,\\[4pt]
		f_{\RT,4}^{(1)}=0\,,\\[4pt]
		f_{\RT,5}^{(1)}=\frac{3}{4}\log(- x)\,,\\[4pt]
		f_{\RT,6}^{(1)}=0\,,\\[4pt]
		f_{\RT,7}^{(1)}=0\,,
	\end{array}\,
	\begin{array}{rl}
		f_{\RT,1}^{(2)}=&\!\!\!\frac{\pi^2}{6}\,,\\[4pt]
		f_{\RT,2}^{(2)}=&\!\!\!0\,,\\[4pt]
		f_{\RT,3}^{(2)}=&\!\!\!\frac{1}{2}\log^2(- x)-\log(-x)\log(1-x^2)\\[4pt]
		&\!\!\!+\frac{\pi^2}{12}-\frac{1}{2}\operatorname{Li}_2(x^2)\,,\\[4pt]
		f_{\RT,4}^{(2)}=&\!\!\!\frac{1}{2}\log^2(-x)+\frac{\pi^2}{12}\,,\\[4pt]
		f_{\RT,5}^{(2)}=&\!\!\!0\,,\\[4pt]
		f_{\RT,6}^{(2)}=&\!\!\!-\frac{\pi^2}{6}\,,\\[4pt]
		f_{\RT,7}^{(2)}=&\!\!\!-\frac{1}{2}\log^2(-x)
	\end{array}
\end{equation}
and
\begin{equation}\label{}
	\begin{split}
		f_{\RT,1}^{(3)}&=\frac{32}{3}\,\zeta_3\,,\\
		f_{\RT,2}^{(3)}&=-\frac{1}{6}\log^3(-x)-\frac{\pi^2}{6}\log(-x)\,,\\
		f_{\RT,4}^{(3)}&=\frac{1}{3}\log^3(-x)+\log(-x)\left(\frac{\pi^2}{6}+\operatorname{Li}_2(x^2)\right)-\operatorname{Li}_3(x^2)-\frac{13}{3}\,\zeta_3\,,\\
		f_{\RT,5}^{(3)}&=-\frac{\pi^2}{8}\log(-x)\,,\\
		f_{\RT,6}^{(3)}&=0\,,\\
		f_{\RT,7}^{(3)}&=\frac{1}{6}\log^3(-x)+\frac{1}{2}\log(-x)\left(\frac{\pi^2}{6}+\operatorname{Li}_2(x^2)\right)-\frac{1}{2}\operatorname{Li}_3(x^2)+\frac{1}{2}\,\zeta_3\,.\\
	\end{split}
\end{equation}
In these expressions $x$ is understood to have a small \emph{negative} imaginary part, so that $\log(-x)=\log x+i\pi$.
For the odd sector, we have instead 
\begin{equation}\label{}
	\begin{array}{l}
		f_{\RT,8}^{(2)}=\frac{\pi^2}{2}\,,\\[4pt]
		f_{\RT,9}^{(2)}=-\frac{\pi^2}{4}\,,\\[4pt]
		f_{\RT,10}^{(2)}=0\,,\\
	\end{array}\qquad
	\begin{array}{l}
		f_{\RT,8}^{(3)}={\pi^2} \log 2\,,\\[4pt]
		f_{\RT,9}^{(3)}=\frac{\pi^2}{2}\,\log(-2x)-\pi^2 \log(1-x)\,,\\[4pt]
		f_{\RT,10}^{(3)}=-\frac{\pi^2}{4}\,\log (-x)\,.\\
	\end{array}
\end{equation}

These explicit expressions for the master integrals in their turn determine the decomposition coefficients defined in \eqref{softexpRT}, for which we find
\begin{equation}
	\begin{split}
I_\RT^{(2)}&=\frac{1}{\epsilon^2} \frac{2 z^2 \left(\log(- z)\right){}^2}{m_1^2 m_2^2
	\left(z^2-1\right){}^2(q^2)^{2\epsilon}}\\ 
&+ \frac{z^2}{\epsilon}\frac{6 \operatorname{Li}_3\left(z^2\right)-\left(6
	\operatorname{Li}_2\left(z^2\right)+\pi ^2\right) \log(- z)-2 \left(\log(- z)\right){}^3-6 \zeta (3)}{3 m_1^2
	m_2^2 \left(z^2-1\right){}^2(q^2)^{2\epsilon}}\,,
	\end{split}
\end{equation}
up to $\mathcal O(\epsilon^0)$, while $I_\RT^{(1)}=\mathcal O(\epsilon^0)$ and
\begin{equation}
	\begin{split}
		I_\RT^{(0)} &= \frac{1}{\epsilon}
		\frac{-z^2 \log \left(-z\right)}{2 m_1^4 m_2^4
			\left(z^2-1\right){}^4 (q^2)^{2\epsilon}}
		\Big[-\left(m_1^2+m_2^2\right) z^4+4 m_1 m_2 z^3 \left(\log \left(-z\right)-1\right)\\&+4
		\left(m_1^2+m_2^2\right) z^2 \log \left(-z\right)
		+4 m_1 m_2 z
		\left(\log \left(-z\right)+1\right)+m_1^2+m_2^2\Big]
	\end{split}
\end{equation}
up to $\mathcal O(\epsilon^0)$. In these equations, $z$ is understood to have a small negative imaginary part, so that $\log(-z)=\log z+i\pi$.

The corresponding expressions for the crossed double box diagram, which is obtained from the double box by exchanging $p_1\leftrightarrow p_4$, can be retrieved via analytic continuation in the same manner as described for the crossed box. Applying therefore \eqref{crossingxsigma} we find	
\begin{equation}\label{softexpRTbar}
	I_{\overline\RT}
	=
	\frac{ I_{\overline\RT}^{(2)}}{q^2} + \frac{I_{\overline\RT}^{(1)}}{q}+I_{\overline\RT}^{(0)}+\cdots
\end{equation}
with 
\begin{equation}
	\begin{split}
		&I_{\overline\RT}^{(2)}=\frac{1}{\epsilon^2}
		\frac{2 z^2 \log ^2z}{m_1^2 m_2^2 \left(z^2-1\right){}^2 (q^2)^{2\epsilon}}\\ 
		&- \frac{z^2}{\epsilon}\frac{ \left[-6 \operatorname{Li}_3\left(z^2\right)+6 \operatorname{Li}_2\left(z^2\right) \log z+2 \log ^3z+\pi ^2 \log
			z+6 \zeta (3)\right]}{3 m_1^2 m_2^2 \left(z^2-1\right){}^2 (q^2)^{2\epsilon}}\,,
	\end{split}
\end{equation}
up to $\mathcal O(\epsilon^0)$, while $I_{\overline\RT}^{(1)}=\mathcal O(\epsilon^0)$ and
\begin{align}
	\begin{split}
	I_{\overline\RT}^{(0)} &= 
	\frac{1}{\epsilon^2}
	\frac{4 z^3 \log z \left(z^2 \left(\log
		z-1\right)+\log \left(z\right)+1\right)}{m_1^3 m_2^3
		\left(z^2-1\right){}^4 (q^2)^{2\epsilon}}
	\\
		&+\frac{1}{\epsilon} \frac{z^2}{{6 m_1^4 m_2^4 \left(z^2-1\right){}^4}}\\
		&\Big[
		-2 m_1 m_2 z \Big(-12 \operatorname{Li}_3\left({z
		}^2\right)+6 \operatorname{Li}_2\left(z^2\right) \left(2 \log {z
		}-1\right)\\
		&+2 \log z \left(2 \log ^2z-6
		\log \left(1-z^2\right)+\pi ^2-3\right)+12 \zeta (3)+\pi ^2\Big)\\
		&-2 m_1 m_2
		z^3 \Big(-12 \operatorname{Li}_3\left(z^2\right)+6 \operatorname{Li}_2\left({z
		}^2\right) \left(2 \log z+1\right)\\
		&+2 \log z
		\left(2 \left(\log z-3\right) \log z+6 \log
		\left(1-z^2\right)+\pi ^2+3\right)+12 \zeta (3)-\pi ^2\Big)\\
		&-12
		\left(m_1^2+m_2^2\right) z^2 \log ^2z+3
		\left(m_1^2+m_2^2\right) z^4 \log \left(z\right)-3
		\left(m_1^2+m_2^2\right) \log z
		\Big]
	\end{split}
\end{align}
up to $\mathcal O(\epsilon^0)$.

\subsection{Non-planar double box}
\label{ssec:Non-planar double box}

In this case the relevant soft integrals are
\begin{equation}\label{}
	\tilde G_{i_{1}, i_{2}, \ldots, i_{9}}=\int_{\ell_1} \int_{\ell_2} \frac{1}{\tilde\rho_{1}^{i_{1}}\tilde \rho_{2}^{i_{2}} \ldots\tilde \rho_{9}^{i_{9}}}\,,
\end{equation}
where
\begin{equation}\label{rho1234IX}
	\tilde\rho_{1}=-2 \ell_{1} \cdot u_{1}, \qquad \tilde\rho_{2}=2 \ell_{1} \cdot u_{2}, \qquad \tilde\rho_{3}=2 \ell_{2} \cdot u_{1}, \qquad \tilde\rho_{4}=2 (\ell_1+\ell_{2}) \cdot u_{2},
\end{equation}
while 
\begin{equation}\label{rho567IX}
	\tilde\rho_{5}=\ell_{1}^{2}\,,\qquad
	\tilde\rho_{6}=\ell_{2}^{2}\,, \qquad
	\tilde\rho_{7}=\left(\ell_{1}+\ell_{2}+q\right)^{2}
\end{equation}
and
\begin{equation}\label{rho89IX}
	\tilde\rho_{8}=\left(\ell_{1}+q\right)^{2}\,,
	\qquad  
	\tilde\rho_{9}=\left(\ell_{2}+q\right)^{2}\,.
\end{equation}
The Feynman prescription $-i0$ is left implicit. Aside from some different sign choices compared to the planar case, an important novelty is that $\tilde\rho_4$ now depends on both $\ell_1$ and $\ell_2$.

\begin{figure}[h!]
	\centering
	\begin{tikzpicture}
		\draw[color=green!60!black,ultra thick] (-3,0)--(3,0);
		\draw[color=blue,ultra thick] (-3,3)--(3,3);
		\draw (-2,0)--(-2,3);
		\draw (0,0)--(.9,1.3);
		\draw (1.1,1.65)--(2,3);
		\draw (2,0)--(0,3);
		\node at (-1,3)[above]{$1$};
		\node at (-1,0)[below]{$2$};
		\node at (1,3)[above]{$3$};
		\node at (1,0)[below]{$4$};
		\node at (-2,1.5)[left]{$5$};
		\node at (0,1)[right]{$6$};
		\node at (2,1)[left]{$7$};
		\node at (-3,3)[left]{$p_1$};
		\node at (-3,0)[left]{$p_2$};
		\node at (3,0)[right]{$p_3$};
		\node at (3,3)[right]{$p_4$};
	\end{tikzpicture}
	\caption{The nonplanar box diagram.}
	\label{NPDoubleBox}
\end{figure}
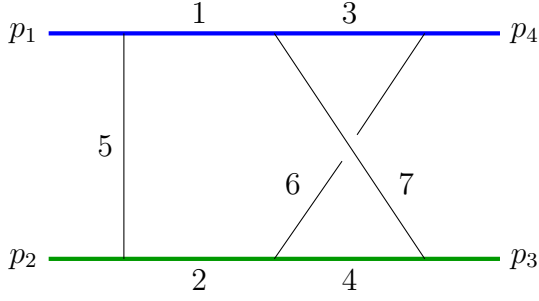

A pure basis of master integrals is given by
\begin{subequations}
	\begin{align}
		f_{\RN,1}={}&-\epsilon^2q^2 \tilde G_{0,0,0,0,2,2,1,0,0}\,,\\ 
		f_{\RN,2}={}&-\epsilon^4\tau\, \tilde G_{0,0,1,1,1,1,1,0,0}\,,\\ 
		f_{\RN,3}={}&\epsilon^3 q^2 \tau\, \tilde G_{0,0,1,1,2,1,1,0,0}\,,\\ 
		f_{\RN,4}={}&-\epsilon^2 q^2 \tilde G_{0,0,2,2,1,1,1,0,0} + \epsilon ^3 q^2 y \,\tilde G_{0,0,1,1,2,1,1,0,0}\,,\\ 
		f_{\RN,5}={}&-\epsilon ^4 \tau\, \tilde G_{0,1,1,0,1,1,1,0,0}\,,\\ 
		f_{\RN,6}={}&\epsilon ^3 q^2 \tau\, \tilde G_{0,1,1,0,1,1,2,0,0}\,,\\ 
		f_{\RN,7}={}&-\epsilon ^2 q^2 \tilde G_{0,2,2,0,1,1,1,0,0} -\epsilon ^3 q^2 y \, \tilde G_{0,1,1,0,1,1,2,0,0}\,,\\ 
		f_{\RN,8}={}&-\epsilon^3(1-6\epsilon) \tilde G_{1,0,1,0,1,1,1,0,0}\,,\\ 
		f_{\RN,9}={}&\epsilon ^3 q^2 \tau\, \tilde G_{1,1,0,0,1,1,2,0,0}\,,\\ 
		f_{\RN,10}={}&-\epsilon ^4 q^2  \tau^2 \tilde G_{1,1,1,1,1,1,1,0,0}\, ,
	\end{align}
\end{subequations}
together with
\begin{subequations}
	\begin{align}
		f_{\RN,11}={}&-\epsilon ^3 q\, \tilde G_{1,0,0,0,1,1,2,0,0}\,,\\ 
		f_{\RN,12}={}&\epsilon ^3 q\, \tilde G_{0,2,1,0,1,1,1,0,0}\,,\\ 
		f_{\RN,13}={}&\epsilon ^3 q\,\tilde G_{0,0,2,1,1,1,1,0,0}\,,\\ 
		f_{\RN,14}={}&\epsilon ^4 q\, \tau \,\tilde G_{1,0,1,1,1,1,1,0,0}\,,\\ 
		f_{\RN,15}={}&\epsilon ^4 q\,\tau \,\tilde G_{1,1,1,0,1,1,1,0,0}\,.
	\end{align}
\end{subequations}
This family of master integrals embeds the nonplanar double box as in Fig.~\ref{NPDoubleBox}.

The scalar non-planar double box integral
\begin{equation}\label{}
	\begin{split}
		I_{\RN} &= \int_{\ell_1} \int_{\ell_2}
		\frac{1}{\left[\bar m_1 \tilde\rho_1 + (\ell_1^2+\ell_1\cdot q)\right]
			\left[\bar m_2 \tilde\rho_2 + (\ell_1^2+\ell_1\cdot q)\right]}\\
		&\frac{1}{\left[\bar m_1 \tilde\rho_3 + (\ell_2^2+\ell_2\cdot q)\right]\left[\bar m_2 \tilde\rho_4 + ((\ell_1+\ell_2)^2+(\ell_1+\ell_2)\cdot q)\right]\tilde\rho_5\tilde\rho_6\tilde\rho_7}
	\end{split}
\end{equation}
can be decomposed as follows in the soft limit,
\begin{equation}\label{softexpRN}
	I_{\RN}
	=
	\frac{ I_{\RN}^{(2)}}{q^2} + \frac{I_{\RN}^{(1)}}{q}+I_{\RN}^{(0)}+\cdots
\end{equation}
with
\begin{align}
	I_{\RN}^{(2)}=-\frac{f_{\RN,10}}{q^2 \left(y^2-1\right) \epsilon ^4 \bar{m}_1^2 \bar{m}_2^2}\,,
\end{align}
\begin{align}\label{key}
	I_{\RN}^{(1)}
	&=
	f_{\RN,12}\,
	\frac{\bar{m}_1+\bar{m}_2}{2 \sqrt{q^2} (-y-1) (1-y) \epsilon ^3 \bar{m}_1^3 \bar{m}_2^3}
	\\
	&-f_{\RN,13}\,
	\frac{\bar{m}_1+\bar{m}_2}{2 \sqrt{q^2} (-y-1) (1-y) \epsilon ^3 \bar{m}_1^3 \bar{m}_2^3}
	\\
	&+f_{\RN,14}\,
	\frac{\left(\bar{m}_1+\bar{m}_2\right) (-4 y \epsilon -y-2 \epsilon )}{2 \sqrt{q^2} (-y-1)
		(1-y) \sqrt{y^2-1} \epsilon ^4 \bar{m}_1^3 \bar{m}_2^3}
	\\
	&-f_{\RN,15}\,
	\frac{\bar{m}_1+\bar{m}_2}{\sqrt{q^2} (-y-1) \sqrt{y^2-1} \epsilon ^3 \bar{m}_1^3 \bar{m}_2^3}\,,
\end{align}
and
\begingroup
\allowdisplaybreaks
\begin{subequations}
	\begin{align}\label{}
		I_{\RN}^{(0)}
		&= - f_{\RN,1}\,
		\frac{\epsilon (\bar{m}_1^2 + \bar{m}_2^2)\left(3 \left(y^2+1\right) \epsilon +1\right)+4 y (\epsilon +1) (6 \epsilon +1)
			\bar{m}_2 \bar{m}_1}{48 \left(y^2-1\right)^2 \epsilon ^4 \bar{m}_1^4 \bar{m}_2^4}
		\\
		&+ f_{\RN,2}\,
		\frac{y \epsilon  \bar{m}_1^2+y \epsilon  \bar{m}_2^2+4 (\epsilon +1) \bar{m}_2 \bar{m}_1}{4
			\left(y^2-1\right)^{3/2} \epsilon ^3 \bar{m}_1^4 \bar{m}_2^4}	
		\\
		&+f_{\RN,3}\,
		\frac{y (1-4 \epsilon  (3 \epsilon +1)) \bar{m}_1^2+y (1-4 \epsilon  (3 \epsilon +1))
			\bar{m}_2^2+2 (1-2 \epsilon ) \bar{m}_2 \bar{m}_1}{16 \left(y^2-1\right)^{3/2} \epsilon ^3
			(2 \epsilon -1) \bar{m}_1^4 \bar{m}_2^4}
		\\
		&+f_{\RN,4}\,
		\Big[
		\frac{\epsilon  (\bar{m}_1^2 +\bar{m}_2^2)\left(4 \epsilon  \left(\left(6 y^2-3\right) \epsilon
			-1\right)-1\right)}{48 \left(y^2-1\right)^2 \epsilon ^4 (2 \epsilon -1) \bar{m}_1^4 \bar{m}_2^4} \nonumber \\
		&\qquad \qquad
		+\frac{4 y (\epsilon +1) (2 \epsilon -1) (6 \epsilon +1) \bar{m}_2
			\bar{m}_1}{48 \left(y^2-1\right)^2 \epsilon ^4 (2 \epsilon -1) \bar{m}_1^4 \bar{m}_2^4}
		\Big]
		\\
		&+f_{\RN,5}\,
		\frac{-\epsilon  \left(y \left(\bar{m}_1^2+\bar{m}_2^2\right)+4 \bar{m}_1 \bar{m}_2\right)-4
			\bar{m}_1 \bar{m}_2}{4 \left(y^2-1\right)^{3/2} \epsilon ^3 \bar{m}_1^4 \bar{m}_2^4}
		\\
		&+f_{\RN,6}\,
		\frac{y (1-3 \epsilon  (2 \epsilon +1)) \bar{m}_1^2+y (1-3 \epsilon  (2 \epsilon +1))
			\bar{m}_2^2+2 \left(1-4 \epsilon ^2\right) \bar{m}_2 \bar{m}_1}{8 \left(y^2-1\right)^{3/2}
			\epsilon ^3 (2 \epsilon -1) \bar{m}_1^4 \bar{m}_2^4}
		\\
		&+f_{\RN,7}\,
		\Big[
		\frac{\epsilon (\bar{m}_1^2+\bar{m}_2^2) \left(2 \epsilon  \left(-3 y^2 (2 \epsilon +1)+12 \epsilon
			+1\right)-1\right)}{48 \left(y^2-1\right)^2 \epsilon ^4 (2 \epsilon -1) \bar{m}_1^4 \bar{m}_2^4} \nonumber \\
		&\qquad \qquad 
		\frac{4 y (\epsilon +1) (2 \epsilon -1) (6 \epsilon +1) \bar{m}_2
			\bar{m}_1}{48 \left(y^2-1\right)^2 \epsilon ^4 (2 \epsilon -1) \bar{m}_1^4 \bar{m}_2^4}
		\Big]
		\\
		&+f_{\RN,8}\,
		\frac{\epsilon  (\bar{m}_1^2 + \bar{m}_2^2 )\left(1-6 \left(y^2-2\right) \epsilon \right)+4 y (\epsilon +1) (6 \epsilon +1) \bar{m}_2
			\bar{m}_1}{48 \left(y^2-1\right)^2 \epsilon ^4 \bar{m}_1^4 \bar{m}_2^4}
		\\
		&+f_{\RN,9}\,
		\frac{y (2 \epsilon +3) \bar{m}_1^2+y (2 \epsilon +3) \bar{m}_2^2+(2-4 \epsilon ) \bar{m}_2
			\bar{m}_1}{24 \left(y^2-1\right)^{3/2} \epsilon ^3 \bar{m}_1^4 \bar{m}_2^4}
		\\
		&+f_{\RN,10}\,
		\frac{4 y (\epsilon +1) \bar{m}_2 \bar{m}_1+(4 \epsilon +3) \bar{m}_1^2+(4 \epsilon +3)
			\bar{m}_2^2}{12 \left(y^2-1\right)^2 \epsilon ^4 \bar{m}_1^4 \bar{m}_2^4}\,.
	\end{align}
\end{subequations}
\endgroup

The master integrals satisfy the following differential equations
\begin{equation}
	\mathrm{d}\vec{f}_\RN=\epsilon\left[A_{\RN,0}\,\operatorname{dlog}(x)+A_{\RN,+1}\,\operatorname{dlog}(x+1)+A_{\RN,-1}\,\operatorname{dlog}(x-1)\right]\vec{f}_\RN.\label{eq:XB2loopSoftDE}
\end{equation}
The even- and odd-$q$ systems once again decouple and we can write 
\begin{align}
	A_{\RN,i}=\left(
	\begin{array}{cc}
		A_{\RN,i}^{(\mathrm{e})} & 0 \\
		0 & A_{\RN,i}^{(\mathrm{o})} \\
	\end{array}
	\right)\,,
\end{align}
where the  matrices are given by
\begin{align} \label{eq:IXmateven}
	A_{\RN,0}^{(\mathrm{e})}={}&
	\left(
	\begin{smallmatrix}
		0 & 0 & 0 & 0 & 0 & 0 & 0 & 0 & 0 & 0 \\
		\frac{1}{2} & -6 & 0 & -1 & 0 & 0 & 0 & 0 & 0 & 0 \\
		\frac{3}{2} & 0 & 2 & -2 & 0 & 0 & 0 & 0 & 0 & 0 \\
		0 & 12 & 2 & 0 & 0 & 0 & 0 & 0 & 0 & 0 \\
		-\frac{1}{2} & 0 & 0 & 0 & -6 & 0 & 1 & 0 & 0 & 0 \\
		-\frac{3}{2} & 0 & 0 & 0 & 0 & 2 & 2 & 0 & 0 & 0 \\
		0 & 0 & 0 & 0 & -12 & -2 & 0 & 0 & 0 & 0 \\
		0 & 0 & 0 & 0 & 0 & 0 & 0 & 0 & 0 & 0 \\
		-\frac{3}{4} & 0 & 0 & 0 & 0 & 0 & 0 & 0 & 0 & 0 \\
		0 & 0 & -\frac{1}{2} & 0 & 0 & -1 & 0 & 0 & 1 & 0 \\
	\end{smallmatrix}
	\right),\quad
	A_{\RN, \pm1}^{(\mathrm{e})}={}\left(
	\begin{smallmatrix}
		0 & 0 & 0 & 0 & 0 & 0 & 0 & 0 & 0 & 0 \\
		0 & 6 & 0 & 0 & 0 & 0 & 0 & 0 & 0 & 0 \\
		0 & 0 & -2 & 0 & 0 & 0 & 0 & 0 & 0 & 0 \\
		0 & 0 & 0 & 0 & 0 & 0 & 0 & 0 & 0 & 0 \\
		0 & 0 & 0 & 0 & 6 & 0 & 0 & 0 & 0 & 0 \\
		0 & 0 & 0 & 0 & 0 & -2 & 0 & 0 & 0 & 0 \\
		0 & 0 & 0 & 0 & 0 & 0 & 0 & 0 & 0 & 0 \\
		0 & 0 & 0 & 0 & 0 & 0 & 0 & 0 & 0 & 0 \\
		0 & 0 & 0 & 0 & 0 & 0 & 0 & 0 & 0 & 0 \\
		0 & 0 & 0 & 0 & 0 & 0 & 0 & 0 & 0 & 0 \\
	\end{smallmatrix}
	\right)\,.
\end{align}
\begin{align} \label{eq:IXmatodd}
	A_{\RN,0}^{(\mathrm{o})}=\left(
	\begin{smallmatrix}
		0 & 0 & 0 & 0 & 0 \\
		0 & -2 & 0 & 0 & 0 \\
		0 & 0 & -2 & 0 & 0 \\
		0 & -1 & 1 & 0 & 0 \\
		0 & 1 & 0 & 0 & 0 \\
	\end{smallmatrix}
	\right),\
	A_{\RN,+1}^{(\mathrm{o})}=\left(
	\begin{smallmatrix}
		0 & 0 & 0 & 0 & 0 \\
		3 & 6 & 0 & 0 & 0 \\
		-3 & 0 & -2 & 0 & 0 \\
		0 & 0 & 0 & 0 & 0 \\
		0 & 0 & 0 & 0 & 0 \\
	\end{smallmatrix}
	\right),\
	A_{\RN,-1}^{(\mathrm{o})}=\left(
	\begin{smallmatrix}
		0 & 0 & 0 & 0 & 0 \\
		-3 & -2 & 0 & 0 & 0 \\
		3 & 0 & 6 & 0 & 0 \\
		0 & 0 & 0 & 0 & 0 \\
		0 & 0 & 0 & 0 & 0 \\
	\end{smallmatrix}
	\right).
\end{align}

Most master integrals can be obtained from the ones discussed in the case of the $\RT$ topology. Indeed, some integrals of the $\RN$ family are identical to their $\RT$ counterparts, possibly up to the overall sign, while others are related to them by crossing symmetry, here denoted by ``$\leftrightarrow$'', 
\begin{align}\label{III-IXrelations}\
	\begin{array}{l}
		f_{\RN,1}=f_{\RT,1}\,,\\[4pt]
		f_{\RN,5}=f_{\RT,2}\,,\\[4pt]
		f_{\RN,6}=f_{\RT,3}\,,\\[4pt]
		f_{\RN,7}=-f_{\RT,4}\,,\\[4pt]
		f_{\RN,8}=f_{\RT,6}\,,\\[4pt]
		f_{\RN,9}=f_{\RT,5}\,,\\[4pt]
		f_{\RN,11}=f_{\RT,8}\,,\\[4pt]
		f_{\RN,12}=f_{\RT,9}\,,\\[4pt]
		f_{\RN,15}=f_{\RT,10}\,,
	\end{array}
	\qquad
	\begin{array}{l}
		f_{\RN,2} \leftrightarrow f_{\RT,2}\,,\\[4pt]
		f_{\RN,3} \leftrightarrow f_{\RT,3}\,,\\[4pt]
		f_{\RN,4} \leftrightarrow -f_{\RT,4}\,,\\[4pt]
		f_{\RN,13} \leftrightarrow f_{\RT,9}\,,
	\end{array}
\end{align}
Crossing symmetry can be
effectively implemented by the replacement
\begin{equation}\label{}
	x \mapsto -\frac{1}{x}-i0\,.
\end{equation}
The only genuinely new integrals are therefore $f_{\RN,10}$, for the even sector, and $f_{\RN,14}$, for the odd sector, whose boundary conditions are given by
\begin{equation}\label{otherbc}
	f_{\RN,10}\big|_{x=1}=0\,,\qquad f_{\RN,14}\big|_{x=1}=0\,.
\end{equation}
More details about this point are provided in Appendix~\ref{app:SoftBC}.
The first few terms of $f_{\RN,10}$ in the $\epsilon$ expansion, obtained by solving the differential equations perturbatively, are as follows:
\begin{align}\label{}
	\begin{split}
	f_{\RN,10}^{(0)}&=0\,,\\
	f_{\RN,10}^{(1)}&=0\,,\\
	f_{\RN,10}^{(2)}&=\frac{1}{4}\,\log x \, \log(-x)\,,\\
		f_{\RN,10}^{(3)}&=-\frac{1}{12}\,\log^3(-x)-\frac{\pi^2}{24}(\log(-x)+2 i\pi)+\frac{i\pi}{2}\log x\, \log(-x)\\  
		&-\frac{1}{4}(\log x -i\pi)\operatorname{Li}_2(x^2)+\frac{1}{4}\operatorname{Li}_3(x^2)-\frac{1}{4}\zeta(3)\,.	
	\end{split}
\end{align}
On the other hand $f_{\RN,14}$ vanishes up to $\mathcal O(\epsilon^3)$ included.
As before, in the above expressions $x$ is implicitly understood to have a small negative imaginary part, so that $\log(-x)=\log x + i\pi$.

The final result for the decomposition coefficients defined in \eqref{softexpRN} is
\begin{align}
	\begin{split}
	&I_\RN^{(2)}=-\frac{1}{\epsilon^2} 
	\frac{z^2 \log z \log \left(-z\right)}{m_1^2 m_2^2 \left(z^2-1\right)^2(q^2)^{2\epsilon}}
	\\ 
		&+ \frac{z^2 }{\epsilon\ 6 m_1^2 m_2^2 \left({z
			}^2-1\right)^2 (q^2)^{2\epsilon}}
		\Big[
		-6 \operatorname{Li}_3\left(z^2\right)+6 \operatorname{Li}_2\left({z
		}^2\right) \left(\log z+2 i \pi \right)\\
		&+\log z
		\left(-11 \pi ^2+2 \log z \left(\log z+6 i \pi
		\right)\right)+6 \zeta (3)-2 i \pi ^3
		\Big]\,,
	\end{split}
\end{align}
up to $\mathcal O(\epsilon^0)$, while $I_\RN^{(3)}=\mathcal O(\epsilon^0)$
and
\begin{align}
		\begin{split}
	I_\RN^{(0)} &= -\frac{1}{\epsilon^2}\frac{z^3 \log \left(-z\right) \left(z^2
		\left(\log z-1\right)+\log z+1\right)}{m_1^3
		m_2^3 \left(z^2-1\right)^4 (q^2)^{2\epsilon}}\\
		&\frac{z^2}{\epsilon\, 12
			m_1^4 m_2^4 \left(z^2-1\right)^4 (q^2)^{2\epsilon}}
		\Big[-2 m_1 m_2 z^3 \Big(-6 \operatorname{Li}_3\left({z
		}^2\right)+6 \operatorname{Li}_2\left(z^2\right) \log z\\
		&+\log
		z \left(2 \log z \left(\log \left({z
		}\right)+12\right)+\pi  (\pi +24 i)-18\right)+6 \zeta (3)-18 i \pi \Big)\\
		&-2 m_1 m_2
		z \Big(-6 \operatorname{Li}_3\left(z^2\right)+6 \operatorname{Li}_2\left({z
		}^2\right) \log z\\
		&+\log z \left(2 \log
		z \left(\log z+12\right)+\pi  (\pi +24
		i)+18\right)+6 \zeta (3)+18 i \pi \Big)\\
		&-3 i \left(m_1^2+m_2^2\right) z^4 \left(2
		\pi -i \log z\right)+12 \left(m_1^2+m_2^2\right) z^2 \log
		z \log(- z) \\
		&+3
		\left(m_1^2+m_2^2\right) \left(\log z+2 i \pi \right)\Big]
	\end{split}
\end{align}
up to $\mathcal O(\epsilon^0)$. 

The corresponding expressions for the crossed non-planar double box diagram can be retrieved applying \eqref{crossingxsigma}, which yields
\begin{equation}\label{softexpRNbar}
	I_{\overline\RN}
	=
	\frac{ I_{\overline\RN}^{(2)}}{q^2} + \frac{I_{\overline\RN}^{(1)}}{q}+I_{\overline\RN}^{(0)}+\cdots
\end{equation}
with 
\begin{align}
	\begin{split}
	&I_{\overline\RN}^{(2)}=-\frac{1}{\epsilon^2}
	\frac{z^2 \log z \log \left(-z\right)}{m_1^2 m_2^2 \left(z^2-1\right)^2 (q^2)^{2\epsilon}}
	\\ 
		&+\frac{z^2}{\epsilon\, 6 m_1^2 m_2^2
			\left(z^2-1\right)^2(q^2)^{2\epsilon}}\Big[
		-6 i \pi  \operatorname{Li}_2\left(z^2\right)-6
		\operatorname{Li}_3\left(z^2\right)+6 \operatorname{Li}_2\left(z^2\right) \log
		z\\
		&+2 \log ^3z-6 i \pi  \log ^2{z
		}+7 \pi ^2 \log z+6 \zeta (3)+i \pi ^3
		\Big]\,,
	\end{split}
\end{align}
up to $\mathcal O(\epsilon^0)$, while 
$I_{\overline\RN}^{(1)}=\mathcal O(\epsilon^0)$
and
\begin{align}
		\begin{split}
	&I_{\overline\RN}^{(0)} = 
	-\frac{1}{\epsilon^2}
	\frac{z^3 \left(\log z+i \pi \right) \left(z^2
		\left(\log z-1\right)+\log z+1\right)}{m_1^3
		m_2^3 \left(z^2-1\right)^4 (q^2)^{2\epsilon}}
	\\
		&+\frac{z^2}{\epsilon\, 12 m_1^4 m_2^4 \left({z
			}^2-1\right)^4 (q^2)^{2\epsilon}}
		\Big[
		2 m_1 m_2 z^3 \Big(-18 \operatorname{Li}_3\left({z
		}^2\right)\\
		&+6 \operatorname{Li}_2\left(z^2\right) \left(3 \log z-i
		\pi +1\right)+6 \log ^3z+6 (3-i \pi ) \log ^2{z
		}\\
		&+9 (-2+\pi  (\pi +4 i)) \log z+12 \left(\log {z
		}-i \pi \right) \log \left(1-z^2\right)\\
		&+18 \zeta (3)+i \pi ^2 (\pi +7
		i)\Big)\\
		&+2 m_1 m_2 z \Big(-18 \operatorname{Li}_3\left(z^2\right)+6
		\operatorname{Li}_2\left(z^2\right) \left(3 \log z-i \pi
		-1\right)\\
		&+3 \log z \left(2 \log z \left(\log
		z-i \pi +5\right)-4 \log \left(1-z^2\right)+3 \pi
		^2+6\right)\\
		&+12 i \pi  \log \left(z-z^3\right)+18 \zeta (3)+(7+i \pi )
		\pi ^2\Big)\\
		&+3 i \left(m_1^2+m_2^2\right) z^4 \left(\pi +i \log {z
		}\right)+12 \left(m_1^2+m_2^2\right) z^2 \log z
		\left(\log z+i \pi \right)\\
		&+3 \left(m_1^2+m_2^2\right) \left(\log
		z-i \pi \right)
		\Big]
	\end{split}
\end{align}
up to $\mathcal O(\epsilon^0)$.

We have checked, where applicable, that our expressions for planar and non-planar double box integrals agree with the soft limit of the exact results in \cite{Smirnov:2001cm,Henn:2013woa} and \cite{Heinrich:2004iq}.

\subsection{$\Ht$ topology}
\label{ssec:H}

\begin{figure}[h!]
	\centering
	\begin{tikzpicture}
		\draw[color=green!60!black,ultra thick] (-2,0)--(2,0);
		\draw[color=blue,ultra thick] (-2,5)--(2,5);
		\draw (-1,0)--(-1,5);
		\draw (1,0)--(1,5);
		\draw (-1,2.5)--(1,2.5);
		\node at (0,0)[below]{$2$};
		\node at (0,5)[above]{$3$};
		\node at (-1,1.125)[left]{$5$};
		\node at (1,3.625)[right]{$6$};
		\node at (0,2.5)[above]{$7$};
		\node at (1,1.125)[right]{$8$};
		\node at (-1,3.625)[left]{$9$};
		\node at (-2,5)[left]{$p_1$};
		\node at (-2,0)[left]{$p_2$};
		\node at (2,0)[right]{$p_3$};
		\node at (2,5)[right]{$p_4$};
	\end{tikzpicture}
	\caption{The H diagram.}
	\label{HDiag}
\end{figure}
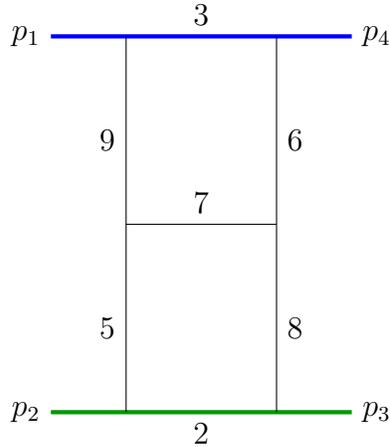

Considering the same family of integrals defined in Eq.~\eqref{bigG}, a pure basis of master integrals for the $\Ht$ topology obtains letting, for the even sector,
\begin{subequations}
	\begin{align}
		f_{\Ht, 1}&=-\epsilon^{2} q^2 G_{0,0,0,0,0,0,1,2,2} \\
		f_{\Ht, 2}&=-\epsilon^{2}(1-4 \epsilon) G_{0,0,2,0,1,0,1,1,0} \\
		f_{\Ht,3}&=\epsilon^{2} q^4 G_{0,0,0,0,2,1,0,1,2} \\
		f_{\Ht, 4}&=\epsilon^{4} q^2 G_{0,1,1,0,1,1,0,1,1} \\
		f_{\Ht, 5}&=-\epsilon^{4} \tau \ G_{0,1,1,0,0,0,1,1,1} \\
		f_{\Ht, 6}&=\epsilon^{3} \tau\, q^2 G_{0,1,1,0,0,0,2,1,1} \\
		f_{\Ht, 7}&=\epsilon^{2} q^2 G_{0,2,2,0,0,0,1,1,1}+\epsilon^{3} y\,q^2 G_{0,1,1,0,0,0,2,1,1} \\
		\label{f8Hdef}
		f_{\Ht, 8}&=-\frac1\tau\, \epsilon^{2}(4 \epsilon-1)\left[(2 \epsilon-1) G_{0,1,1,0,0,1,1,0,1}+y\, G_{0,2,0,0,0,1,1,0,1}\right] \\
		f_{\Ht, 9}&=-\epsilon^{4} \tau q^4 G_{0,1,1,0,1,1,1,1,1} \\
		\label{f10Hdef}
		\begin{split}
			f_{\Ht, 10}&=\epsilon^{4} q^2 G_{-1,1,1,-1,1,1,1,1,1}+\frac{1}{2} \epsilon^{2}(2 \epsilon-1) G_{0,0,0,0,1,1,0,1,1} \\
			&+2 \epsilon^{4} y\, q^2 G_{0,1,1,0,1,1,0,1,1}-\epsilon(3 \epsilon-2)(3 \epsilon-1)\frac{1}{q^2}\, G_{0,0,0,0,1,1,1,0,0}\,.
		\end{split}
	\end{align}
\end{subequations}
For the odd sector, we used the software \texttt{epsilon} \cite{Prausa:2017ltv} to derive the following pure basis of master integrals,
\begin{subequations}
	\begin{align}
		f_{\Ht, 11}&=\epsilon^2 (1-2\epsilon)^2 q^{-1}G_{0,0,0,1,1,0,1,1,0}\\
		f_{\Ht, 12}&=\epsilon^3 q\, G_{0,2,1,0,0,1,1,0,1}\\
		f_{\Ht,13}&=\epsilon^3 (-1+2\epsilon)\,q\, G_{0,0,0,1,1,1,0,1,1}\\
		\begin{split}
			f_{\Ht, 14}&=
			\frac18 (y-1)\epsilon^3 q^5 G_{0,2,1,0,1,1,1,1,1}
			+
			\frac{3\epsilon^3(1+3\epsilon)}{2(1+2\epsilon)}\,q\, G_{1,0,0,0,1,1,2,0,0} \\
			&+\frac{\epsilon^3(1-2\epsilon(2+3y))}{12(1+2\epsilon)}\,q\, G_{0,2,1,0,1,1,1,0,0}
			+\frac{2\epsilon^4}{(1+2\epsilon)(1+y)}\,q\, G_{0,2,1,0,0,1,1,0,1}\\
			&+\frac{1}{2}(1-2\epsilon) \epsilon^3q\, G_{0,0,0,1,1,1,0,1,1}\\
			&+\frac{2\epsilon^2(1-2\epsilon)^2(-1-2\epsilon+y(-1+\epsilon))}{3(1+2\epsilon)(1+y)q}\,G_{0,0,0,1,1,0,1,1,0}
		\end{split}\\
	f_{\mathrm{\Ht,15}}&=-\epsilon^3 q\,
	G_{1,0,0,0,1,1,2,0,0}\\ 
	f_{\mathrm{\Ht,16}}&=\epsilon^3 q\, G_{0,2,1,0,1,1,1,0,0}\,.\\ 
	\end{align}
\end{subequations}
The scalar $\Ht$ diagram is given in particular as in fig.~\ref{HDiag} by
\begin{align}\label{Hf9}
	I_{\Ht}
	&=
	\int_{\ell_1}\int_{\ell_2}
	\frac{1}{[\bar m_2 \rho_2+(\ell_1^2-\ell_1\cdot q)][\bar m_1 \rho_3+(\ell_2^2-\ell_2\cdot q)]\rho_5 \rho_6\rho_7 \rho_8\rho_9}
	\\
	&=
	-\frac{1}{q^4} \frac{f_{\Ht,9}}{\epsilon^{4} \bar{m}_{1} \bar{m}_{2} \tau }\,+\cdots
\end{align}
up to subleading corrections in $q$.

The differential equations read in this case
\begin{equation}\label{Hde}
	\mathrm{d} \vec{f}_{\Ht}=\epsilon\left[A_{\Ht, 0} \operatorname{dlog}(x)+A_{\Ht,+1} \operatorname{dlog}(x+1)+A_{\Ht,-1} \operatorname{dlog}(x-1)\right] \vec{f}_{\Ht}\,,
\end{equation}
with
\begin{equation}\label{}
	A_{\Ht,i}=\left(
	\begin{array}{cc}
		A_{\Ht,i}^{(\mathrm{e})} & 0 \\
		0 & A_{\Ht,i}^{(\mathrm{o})} \\
	\end{array}
	\right),
\end{equation}
\begin{align}\label{}
	A^{(\mathrm e)}_{\Ht, 0}&=\!\left(\begin{smallmatrix}
		0 & 0 & 0 & 0 & 0 & 0 & 0 & 0 & 0 & 0 \\
		0 & 0 & 0 & 0 & 0 & 0 & 0 & 0 & 0 & 0 \\
		0 & 0 & 0 & 0 & 0 & 0 & 0 & 0 & 0 & 0 \\
		0 & 0 & 0 & 0 & 0 & 0 & 0 & 0 & 0 & 0 \\
		-\frac{1}{2} & 0 & 0 & 0 & -6 & 0 & -1 & 0 & 0 & 0 \\
		-\frac{3}{2} & 0 & 0 & 0 & 0 & 2 & -2 & 0 & 0 & 0 \\
		0 & 0 & 0 & 0 & 12 & 2 & 0 & 0 & 0 & 0 \\
		0 & 2 & 0 & 0 & 0 & 0 & 0 & 2 & 0 & 0 \\
		2 & -4 & 0 & 0 & 0 & 4 & 2 & 4 & 2 & -2 \\
		-1 & 0 & -1 & 0 & 12 & 8 & 0 & 8 & 2 & -2
	\end{smallmatrix}\right), \qquad 
	A^{(\mathrm e)}_{\Ht,\pm 1}=\!\left(\begin{smallmatrix}
		0 & 0 & 0 & 0 & 0 & 0 & 0 & 0 & 0 & 0 \\
		0 & 0 & 0 & 0 & 0 & 0 & 0 & 0 & 0 & 0 \\
		0 & 0 & 0 & 0 & 0 & 0 & 0 & 0 & 0 & 0 \\
		0 & 0 & 0 & 0 & 0 & 0 & 0 & 0 & 0 & 0 \\
		0 & 0 & 0 & 0 & 6 & 0 & 0 & 0 & 0 & 0 \\
		0 & 0 & 0 & 0 & 0 & -2 & 0 & 0 & 0 & 0 \\
		0 & 0 & 0 & 0 & 0 & 0 & 0 & 0 & 0 & 0 \\
		0 & 0 & 0 & 0 & 0 & 0 & 0 & -2 & 0 & 0 \\
		0 & 0 & 0 & 0 & 0 & -4 & 0 & -4 & -2 & 0 \\
		1 & 0 & 1 & \pm 4 & 0 & 0 & 0 & 0 & 0 & 2
	\end{smallmatrix}\right)\,,\\
A^{(\mathrm o)}_{\Ht, 0}&\!=\!\!\left(\begin{smallmatrix}
	0 & 0 & 0 & 0 & 0 & 0  \\
	0 & 2 & 0 & 0 & 0 & 0 \\
	0 & 0 & 0 & 0 & 0 & 0 \\
	0 & 0 & \frac12 & 0 & 0 & -\frac76 \\
	0 & 0 & 0 & 0 & 0 & 0\\
	0 & 0 & 0 & 0 & 0 & -2 \
\end{smallmatrix}\right)\!\!,\ 
A^{(\mathrm o)}_{\Ht,+ 1}\!=\!\!\left(\begin{smallmatrix}
	0 & 0 & 0 & 0 & 0 & 0 \\
	2 & -2 & 0 & 0 & 0 & 0 \\
	0 & 0 & 0 & 0 & 0 & 0 \\
	\frac{10}{3} & 2 & 0 & 2 & \frac14 & \frac13 \\
	0 & 0 & 0 & 0 & 0 & 0 \\
	0 & 0 & 0 & 0 & 3 & 6 \\
\end{smallmatrix}\right)\!\!,
\
A^{(\mathrm o)}_{\Ht,- 1}\!=\!\!\left(\begin{smallmatrix}
	0 & 0 & 0 & 0 & 0 & 0 \\
	-2 & -2 & 0 & 0 & 0 & 0 \\
	0 & 0 & 0 & 0 & 0 & 0 \\
	-\frac{10}{3} & -2 & -1 & -2 & -\frac14 & 2 \\
	0 & 0 & 0 & 0 & 0 & 0 \\
	0 & 0 & 0 & 0 & -3 & -2 \\
\end{smallmatrix}\right)\!\!.
\end{align}
Let us note that some integrals coincide with those appearing in Section~\ref{ssec:Double box},
\begin{equation}\label{coinc}
	f_{\Ht,1}=f_{\RT,1}\,,\quad
	f_{\Ht,5}=f_{\RT,2}\,,\quad
	f_{\Ht,6}=f_{\RT,3}\,,\quad
	f_{\Ht,7}=f_{\RT,4}\,,
\end{equation}
and
\begin{equation}
	f_{\Ht,15}=f_{\RT,8}\,,\quad f_{\Ht,16}=f_{\RT,9}\,.
\end{equation}
For the near-static boundary conditions of the remaining integrals, we find (see Appendix~\ref{app:SoftBC})
\begin{subequations}
	\begin{align}\label{fH2}
		f_{\Ht,2}&=-\frac{e^{2 \gamma_E  \epsilon }}{(q^2)^{2\epsilon}}\frac{\epsilon ^2 \Gamma (1-2 \epsilon )^2 \Gamma (1-\epsilon ) \Gamma
			(\epsilon ) \Gamma (2 \epsilon )}{2 \Gamma (1-4 \epsilon )}\,,\\
		\label{fH3}
		f_{\Ht,3}&=\frac{e^{2 \gamma_E  \epsilon }}{(q^2)^{2\epsilon}}\left[\frac{\epsilon \Gamma (1-\epsilon ) \Gamma (-\epsilon ) \Gamma (\epsilon +1)}{\Gamma
			(1-2 \epsilon )}\right]^2\,,\\
		\label{fH4}
		f_{\Ht,4}&=\frac{e^{2 \gamma_E  \epsilon }}{(q^2)^{2\epsilon}}\left[\frac{\epsilon^2\sqrt{\pi }\, \Gamma \left(\frac{1}{2}-\epsilon \right)^2 \Gamma \left(\epsilon +\frac{1}{2}\right)}{2 \Gamma (1-2
			\epsilon)}\right]^2\,,\\
		\label{fH8}
		f_{\Ht,8}\big|_{x\to1}&\sim - i\pi \bigg(\frac{e^{i\pi/2}e^{\gamma_E}}{(1-x)q^2}\bigg)^{2\epsilon} \frac{\epsilon \Gamma(1+2 \epsilon) \Gamma(1-2 \epsilon)^{2} \Gamma(1-\epsilon)\Gamma\left(\frac{1}{2}+\epsilon\right)}{2 \Gamma(1-4 \epsilon) \Gamma\left(\frac{1}{2}\right)}\,,\\
		\label{fH9}
		f_{\Ht,9}\big|_{x=1}&=0\,,\\
		\label{fH10}
		f_{\Ht,10}\big|_{x=1}&=\frac{e^{2 \gamma_E  \epsilon }}{(q^2)^{2\epsilon}}\frac{\epsilon}{2} \left[\frac{2 \Gamma (2 \epsilon ) \Gamma (1-\epsilon )^3}{\Gamma (1-3
			\epsilon )}+\pi ^3 16^{\epsilon } \epsilon  \left(\frac{\pi  \sec ^2(\pi  \epsilon )}{\Gamma (-\epsilon
			)^2}-\frac{\csc ^2(\pi  \epsilon )}{\Gamma \left(\frac{1}{2}-\epsilon \right)^2}\right)\right]
	\end{align}
\end{subequations}
and
\begin{subequations}
	\begin{align}
		\label{fH11}
			f_{\Ht,11}&=\epsilon^2(1-2\epsilon)^2\frac{e^{2 \gamma_E\epsilon}}{(q^2)^{2\epsilon}}
			\frac{\Gamma\left(-\frac12+2\epsilon\right)\Gamma\left(\frac32-2\epsilon\right)^2\Gamma\left(1-\epsilon\right)\Gamma\left(-\frac12+\epsilon\right)}{2\Gamma\left(3-4\epsilon\right)}\,,\\
		\begin{split}\label{fH12}
			f_{\Ht,12}	\big|_{x\to1}&
				\sim 
				\epsilon^3\frac{e^{2 \gamma_E\epsilon}}{(q^2)^{2\epsilon}}
				\frac{\Gamma \left(\frac{1}{2}-2 \epsilon \right)^2 \Gamma (-\epsilon ) \Gamma
					\left(\epsilon +\frac{1}{2}\right) \Gamma \left(2 \epsilon +\frac{1}{2}\right)}{4 \Gamma (1-4 \epsilon )}\\
			&+\bigg(\frac{e^{i\pi/2}e^{\gamma_E}}{(1-x)q^2}\bigg)^{2\epsilon} \frac{\sqrt{\pi}\,\epsilon^2\Gamma(1+\epsilon)\Gamma(1-\epsilon)\Gamma\left(\frac{1}{2}+2\epsilon\right)\Gamma\left(\frac{1}{2}-2\epsilon\right)^2}{2\Gamma(1-4\epsilon)}\,.
		\end{split}
	\end{align}
\end{subequations}
We omit the boundary condition for $f_{\Ht,13}$, which we shall not need in the following, while postponing the discussion of the static limit of $f_{\Ht,14}$ to its cut counterpart. For the remainder of this subsection, we shall focus on even master integrals only.

Eqs.~\eqref{fH2}, \eqref{fH3}, \eqref{fH4} and \eqref{fH11} hold for any $x$, and are independent of it. For $f_{\Ht,8}$, $f_{\Ht,9}$ we find instead
\begin{align}\label{}
	f_{\Ht,8}^{(0)}&=0\,,\\
	f_{\Ht,8}^{(1)}&=-\frac{1}{2}\log(-x)\,,\\
	f_{\Ht,8}^{(2)}&=-\frac{1}{2}\log^2(- x)+\log(-x)\log(1-x^2)-\frac{\pi^2}{12}+\frac{1}{2}\operatorname{Li}_2(x^2)
\end{align} 
and
\begin{align}\label{Hf9sol}
	f_{\Ht,9}^{(0)}=0\,,\qquad
	f_{\Ht,9}^{(1)}=0\,,\qquad
	f_{\Ht,9}^{(2)}=\frac{1}{3}\log(-x)\left(
	\pi^2+\log^2(-x)
	\right)\,,
\end{align} 
while 
\begin{equation}\label{}
	f_{\Ht,10}^{(0)}=0\,,\qquad f_{\Ht,10}^{(0)}=0\,,\qquad 
	f_{\Ht,10}^{(3)}=-3\zeta(3)\,.
\end{equation}
Substituting Eq.~\eqref{Hf9sol} into Eq.~\eqref{Hf9}, we thus find
\begin{equation}\label{H}
	I_{\Ht}=-\frac{2z \log z}{3m_1 m_2(1-z^2)\epsilon(q^2)^{2+2\epsilon}}\left(-2\pi^2 + \log^2z+3i\pi\log z\right)+\cdots\,.
\end{equation}
up to subleading orders in $q$. Applying Eq.~\eqref{crossingxsigma} we also obtain the corresponding result for the crossed topology,
\begin{equation}\label{Hbar}
	I_{\overline\Ht}=\frac{2z \log z}{3m_1 m_2(1-z^2)\epsilon(q^2)^{2+2\epsilon}}\left(\pi^2 + \log^2z\right)+\cdots\,.
\end{equation}
Results \eqref{H} and \eqref{Hbar} agree in particular with the exact expressions obtained in Ref.~\cite{Bianchi:2016yiq} for the equal mass case without expanding in $q$. The real part of $I_{\Ht}+I_{\overline\Ht}$ also agrees with the result obtained via velocity resummation of the potential-region expansion \cite{Bern:2019crd,Parra-Martinez:2020dzs}.

\subsection{Cut topologies}
\label{sec:cutto}

In the calculation of inclusive quantities associated to the three-particle cut, which will be discussed in Section~\ref{pspace}, it is useful to consider master integrals where two massive propagators and one massless propagator are replaced with delta functions according to 
\begin{equation}
	(-1)^n\frac{i2\pi}{n!}\, \delta^{(n)}(p^2+m^2) = \frac{1}{(p^2+m^2-i0)^{n+1}}-\frac{1}{(p^2+m^2+i0)^{n+1}}\,,
\end{equation}
the so-called cut master integrals. The advantage of this step is to make differential equations and IBP reduction techniques directly available for the calculation of otherwise complicated phase space integrals \cite{Anastasiou:2002yz,Anastasiou:2002qz,Anastasiou:2003yy,Anastasiou:2015yha,Herrmann:2021lqe}. Indeed, cut integrals can be handled just as the standard integrals discussed in the previous subsections above up to two novelties:
\begin{itemize}
	\item while performing the IBP reduction, integrals where any of the cut propagators cancel out should be dropped at all intermediate steps\footnote{This can be done via the ``CutDs" option in \texttt{LiteRed}, for instance.},
	\item once the corresponding differential equations are obtained, they should be solved imposing appropriate cut boundary conditions.
\end{itemize}
The needed cut boundary conditions can be directly determined via Cutkosky rules, taking twice the imaginary part of the standard boundary conditions, whenever the integral's topology only involves one cut. When more than one cut is present, we select imaginary parts arising from the singular static region discussed in Appendix~\ref{app:SoftBC}.

Let us now present the cut integral families topology by topology. For the cut $\RT$ family, $\cRT$, we choose to cut propagators number 1, 4 and 7, which highlight by underlining the corresponding entries. A basis of master integrals is given by: 
\begin{subequations}
	\begin{align}
		f_{\cRT,2}={}&-\epsilon ^4 \tau\, G_{\underline 1,0,0,\underline 1,0,0,\underline 1,1,1}\,, \label{eq:cfRT2}\\ 
		f_{\cRT,3}={}&\epsilon ^3 q^2 \tau\, G_{\underline 1,0,0,\underline 1,0,0,\underline2,1,1}\,,\\ 
		f_{\cRT,4}={}&\epsilon ^2 q^2 G_{\underline2,0,0,\underline2,0,0,\underline1,1,1} +\epsilon ^3 y \, q^2 G_{\underline1,0,0,\underline1,0,0,\underline2,1,1}\,,\\ 
		f_{\cRT,7}={}&-\epsilon ^4 \tau^2 q^2 G_{\underline1,1,1,\underline1,0,0,\underline1,1,1}\,, 
	\end{align}
\end{subequations}
and 
\begin{subequations}
	\begin{align}
		f_{\mathrm{\cRT,9}}={}&\epsilon^3 q\, G_{\underline2,0,0,\underline1,0,0,\underline1,1,1}\,,\\ 
		f_{\mathrm{\cRT,10}}={}&\epsilon^4\tau \, q\, G_{\underline1,0,1,\underline1,0,0,\underline 1,1,1}\,,\label{eq:cfRT10}\,.
	\end{align}
\end{subequations}
These obey the differential equations
\begin{equation}
	\mathrm{d} \vec {f}_{\cRT}= \epsilon\left[A_{\cRT,0}\,\operatorname{dlog}(x)+A_{\cRT,+1}\,\operatorname{dlog}(x+1)+A_{\cRT,-1}\,\operatorname{dlog}(x-1)\right]\vec{f}_{\cRT}\label{eq:cRTSoftDE}
\end{equation}
with
\begin{align}
	A_{\cRT,i}=\left(
	\begin{matrix}
		A_{\cRT, i}^{(\mathrm{e})} & 0 \\
		0 & A_{\cRT,i}^{(\mathrm{o})} \\
	\end{matrix}
	\right)\,,
\end{align}
\begin{align}\label{eq:cRTmateven}
	A_{\cRT,0}^{(\mathrm{e})}={}&\left(
	\begin{matrix}
		-6 & 0 & -1 & 0 \\[2pt]
	    0 & 2 & -2 & 0 \\
		12 & 2 & 0 & 0 \\
		0 & \tfrac12 & 0 & 0 \\
	\end{matrix}
	\right),\quad 
	A_{\cRT,\pm1}^{(\mathrm{e})} = \left(
	\begin{matrix}
		6 & 0 & 0 & 0 \\
		0 & -2 & 0 & 0 \\
		0 & 0 & 0 & 0 \\
		0 & 0 & 0 & 0 \\
	\end{matrix}
	\right),
\end{align}
\begin{align} \label{eq:cRTmatodd}
	A_{\cRT,0}^{(\mathrm{o})}={}\left(
	\begin{matrix}
		-2 & 0 \\
		1 & 0 \\
	\end{matrix}
	\right),\quad
	A_{\cRT,+1}^{(\mathrm{o})}={}\left(
	\begin{matrix}
		6 & 0 \\
		0 & 0 \\
	\end{matrix}
	\right),\quad
	A_{\cRT,-1}^{(\mathrm{o})}={}\left(
	\begin{matrix}
		-2 & 0 \\
		0 & 0 \\
	\end{matrix}
	\right).
\end{align}
The new equations can be thus obtained from the old ones by  erasing all integrals that do not posses the needed three-particle cut. The only exception to this rule is the differential equation for $f_{\cRT,7}$, an integral with two distinct three-particle cuts, which acquires an extra factor of $\frac12$, as can be seen from the last line of $A_{\cRT,0}^{(e)}$.
The nontrivial boundary conditions read
\begin{subequations}
	\begin{align}
		\label{fcIII3}
		f_{\cRT,3}\big|_{x\to1}& \sim \pi \cos(\pi\epsilon) \bigg(\frac{e^{\gamma_E}}{(1-x)q^2}\bigg)^{2\epsilon} \frac{\epsilon \Gamma(1+2 \epsilon) \Gamma(1-2 \epsilon)^{2} \Gamma(1-\epsilon)\Gamma\left(\frac{1}{2}+\epsilon\right)}{\Gamma(1-4 \epsilon) \Gamma\left(\frac{1}{2}\right)}\,, \\
		\label{fcIII9}
		f_{\cRT,9}\big|_{x\to1}
		&\sim c_{BC}(x)\,,
	\end{align}
\end{subequations}
where
\begin{equation}\label{cBC}
	c_{BC}(x)
	=
	 \sin(\pi\epsilon) \bigg(\frac{e^{\gamma_E}}{(1-x)q^2}\bigg)^{2\epsilon} \frac{\sqrt{\pi}\,\epsilon^2\Gamma(1+\epsilon)\Gamma(1-\epsilon)\Gamma\left(\frac{1}{2}+2\epsilon\right)\Gamma\left(\frac{1}{2}-2\epsilon\right)^2}{\Gamma(1-4\epsilon)}\,.
\end{equation}
Note that the imaginary part in the boundary condition of $f_{\RT,10}$ in eq.~\eqref{fIII10} arises from the two-particle cut (i.e. from the ordinary static region of Appendix~\ref{app:SoftBC}) and must be discarded. The solutions are, for the even sector,
\begin{equation}\label{}
	\begin{array}{l}
		f_{\cRT,2}^{(1)}=0\,,\\[4pt]
		f_{\cRT,3}^{(1)}=\pi\,,\\[4pt]
		f_{\cRT,4}^{(1)}=0\,,\\[4pt]
		f_{\cRT,7}^{(1)}=0\,,
	\end{array}\qquad
	\begin{array}{rl}
		f_{\cRT,2}^{(2)}=&\!\!\!0\,,\\[4pt]
		f_{\cRT,3}^{(2)}=&\!\!\!2\pi\log x-2\pi \log(1-x^2)\\[4pt]
		f_{\cRT,4}^{(2)}=&\!\!\!2\pi\log x\,,\\[4pt]
		f_{\cRT,7}^{(2)}=&\!\!\!\frac{\pi}{2}\log x\,,
	\end{array}
\end{equation}
together with
\begin{equation}\label{}
	\begin{split}
		f_{\cRT,2}^{(3)}&=-\pi  \log ^2(x)\,,\\
		f_{\cRT,3}^{(3)}&=2 \pi  \log ^2\left(1-x^2\right)-4 \pi  \log (x) \log \left(1-x^2\right)-\frac{\pi ^3}{2}\,,\\
		f_{\cRT,4}^{(3)}&=2 \pi  \text{Li}_2\left(x^2\right)+2 \pi  \log ^2(x)-\frac{\pi ^3}{3}\,,\\
		f_{\cRT,7}^{(3)}&=\frac{\pi  \text{Li}_2\left(x^2\right)}{2}+\frac{1}{2} \pi  \log ^2(x)-\frac{\pi ^3}{12}\,,
	\end{split}
\end{equation}
and
\begin{equation}\label{}
	\begin{split}\label{}
		f_{\cRT,4}^{(4)}&=-2 \pi  \text{Li}_3\left(x^2\right)-4 \pi  \text{Li}_3\left(1-x^2\right)+4 \pi  \text{Li}_2\left(1-x^2\right) \log \left(\frac{1}{x}-x\right)\\
		&-4 \pi  \log \left(\frac{x^2}{1-x^2}\right) \log \left(1-x^2\right) \log (x)-4 \pi  \log ^3(x)-\frac{1}{3} \pi ^3 \log (x)+2 \pi  \zeta (3)\,,\\
		f_{\cRT,7}^{(4)}&=-\frac{\pi  \text{Li}_3\left(x^2\right)}{2}-\pi  \text{Li}_3\left(1-x^2\right)+\pi  \text{Li}_2\left(1-x^2\right) \log \left(\frac{1}{x}-x\right)\\
		&-\pi  \log \left(\frac{x^2}{1-x^2}\right) \log \left(1-x^2\right) \log (x)-\frac{1}{12} \pi ^3 \log (x)+\frac{\pi  \zeta (3)}{2}\,.
	\end{split}
\end{equation}
For the odd sector,
\begin{equation}\label{}
	\begin{array}{l}
		f_{\cRT,9}^{(3)}=\pi ^3\,,\\[4pt]
		f_{\cRT,10}^{(3)}=0\,,\\
	\end{array}\qquad
	\begin{array}{l}
		f_{\cRT,9}^{(4)}=-2 \pi ^3 \log \left(\frac{1-x}{4}\right)-2 \pi ^3 \log (x)+6 \pi ^3 \log \left(\frac{x+1}{2}\right)\,,\\[4pt]
		f_{\cRT,10}^{(4)}=\pi ^3 \log (x)\,.\\
	\end{array}
\end{equation}

For the cut $\RN$ family, $\cRN$, we consider,
\begin{subequations}
	\begin{align}
		f_{\cRN,5}={}&-\epsilon ^4 \tau\, \tilde G_{\underline 1,0,0,\underline 1,1,\underline1,1,0,0}\,,\\ 
		f_{\cRN,6}={}&\epsilon ^3 q^2 \tau\, \tilde G_{\underline1,0,0\underline1,1,\underline2,1,0,0}\,,\\ 
		f_{\cRN,7}={}&-\epsilon ^2 q^2 \tilde G_{\underline2,0,0,\underline2,1,\underline1,1,0,0} -\epsilon ^3 q^2 y \, \tilde G_{\underline1,0,0\underline1,1,\underline2,1,0,0}\,,\\
		f_{\cRN,10}={}&-\epsilon ^4 q^2  \tau^2 \tilde G_{\underline1,1,1\underline1,1,\underline1,1,0,0}\, ,
	\end{align}
\end{subequations}
and
\begin{subequations}
	\begin{align}
		f_{\cRN,12}={}&\epsilon ^3 q\, \tilde G_{\underline2,0,0,\underline1,1,\underline1,1,0,0}\,,\\ 
		f_{\cRN,14}={}&\epsilon ^4 q\, \tau \,\tilde G_{\underline1,0,1,\underline 1,1,\underline1,1,0,0}\,,\\ 
		f_{\cRN,15}={}&\epsilon ^4 q\,\tau \,\tilde G_{\underline1,1,0,\underline1,1,\underline1,1,0,0}\,,
	\end{align}
\end{subequations}
which satisfy the differential equations
\begin{equation}
	\mathrm{d}\vec{f}_\cRN=\epsilon\left[A_{\cRN,0}\,\operatorname{dlog}(x)+A_{\cRN,+1}\,\operatorname{dlog}(x+1)+A_{\cRN,-1}\,\operatorname{dlog}(x-1)\right]\vec{f}_\cRN\label{eq:cXB2loopSoftDE}
\end{equation}
with
\begin{align}
	A_{\cRN,i}=\left(
	\begin{array}{cc}
		A_{\cRN,i}^{(\mathrm{e})} & 0 \\
		0 & A_{\cRN,i}^{(\mathrm{o})} \\
	\end{array}
	\right)
\end{align}
and
\begin{align} \label{eq:cIXmateven}
	A_{\cRN,0}^{(\mathrm{e})}={}&
	\left(
	\begin{matrix}
		-6 & 0 & 1 & 0 \\
		0 & 2 & 2 & 0 \\
		-12 & -2 & 0 & 0 \\
		0 & -\frac12 & 0 & 0 \\
	\end{matrix}
	\right),\quad
	A_{\cRN, \pm1}^{(\mathrm{e})}={}\left(
	\begin{matrix}
		6 & 0 & 0 & 0 \\
		0 & -2 & 0 & 0 \\
		0 & 0 & 0 & 0 \\
		0 & 0 & 0 & 0 \\
	\end{matrix}
	\right),
\end{align}
\begin{align} \label{eq:cIXmatodd}
	A_{\cRN,0}^{(\mathrm{o})}=\left(
	\begin{matrix}
		-2 & 0 & 0 \\
		-1 & 0 & 0 \\
		1 & 0 & 0 \\
	\end{matrix}
	\right),\
	A_{\cRN,+1}^{(\mathrm{o})}=\left(
	\begin{matrix}
		6 & 0 & 0 \\
		0 & 0 & 0 \\
		0 & 0 & 0 \\
	\end{matrix}
	\right),\
	A_{\cRN,-1}^{(\mathrm{o})}=\left(
	\begin{matrix}
		-2 & 0 & 0 \\
		0 & 0 & 0 \\
		0 & 0 & 0 \\
	\end{matrix}
	\right).
\end{align}
The boundary conditions can be determined via \eqref{III-IXrelations}, which also hold for cut integrals.
Since this specific family does not enter any of the calculations discussed below, we refrain from displaying the corresponding solutions explicitly.  

We can encompass the three-particle cut of the $\overline\RN$ topology by using crossing and cutting a different set of propagators in the $\RN$ family, as follows:
\begin{subequations}
	\begin{align}
		f_{\cRNbar,2}={}&-\epsilon^4\tau\, \tilde G_{0,0,\underline1,\underline1,\underline1,1,1,0,0}\,,\\ 
		f_{\cRNbar,3}={}&\epsilon^3 q^2 \tau\, \tilde G_{0,0,\underline1,\underline1,\underline2,1,1,0,0}\,,\\ 
		f_{\cRNbar,4}={}&-\epsilon^2 q^2 \tilde G_{0,0,\underline2,\underline2,\underline1,1,1,0,0} + \epsilon ^3 q^2 y \,\tilde G_{0,0,\underline1,\underline1,\underline2,1,1,0,0}\,,\\ 
		f_{\cRNbar,10}={}&-\epsilon ^4 q^2  \tau^2 \tilde G_{1,1,\underline1,\underline1,\underline1,1,1,0,0}\, ,
	\end{align}
\end{subequations}
together with
\begin{subequations}
	\begin{align}
		f_{\cRNbar,13}={}&\epsilon ^3 q\,\tilde G_{0,0,\underline2,\underline1,\underline1,1,1,0,0}\,,\\ 
		f_{\cRNbar,14}={}&\epsilon ^4 q\, \tau \,\tilde G_{0,1,\underline1,\underline1,\underline1,1,1,0,0}\,.
	\end{align}
\end{subequations}
These master integrals obey
\begin{equation}
	\mathrm{d}\vec{f}_{\cRNbar}=\epsilon\left[A_{\cRNbar,0}\,\operatorname{dlog}(x)+A_{\cRNbar,+1}\,\operatorname{dlog}(x+1)+A_{\cRNbar,-1}\,\operatorname{dlog}(x-1)\right]\vec{f}_{\cRNbar},\label{eq:cXBbar2loopSoftDE}
\end{equation}
with
\begin{align}
	A_{\cRNbar,i}=\left(
	\begin{array}{cc}
		A_{\cRNbar,i}^{(\mathrm{e})} & 0 \\
		0 & A_{\cRNbar,i}^{(\mathrm{o})} \\
	\end{array}
	\right)\,,
\end{align}
and
\begin{align} \label{eq:cIXbarmateven}
	A_{\cRNbar,0}^{(\mathrm{e})}={}&
	\left(
	\begin{matrix}
		-6 & 0 & -1 & 0 \\
		0 & 2 & -2 & 0 \\
		12 & 2 & 0 & 0 \\
		0 & -\frac12 & 0 & 0 \\
	\end{matrix}
	\right),\qquad
	A_{\cRNbar, \pm1}^{(\mathrm{e})}={}\left(
	\begin{matrix}
	6 & 0 & 0 & 0 \\
	0 & -2 & 0 & 0 \\
	0 & 0 & 0 & 0 \\
	0 & 0 & 0 & 0 \\
	\end{matrix}
	\right),
\end{align}
\begin{align} \label{eq:cIXbarmatodd}
	A_{\cRNbar,0}^{(\mathrm{o})}=\left(
	\begin{matrix}
		-2 & 0 \\
		1 & 0\\
	\end{matrix}
	\right),\qquad
	A_{\cRNbar,+1}^{(\mathrm{o})}=\left(
	\begin{matrix}
		-2 & 0 \\
		0 & 0 \\
	\end{matrix}
	\right),\qquad
	A_{\cRNbar,-1}^{(\mathrm{o})}=\left(
	\begin{matrix}
		6 & 0 \\
		0 & 0 \\
	\end{matrix}
	\right).
\end{align}
For this family, using the dictionary provided by \eqref{III-IXrelations}, the solutions of the differential equations can be obtained from those of the $\cRT$ family using crossing symmetry
\begin{equation}\label{}
\begin{array}{l}
	f_{\cRNbar,2} \leftrightarrow f_{\cRT,2}\,,\\[4pt]
	f_{\cRNbar,3} \leftrightarrow f_{\cRT,3}\,,\\[4pt]
	f_{\cRNbar,4} \leftrightarrow -f_{\cRT,4}\,,\\[4pt]
	f_{\cRNbar,13} \leftrightarrow f_{\cRT,9}\,.
\end{array}
\end{equation}
The only new integrals are thus $f_{\cRNbar,10}$, $f_{\cRNbar,14}$, whose static boundary conditions are zero.
Since they always enter the calculations discussed below via crossing symmetry, it is natural to display their crossed expressions directly:
\begin{equation}\label{}
\begin{split}
		f_{\cRNbar,10}^{(2)} \leftrightarrow \frac{\pi}{2}\,\log(x)\,,\qquad 
	f_{\cRNbar,10}^{(3)} \leftrightarrow
	\frac{\pi  \text{Li}_2\left(x^2\right)}{2}+\frac{1}{2} \pi  \log ^2(x)-\frac{\pi ^3}{12}
\end{split}
\end{equation}
and
\begin{equation}
	f_{\cRNbar,14}^{(4)} \leftrightarrow
	-\pi ^3 \log (x)\,.
\end{equation}

Finally, for the cut $\Ht$ topology, $\cHt$, we choose
\begin{subequations}
	\begin{align}
		f_{\cHt, 5}&=-\epsilon^{4} \tau \ G_{0,\underline1,\underline1,0,0,0,\underline1,1,1} \\
		f_{\cHt, 6}&=\epsilon^{3} \tau\, q^2 G_{0,\underline1,\underline1,0,0,0,\underline2,1,1} \\
		f_{\cHt, 7}&=\epsilon^{2} q^2 G_{0,\underline2,\underline2,0,0,0,\underline1,1,1}+\epsilon^{3} y\,q^2 G_{0,\underline1,\underline1,0,0,0,\underline2,1,1} \\
		\label{f8cHdef}
		f_{\cHt, 8}&=-\frac1\tau\, \epsilon^{2}(4 \epsilon-1)(2 \epsilon-1) G_{0,\underline1,\underline1,0,0,1,\underline1,0,1} \\
		f_{\cHt, 9}&=-\epsilon^{4} \tau q^4 G_{0,\underline1,\underline1,0,1,1,\underline1,1,1} \\
		\label{f10cHdef}
			f_{\cHt, 10}&=\epsilon^{4} q^2 G_{-1,\underline1,\underline1,-1,1,1,\underline1,1,1} 
	\end{align}
\end{subequations}
and
\begin{subequations}
	\begin{align}
		f_{\cHt, 12}&=\epsilon^3 q\, G_{0,\underline2,\underline1,0,0,1,\underline1,0,1}\\
		\begin{split}\label{fcH14}
			f_{\cHt, 14}&=
			\frac18 (y-1)\epsilon^3 q^5 G_{0,\underline2,\underline1,0,1,1,\underline1,1,1}\\
			&+\frac{\epsilon^3(1-2\epsilon(2+3y))}{12(1+2\epsilon)}\,q\, G_{0,\underline2,\underline1,0,0,0,\underline1,1,1}
			+\frac{2\epsilon^4}{(1+2\epsilon)(1+y)}\,q\, G_{0,\underline2,\underline1,0,0,1,\underline1,0,1}
		\end{split}\\
		f_{\mathrm{\cHt,16}}&=\epsilon^3 q\, G_{0,\underline2,\underline1,0,0,0,\underline1,1,1}\,.
	\end{align}
\end{subequations}
The differential equations read in this case
\begin{equation}\label{cHde}
	\mathrm{d} \vec{f}_{\cHt}=\epsilon\left[A_{\cHt, 0} \operatorname{dlog}(x)+A_{\cHt,+1} \operatorname{dlog}(x+1)+A_{\cHt,-1} \operatorname{dlog}(x-1)\right] \vec{f}_{\cHt}\,,
\end{equation}
with
\begin{equation}\label{}
	A_{\cHt,i}=\left(
	\begin{array}{cc}
		A_{\cHt,i}^{(\mathrm{e})} & 0 \\
		0 & A_{\cHt,i}^{(\mathrm{o})} \\
	\end{array}
	\right)\,,
\end{equation}
\begin{align}\label{}
	A^{(\mathrm e)}_{\cHt, 0}&=\!\left(\begin{matrix}
		-6 & 0 & -1 & 0 & 0 & 0 \\
		0 & 2 & -2 & 0 & 0 & 0 \\
		12 & 2 & 0 & 0 & 0 & 0 \\
		0 & 0 & 0 & 2 & 0 & 0 \\
		0 & 4 & 2 & 4 & 2 & -2 \\
		12 & 8 & 0 & 8 & 2 & -2
	\end{matrix}\right), \qquad 
	A^{(\mathrm e)}_{\cHt,\pm 1}=\!\left(\begin{matrix}
		6 & 0 & 0 & 0 & 0 & 0 \\
		0 & -2 & 0 & 0 & 0 & 0 \\
		0 & 0 & 0 & 0 & 0 & 0 \\
		0 & 0 & 0 & -2 & 0 & 0 \\
		0 & -4 & 0 & -4 & -2 & 0 \\
		0 & 0 & 0 & 0 & 0 & 2
	\end{matrix}\right)\,,\\
	A^{(\mathrm o)}_{\cHt, 0}&=\!\left(\begin{matrix}
		2 & 0 & 0 \\
		0 & 0 & -\frac76 \\
		0 & 0 & -2 
	\end{matrix}\right),\quad
	A^{(\mathrm o)}_{\cHt,+ 1}=\!\left(\begin{matrix}
		-2 & 0 & 0 \\
		2 & 2 & \frac13 \\
		0 & 0 & 6 \\
	\end{matrix}\right),
	\quad
	A^{(\mathrm o)}_{\cHt,- 1}=\!\left(\begin{matrix}
		-2 & 0 & 0 \\
		-2 & -2 & 2 \\
		0 & 0 & -2 
	\end{matrix}\right)\,.
\end{align}
The cut boundary conditions are given by twice the imaginary part of the corresponding  standard ones discussed in the previous subsections. The only integral that requires a separate analysis is $f_{\cHt,14}$, for which one can check using the tools developed in Appendix~\ref{app:SoftBC} that the contribution coming from the first line of \eqref{fcH14} is suppressed in the static limit due to the factor of $y-1$.
Consequently, as $x\to1^-$,
\begin{equation}
	f_{\cHt,12}\sim f_{\cHt,16}\sim c_{BC}(x)\,,\qquad
	f_{\cHt,14}\sim \frac{1}{12}\, c_{BC}(x)
\end{equation}
where $c_{BC}(x)$ is defined in eq.~\eqref{cBC}.
The corresponding solutions are, for the even sector,
\begin{equation}\label{}
	\begin{array}{l}
		f_{\cHt,5}^{(1)}=0\,,\\[4pt]
		f_{\cHt,6}^{(1)}=\pi\,,\\[4pt]
		f_{\cHt,7}^{(1)}=0\,,\\[4pt]
		f_{\cHt,8}^{(1)}=-\pi\,,\\[4pt]
		f_{\cHt,9}^{(1)}=0\,,\\[4pt]
		f_{\cHt,10}^{(1)}=0\,,
	\end{array}\qquad
		\begin{array}{l}
		f_{\cHt,5}^{(2)}=0\,,\\[4pt]
		f_{\cHt,6}^{(2)}=2 \pi  \log (x)-2 \pi  \log \left(1-x^2\right)\,,\\[4pt]
		f_{\cHt,7}^{(2)}=2 \pi  \log (x)\,,\\[4pt]
		f_{\cHt,8}^{(2)}=2 \pi  \log \left(1-x^2\right)-2 \pi  \log (x)\,,\\[4pt]
		f_{\cHt,9}^{(2)}=0\,,\\[4pt]
		f_{\cHt,10}^{(2)}=0\,,
	\end{array}
\end{equation}
together with
\begin{equation}
	\begin{split}
			f_{\cHt,5}^{(3)}&=-\pi  \log ^2(x)\,,\\
		f_{\cHt,6}^{(3)}&=2 \pi  \log ^2\left(1-x^2\right)-4 \pi  \log (x) \log \left(1-x^2\right)-\frac{\pi ^3}{2}\,,\\
		f_{\cHt,7}^{(3)}&=2 \pi  \text{Li}_2\left(x^2\right)+2 \pi  \log ^2(x)-\frac{\pi ^3}{3}\,,\\
		f_{\cHt,8}^{(3)}&=-2 \pi  \log ^2\left(1-x^2\right)+4 \pi  \log \left(1-x^2\right) \log (x)-2 \pi  \log ^2(x)+\frac{\pi ^3}{2}\,,\\
		f_{\cHt,9}^{(3)}&=2 \pi  \log ^2(x)\,,\\
		f_{\cHt,10}^{(3)}&=0\,.
	\end{split}
\end{equation}
For the odd sector, instead,
\begin{equation}\label{}
	\begin{split}
	f_{\cHt,12}^{(3)}=\pi ^3\,,\qquad
	f_{\cHt,14}^{(3)}=\frac{\pi ^3}{12}\,,\qquad
	f_{\cHt,16}^{(3)}=\pi ^3\,,
	\end{split}
\end{equation}
and
\begin{equation}\label{}
	\begin{split}
		f_{\cHt,12}^{(4)}&=-2 \pi ^3 \log \left(\frac{1-x}{4}\right)+2 \pi ^3 \log (x)-2 \pi ^3 \log \left(\frac{x+1}{2}\right)\,,\\
		f_{\cHt,14}^{(4)}&=-\frac{1}{6} \pi ^3 \log \left(\frac{1-x}{4}\right)-\frac{7}{6} \pi ^3 \log (x)+\frac{5}{2} \pi ^3 \log \left(\frac{x+1}{2}\right)\,,\\
		f_{\cHt,16}^{(4)}&=-2 \pi ^3 \log \left(\frac{1-x}{4}\right)-2 \pi ^3 \log (x)+6 \pi ^3 \log \left(\frac{x+1}{2}\right)\,.
	\end{split}
\end{equation}

\section{The ${\cal N}=8$  Eikonal up to 3PM Order}
\label{sec:eikonal3PM}

In this section we combine the above integrals to obtain the scattering amplitude and to reconstruct the eikonal as in Eq.~\eqref{eikonalgen}. After discussing for completeness tree and one-loop level in Subsection~\ref{ssec:tree1L}, we then turn to the two-loop (3PM) level in Subsections~\ref{ssec:tree1L} and \ref{ssec:2L}. 

\subsection{Tree and one loop}
\label{ssec:tree1L}

We consider the elastic $2\to 2$ scattering in ${\cal{N}}=8$ supergravity where the external states are massive thanks to a non-trivial Kaluza-Klein momentum in the compact directions~\cite{Caron-Huot:2018ape,Parra-Martinez:2020dzs}. For later convenience we focus on a $s-u$ symmetric process where the states labelled by $p_1$ and $p_4$ in Fig.~\ref{2-to-2} are dilatons with KK momentum in one compact direction and those labelled by $p_2$ and $p_3$ are axions with KK momentum in another orthogonal compact direction. Then the tree-level amplitude reads
\begin{align}
  \mathcal{A}_0 (s, q^2) = & ~ {8\pi G}  \frac{(s-m_1^2-m_2^2)^4+(u-m_1^2-m_2^2)^4-t^4}{2(s-m_1^2-m_2^2)\, t\, (u-m_1^2-m_2^2)}  \nonumber \\ \label{T1} = &~  
  {4\pi G} \left( \frac{(s-m_1^2 -m_2^2)^2 + (u-m_1^2-m_2^2)^2}{-t}\right. \\ \nonumber & ~ \left. +  \frac{ t^2 - (u-m_1^2-m_2^2)^2}{(s-m_1^2 -m_2^2)}  +  \frac{ t^2 - (s-m_1^2-m_2^2)^2}{(u-m_1^2 -m_2^2)} \right)
\end{align}
in terms of the Mandelstam invariants~\eqref{eq:mandelstam}.

From the scattering amplitude in Eq.~\eqref{T1} we can get the leading eikonal by going to impact parameter space, following the general strategy outlined in Section~\ref{ssec:eikonalgen} and using in particular Eq.~\eqref{impparsp}.
The tree-level amplitude $\mathcal{A}_0$ in the classical limit and the eikonal phase $2\delta_0$ can be thus cast in the form
\begin{eqnarray}
\mathcal{A}_0 =   \frac{8 \pi m_1^2 m_2^2 G (1+ z^2)^2}{q^2 z^2 } =\frac{32\pi G m_1^2 m_2^2 \sigma^2}{q^2}
\label{T6}
\end{eqnarray}
and
\begin{eqnarray}
2\delta_0 =  \frac{m_1m_2 G (\pi b^2)^{\epsilon}\Gamma (-\epsilon) (1+z^2)^2}{z (1-z^2)} =G m_1 m_2 (\pi b^2)^{\epsilon} \Gamma (-\epsilon)  \frac{2\sigma^2}{\sqrt{\sigma^2-1}}
\label{T6a}
\end{eqnarray}
The one-loop amplitude is instead equal to 
\begin{eqnarray}
\mathcal{A}_1 (s, q^2) = (8\pi G)^2 \frac{c(\epsilon)}{2} \Big( (s-m_1^2 -m_2^2)^4 + (u-m_1^2-m_2^2)^4 - q^8\Big)\left( I_{\mathrm{II}} + I_{\overline{\mathrm{II}}}\right) \,,
\label{1L1}
\end{eqnarray}
where
\begin{equation}\label{pre1loop}
c(\epsilon) \equiv
 \frac{e^{-\gamma_E\epsilon}}{(4\pi)^{2-\epsilon}}\,.
\end{equation}
Inserting the results of Section~\ref{ssec:Box} in Eq.~\eqref{1L1} we get
\begin{equation}
\begin{split}
& \frac{\mathcal{A}_1 (s , q^2)}{4pE} = \frac{(8\pi G)^2}{(4\pi)^2}\left( \frac{4\pi}{q^2}\right)^{\epsilon} \Bigg\{ \frac{i\pi m_1^2 m_2^2}{2q^2} \left(\frac{1+z^2}{z}\right)^2 \left( \frac{1+z^2}{1-z^2}\right)^2 \frac{\Gamma^2 (-\epsilon) \Gamma (1+\epsilon)}{\Gamma (-2\epsilon)}  \\
&+ \frac{\sqrt{\pi} m_1m_2 (m_1+m_2)}{\sqrt{q^2}} \left(\frac{1+z^2}{z}\right)\left( \frac{1+z^2}{1-z^2}\right)^3
\frac{\Gamma(\epsilon+\frac{1}{2}) \Gamma^2 (\frac{1}{2} -\epsilon)}{\Gamma (-2\epsilon)} \\
& -  \left( \frac{1+z^2}{1-z^2}\right)^3  \frac{\Gamma^2 (-\epsilon) \Gamma (1+\epsilon)}{\Gamma (-2\epsilon)} \Bigg[ m_1m_2 (1+2\epsilon)\frac{1+z^2}{2z}  \left[ 1 + \frac{1+z^2}{1-z^2}  \log z \right]  \\
& + i \frac{\pi \epsilon}{2}  (m_1^2+m_2^2) \frac{1+z^2}{1-z^2} +i \pi m_1 m_2 \frac{1-z^2}{z} + i \pi \epsilon m_1m_2 \frac{1+z^2}{2z}\frac{1+z^2}{1-z^2} \Bigg]  \Bigg\}
\label{1L9}
\end{split}
\end{equation}
or equivalently in terms of $\sigma$, using the  relations collected in Section~\ref{ssec:conventions},
\begin{align}
&\frac{\mathcal{A}_1 (s , q^2)}{4pE} = \frac{(8\pi G)^2}{(4\pi)^2}\left( \frac{4\pi}{q^2}\right)^{\epsilon} \Bigg\{   \frac{ 2 i \pi m_1^2 m_2^2}{q^2} \frac{\sigma^4}{\sigma^2-1} \frac{\Gamma (1+\epsilon) \Gamma^2 (-\epsilon)}{\Gamma (-2\epsilon)}  \nonumber \\
+ & \frac{2\sqrt{\pi} m_1 m_2 (m_1+m_2)}{\sqrt{q^2}}  \frac{\sigma^4}{(\sigma^2-1)^{\frac{3}{2}}}\frac{\Gamma ( \epsilon+ \frac{1}{2}) \Gamma^2 ( \frac{1}{2}-\epsilon)}{\Gamma (-2\epsilon)}-  \frac{\sigma^3}{(\sigma^2-1)^2}  \frac{\Gamma^2 (-\epsilon) \Gamma (1+\epsilon)}{\Gamma(-2\epsilon)} 
\label{A10a} \\ 
 \times & \Bigg[ m_1m_2 \left[(1+2\epsilon)\left(\sigma^2 \log z +\sigma \sqrt{\sigma^2-1} \right)+i\pi \left(2(\sigma^2-1)+\epsilon \sigma^2 \right) \right] 
  + i\frac{\pi \epsilon}{2} (m_1^2+m_2^2)\sigma \Bigg]\Bigg\}  \,.
\nonumber
\end{align}

The term in the first line of the two previous equations, that we call $\mathcal{A}_1^{(1)}$, is the super-classical term. Its Fourier transform in impact parameter space reproduces the quadratic term of the expansion of the leading eikonal:
\begin{equation}
i{\tilde{\mathcal A}}_1^{(1)} (s, b)= \frac{1}{2} \left( \frac{2 \ii m_1m_2 G (\pi b^2)^{\epsilon}\sigma^2 \Gamma (-\epsilon) }{ \sqrt{\sigma^2-1}}\right)^2= \frac{1}{2} (2 \ii \delta_0)^2\,\,.
\label{A11}
\end{equation}
 The Fourier transform of the term in the second line of Eqs. \eqref{1L9} and \eqref{A10a}, that we call $\mathcal{A}_1^{(2)}$, gives the sub-leading eikonal:
 \begin{eqnarray}
&&\ii {\tilde{\mathcal A}}_1^{(2)} = 2 \ii \delta_1
\nonumber \\
&&
 = \frac{4(\pi b^2)^{2\epsilon} m_1m_2 (m_1+m_2) G^2  }{ \sqrt{\pi}b}
\left[\frac{1+z^2}{z} \left( \frac{1+z^2}{1-z^2}\right)^3 \right] 
\frac{\Gamma ( \frac{1}{2} - 2\epsilon) \Gamma^2 ( \frac{1}{2} -\epsilon)}{ \Gamma (-2 \epsilon)}
\nonumber \\
&&
\nonumber \\
&& = \frac{8 \ii (\pi b^2)^{2\epsilon} G^2 m_1m_2 (m_1+m_2)}{\sqrt{\pi} \sqrt{b^2}}
\frac{\sigma^4}{(\sigma^2-1)^{\frac{3}{2}}} \frac{\Gamma ( \frac{1}{2} - 2\epsilon) \Gamma^2 ( \frac{1}{2} -\epsilon)}{ \Gamma (-2 \epsilon)}\,.
\label{A12}
\end{eqnarray}
Finally, going to impact parameter with the quantum term one gets the quantum part of the eikonal that has a real part given by
\begin{eqnarray}
&&2 \operatorname{Re} \Delta_1 = \frac{4 m_1m_2 G^2 (\pi b^2)^{2\epsilon}}{\pi b^2}  \left(\frac{1+z^2}{1-z^2}\right)^3    \left[ \frac{1+z^2}{z}  + \frac{(1+z^2)^2}{z (1-z^2) } \log z \right] (1+2\epsilon)  \Gamma^2 (1-\epsilon) \nonumber \\
&& =\frac{8 G^2 m_1 m_2 (\pi b^2)^{2\epsilon}}{\pi b^2}
\frac{\sigma^4  \left(\sigma \log z +  \sqrt{\sigma^2-1} \right) }{(\sigma^2-1)^2}  
(1+2\epsilon) \Gamma^2 (1-\epsilon)\, ,
\label{1L7}
\end{eqnarray}
and an imaginary part given by
\begin{eqnarray}
\label{1L71}
&&2\operatorname{Im} \Delta_1 =\frac{4  (\pi b^2)^{2\epsilon} G^2}{b^2} \left( \frac{1+z^2}{1-z^2}\right)^3   \nonumber \\
&& \times   \Bigg[ \epsilon (m_1^2 +m_2^2) \frac{1+z^2}{1-z^2}  
+ 2m_1 m_2 \frac{1-z^2}{z} + \epsilon m_1m_2 \frac{(1+z^2)^2}{z(1-z^2)} \Bigg]\Gamma^2 (1-\epsilon)  \\
&& = \frac{8 G^2 (\pi b^2)^{2\epsilon} \Gamma^2 (1-\epsilon)}{ b^2}\frac{\sigma^3}{(\sigma^2-1)^2} 
\Bigg[ \frac{ \epsilon}{2} (m_1^2+m_2^2)\sigma +  2m_1 m_2 (\sigma^2-1)+\epsilon m_1 m_2 \sigma^2  \Bigg] \,. \nonumber
\end{eqnarray}

\subsection{Two loops}
\label{ssec:2L}

We combine here the results of Sections~\ref{ssec:Double box}, \ref{ssec:Non-planar double box} and \ref{ssec:H} to give the expansion of the two-loop amplitude in powers of $q^2$ and $\epsilon$. As already emphasized, we restrict to the terms that are non-analytic in $q^2$, which are the only ones relevant for the long-range behaviour of the eikonal in $ b$-space.
Furthermore, we will neglect terms of order $\mathcal O(q^2)^{\frac12-2\epsilon}$ or smaller, which are irrelevant for determining $2\delta_2$ and would only enter the calculation of the quantum remainder $2\Delta_2$.
To this effect, the relevant terms of the $s-u$ symmetric two-loop scattering amplitude~\cite{Herrmann:2021lqe} are given by:
\begin{eqnarray}\label{CR1}
&&\mathcal{A}_2 (s, q^2) = (8\pi G)^3 \frac{c(\epsilon)^2}{2} \left[ (s-m_1^2-m_2^2)^4 + (u-m_1^2-m_2^2)^4 \right]
\\
&& \times \Bigg[(s-m_1^2-m_2^2)^2 \left( I_{\RT} +2 I_{\RN}\right) + (u-m_1^2-m_2^2)^2 \left( I_{\overline{\RT}} +2I_{\overline{\RN}} \right) +t^2 (I_{\Ht}+ I_{{\overline{\Ht}}} )\Bigg]\,,
\nonumber
\end{eqnarray}
with $c(\epsilon)$ as in Eq.~\eqref{pre1loop}.
Inserting the previously computed diagrams we get the following result:
\begin{eqnarray}
&&\mathcal{A}_2 (s, q^2) = \frac{(8\pi G)^3}{(4\pi)^4} \left( \frac{4\pi {\rm e}^{-\epsilon \gamma_E}}{q^2} \right)^{2\epsilon} 
\Bigg\{ - \frac{2\pi^2 m_1^4 m_2^4}{\epsilon^2 q^2}\left[ \left(\frac{1+z^2}{1-z^2}\right)^2\left( \frac{1+z^2}{z}\right)^4 \right] \nonumber \\
&& + \frac{4\pi m_1^3 m_2^3 }{\epsilon^2} \frac{(1+z^2)^5}{(1-z^2)^4 z^3}\left[ \pi (1-z^2)^2+i \left( -1-4z^2 \log z +z^4\right)\right] \nonumber \\
&& - \frac{i \pi m_1^2 m_2^2}{3\epsilon} \left[ \left( \frac{1+z^2}{1-z^2}\right)^4 \frac{1}{z^3}\right] \Bigg[ 6i\pi (m_1^2+m_2^2) z(1+z^2)^2 \nonumber \\
&& + 6m_1m_2 \Big[ 4(1+z^2)^2 (1-z^2)+ 2(i \pi +2 \log z)  (1+z^2)^2  \nonumber \\
&& + \left(\frac{\pi^2}{6} - \operatorname{Li}_2 (z^2) \right) (z^2+1)(z^4 -6z^2 +1)
 - 2 (1+z^2)^2 (1-z^2) \log (1-z^2) \nonumber \\
&& +2 ( i\pi \log z +\log^2 z) \left(z^2 +4z^4 - z^6\right) \Big] \Bigg] \Bigg\}  \,.
\label{CR6}
\end{eqnarray}
The above result can be organized in the following convenient form: 
\begin{equation}\label{}
\mathcal{A}_2 = (8\pi G)^3 \frac{c(\epsilon)^2}{(q^2)^{2\epsilon}}
\left[
\frac{\mathcal A_2^{(2,2)}}{\epsilon^2 q^2}
+
\frac{\mathcal A_2^{(2,0)}}{\epsilon^2}
+
\frac{\mathcal A_2^{(1,0)}}{\epsilon}
+\cdots
\right]
\end{equation}
up to terms that are further suppressed in $q^2$, i.e. $\mathcal O((q^2)^{\frac12-2\epsilon})$ or smaller, or that are subleading in $\epsilon$. Note in particular that the coefficient of $\left((q^2)^{1+2\epsilon}\epsilon\right)^{-1}$ vanishes identically. 
A term of $\mathcal{O}(\epsilon^0 (q^2)^{-1-2\epsilon})$ is also needed in order to fully reproduce the ${\cal O}(\delta_0^3)$ exponentiation and will not be discussed further. Similarly, $\mathcal O((q^2)^{-\frac{1}{2}-2\epsilon})$ terms start contributing to order $\mathcal O(\epsilon^0)$, although they are omitted for simplicity. We checked that they reproduce the $\mathcal O(\delta_0\delta_1)$ exponentiation to leading order.
Explicitly, in terms of the variable $z$ and $\sigma$ defined in Section~\ref{ssec:conventions},
the above coefficients read
\begin{align}
\label{A22}
	\mathcal A_2^{(2,2)}
&=-\frac{2 \pi ^2 m_1^4 m_2^4 \left(z^2+1\right){}^6}{z^4 \left(z^2-1\right){}^2}  = -\frac{32 \pi ^2 m_1^4 m_2^4 \sigma^6}{(\sigma^2-1)}\,,\\
\mathcal A_2^{(2,0)}
&=
\frac{4 \pi ^2 m_1^3 m_2^3 \left(z^2+1\right){}^5}{z^3
	\left(z^2-1\right){}^2}	
+
i
\,
\frac{4 \pi  m_1^3 m_2^3 \left(z^2+1\right){}^5 \left(z^4-4 z^2 \log \left(z\right)-1\right)}{z^3
	\left(z^2-1\right){}^4} \, \nonumber \\
& = \frac{32 \pi^2 m_1^3 m_2^3 \sigma^5}{(\sigma^2-1)}	
-
i
\,
32 \pi  m_1^3 m_2^3 \left( \frac{ \sigma^6 }{
	(\sigma^2-1)^{3/2}} +  \frac{ \sigma^5 }{
	(\sigma^2-1)^2} \log(z) \right) \,,	
	\label{A22bis}
\end{align}
and
\begin{equation}
\label{RA10}
\begin{aligned}
\operatorname{Re}\mathcal A_2^{(1,0)}
&=
\frac{2 \pi ^2 m_1^2 m_2^2 \left(z^2+1\right){}^6
	\left(\left(m_1^2+m_2^2\right) z+2 m_1 m_2\right)}{z^3
	\left(z^2-1\right){}^4}\\
&-\frac{4 \pi ^2 m_1^3 m_2^3
	\left(z^2+1\right){}^4 \left(z^4-4 z^2-1\right) \log \left(z\right)}{z \left(z^2-1\right){}^4}\,, \\
	&=
8 \pi ^2 m_1^2 m_2^2 
	\left[(m_1^2+m_2^2)  \frac{\sigma^6}{(\sigma^2-1)^2} + 2 m_1 m_2 \left( \frac{\sigma^7}{(\sigma^2-1)^2}+ \frac{\sigma^6}{(\sigma^2-1)^{3/2}} \right)  \right]\\
&+16 \pi ^2 m_1^3 m_2^3
	\sigma^4 \left(\frac{1}{(\sigma^2-1)^{1/2}} - \frac{\sigma (\sigma^2-2)}{(\sigma^2-1)^2} \right) \log (z) \,,
\end{aligned}
\end{equation}
together with
\begin{equation} \label{IA10}
\begin{aligned}
\operatorname{Im}\mathcal A_2^{(1,0)}
&=
\frac{2 \pi  m_1^3 m_2^3 \left(z^4-6 z^2+1\right)
	\left(z^2+1\right){}^5 \operatorname{\operatorname{Li}}_2\left(z^2\right)}{z^3 \left(z^2-1\right){}^4}
\\
&-\frac{\pi 
	m_1^3 m_2^3 \left(-24(z^4-1) +\pi^2 (z^4 -6z^2 +1)
	\right) 
	\left(z^2+1\right){}^5}{3 z^3
	\left(z^2-1\right){}^4}\\
&+\frac{4 \pi  m_1^3 m_2^3 \left(z^4-4 z^2-1\right) \left(z^2+1\right){}^4 \log
	^2\left(z\right)}{z \left(z^2-1\right){}^4}
-\frac{4 \pi  m_1^3 m_2^3 \left(z^2+1\right){}^6
	\log \left(1-z^2\right)}{z^3 \left(z^2-1\right){}^3}\\
&-\frac{8 \pi  m_1^3 m_2^3 \left(z^2+1\right){}^6
	\log \left(z \right)}{z^3 \left(z^2-1\right){}^4}\,, \\
	&=
\frac{16 \pi  m_1^3 m_2^3 \sigma^5(\sigma^2 -2)
	 \text{Li}_2\left(z^2\right)}{(\sigma^2-1){}^2}
-64 \pi 
	m_1^3 m_2^3 \frac{\sigma^6}{(\sigma^2-1)^{3/2}} - \frac83 \pi^3 m_1^3 m_2^3 \frac{\sigma^5(\sigma^2-2)}{\left(\sigma^2-1\right){}^{2}}\\
&- 16 \pi  m_1^3 m_2^3 \sigma^4 \left(\frac{1}{(\sigma^2-1)^{1/2}} - \frac{\sigma (\sigma^2-2)}{(\sigma^2-1)^2} \right) \log^2\left(z\right)
+\frac{32 \pi  m_1^3 m_2^3 \sigma^6
	\log \left(1-z^2\right)}{\left(\sigma^2-1\right){}^{3/2}}\\
&-32 \pi  m_1^3 m_2^3 \sigma^6
\left(
\frac{\sigma}{\left(\sigma^2-1\right)^2}
+
\frac{1}{(\sigma^2-1)^{3/2}}
\right)\log \left(z \right)\,.
\end{aligned}
\end{equation}

\subsection{Eikonal and deflection angle to 3PM order}
\label{ssec:deltachi3PM}

In Subsection~\ref{ssec:tree1L} we have  computed $2\delta_0$ and $2\Delta_1$ that can now  be used to verify that the eikonal exponentiation takes place as expected up to two loops and to extract $2\delta_2$ from the previous amplitude through the following relations: 
\begin{eqnarray}
&&\operatorname{Re} (2 \delta_2) = \operatorname{Re} {\tilde{A}}_2 + \frac{1}{6} (2 \delta_0)^3 + 2\delta_0 		\operatorname{Im} 2\Delta_1  \nonumber \\
&& \operatorname{Im} (2 \delta_2) = \operatorname{Im} {\tilde{A}}_2 - 2\delta_0 \operatorname{Re} 2 \Delta_1
\label{CR6a}
\end{eqnarray}
where with ${\tilde{A}}_2 (s, b)$ we mean the Fourier transform in impact parameter space of the two-loop amplitude divided by the factor $4pE$ as in Eq.~\eqref{impparsp}. We get :
\begin{equation}
\begin{split}
	\operatorname{Re} (2\delta_2) &=\frac{4 m_1^2 m_2^2 G^3 (1+z^2)^4}{b^2 z^2 (1-z^2)^5} 
	\Bigg[ (1+z^2)^2 (1-z^2) + 2 z^2 (1 + 4z^2 -z^4 )\log z \Bigg]  \\
	&= \frac{16 m_1^2 m_2^2 G^3 \sigma^6}{b^2 (\sigma^2 -1)^2} - \frac{16 m_1^2 m_2^2 \sigma^4 G^3}{b^2 (\sigma^2-1)} \cosh^{-1} (\sigma) \Bigg[ 1 - \frac{\sigma (\sigma^2-2)}{(\sigma^2-1)^{\frac{3}{2}}}   \Bigg]
	\label{CR8}
\end{split}
\end{equation}
and
\begin{align}
	\begin{split}
\operatorname{Im} 2 \delta_2 =&~  \frac{4 m_1^2 m_2^2 G^3 }{\pi b^2 \epsilon} \frac{(1+z^2)^5}{ (1-z^2)^5 z^2}\left[ -1 +z^4 + \left(1-6z^2 +z^4\right) \log z \right]  \\
& - \frac{4m_1^2 m_2^2 G^3}{\pi b^2}  \frac{(1+z^2)^4}{ z^2 (1-z^2 )^5} \Bigg\{ 2 z^2 
\left( 1+4z^2 -z^4\right) \log^2 z
 \\
& + (1+z^2)^2 (1-z^2) 
 \left[ 2  + 3 \log (\pi b^2 {\rm e}^{\gamma_E})- 2 \log (1-z^2) +2 \log z \right]    \\
& + \left[\frac{\pi^2}{6} -  \operatorname{Li}_2 (z^2) - 3 \log z  \log (\pi b^2 {\rm e}^{\gamma_E}) \right] (z^2+1)(z^4 -6z^2 +1)  \\ 
=& - \frac{16 m_1^2m_2^2 G^3}{\pi b^2}
 \frac{\sigma^4}{(\sigma^2-1)^2}  \Bigg\{\frac{1}{\epsilon}\left( \sigma^2  + \frac{\sigma(\sigma^2-2)}{(\sigma^2-1)^{\frac{1}{2}}} \cosh^{-1} (\sigma)\right)
  \\
 & - \left(\log (4(\sigma^2-1)) -3 \log (\pi b^2 {\rm e}^{\gamma_E})  \right) \left[\sigma^2 + \frac{\sigma (\sigma^2-2)}{(\sigma^2-1)^{\frac{1}{2}}} \cosh^{-1} (\sigma) \right]  \\ & + (\sigma^2-1)\left[ 1  + \frac{\sigma (\sigma^2 -2) }{(\sigma^2-1)^{\frac{3}{2}}}  \right](\cosh^{-1} (\sigma))^2  \\
 & + \frac{\sigma (\sigma^2-2)}{(\sigma^2-1)^{\frac{1}{2}}}  \operatorname{Li}_2 (1-z^2) + 2 \sigma^2  \Bigg\},
\label{CR15}
\end{split}
\end{align}
where we made the branch-cut singularity for $z>1$ more explicit via
\begin{eqnarray}
	\operatorname{Li}_2 (z^2)
 = -\operatorname{Li}_2 (1-z^2) +\frac{\pi^2}{6} - \log(z^2) \log (1-z^2)\,.
\label{enha3}
\end{eqnarray}
and used
\begin{equation}
	- \log z =\cosh^{-1} (\sigma) = \log \left( \sigma + \sqrt{\sigma^2-1} \right) = 2 \sinh^{-1} \sqrt{\frac{\sigma-1}{2}}\;,
	\label{CR16}
\end{equation}
\begin{equation}
    \label{eq:l1mzsi}
    \log(1-z^2) = \frac{1}{2} \log(4(\sigma^2-1)) - \cosh^{-1} (\sigma)\;.
  \end{equation}
Notice that in the ultrarelativistic limit $\sigma \gg 1$ the terms proportional to $(\sigma \log(2\sigma))^2$ present in the second and third last lines separately cancel in agreement with the general pattern discussed in~\cite{DiVecchia:2020ymx}.

Let us  finally turn to the calculation of the deflection angle $\chi$ as outlined in Section~\ref{ssec:eikonalgen}. To 3PM order, we have
\begin{eqnarray}
2 \sin  \frac{\chi}{2} = - \frac{1}{p} \frac{\partial ~ \operatorname{Re} 2\delta (s, b)}{\partial b} \,, \qquad \delta = \delta_0 +  \delta_1 +  \delta_2 
\label{DA1}
\end{eqnarray}
where $p$ is the momentum of either particle in the centre-of-mass frame as in Eq.~\eqref{T4}. Moving from $b$ to the orthogonal impact parameter $b_J$, we have 
\begin{equation}
2 \tan\left(\frac{\chi}{2}\right) = \chi \left(1+ \frac{1}{12} \chi^2 + \cdots \right) = - \frac{1}{p} \frac{\partial  \operatorname{Re} 2\delta }{\partial b_J} = \frac{4 G m_1 m_2 \sigma^2}{J (\sigma^2-1)^{1/2}} + \dots
\end{equation}
from which we obtain:
\begin{equation}
\chi  =\chi_{1\mathrm{PM}} + \chi_{2\mathrm{PM}}+ \chi_{3\mathrm{PM}}
\end{equation}
with (to leading non-vanishing order in $\epsilon$ at each PM order):
\begin{align}
\chi_{1\mathrm{PM}} &= \frac{4 G m_1 m_2 \sigma^2}{J (\sigma^2-1)^{1/2}}  \\
\chi_{2\mathrm{PM}} &= -\frac{8 \pi  m_1^2 m_2^2 (m_1+m_2)  G^2\epsilon \sigma^4 }{J^2\sqrt{m_1^2+m_2^2+2m_1m_2 \sigma} \,\,(\sigma^2-1)}  \\
\begin{split}
\chi_{3\mathrm{PM}} &= - \frac{16 m_1^3 m_2^3 \sigma^6 G^3}{3 J^3(\sigma^2-1)^{3/2}} +  \frac{32 m_1^4 m_2^4 \sigma^6 G^3}{J^3(\sigma^2-1) (m_1^2 +m_2^2 +2m_1m_2 \sigma)}  \\
& -\frac{ 32 m_1^4 m_2^4 \sigma^4 G^3}{J^3(m_1^2 +m_2^2 +2m_1 m_2 \sigma)} 
\Bigg [ 1 - \frac{\sigma (\sigma^2-2)}{(\sigma^2-1)^{\frac{3}{2}}}   \Bigg]  \cosh^{-1}(\sigma)
\label{DA2}
\end{split}
\end{align}
The 1PM contribution corresponds to a tree diagram where both the graviton and the dilaton are exchanged, the 2PM contribution is absent for $\epsilon=0$ in agreement with the results of Ref.\cite{Caron-Huot:2018ape} at one loop. The first term in $\chi_{3\mathrm{PM}}$ comes from the expansion of $\tan(\frac{\chi}{2})$ at small $\chi$ while the remaining terms are genuine new contributions from  $\operatorname{Re} \delta_2$. 

Anticipating a more complete discussion in Subsection \ref{connection} we remark that in the second form of equation \eqref{CR8} (already presented together with \eqref{DA2} in \cite{DiVecchia:2020ymx}) the first and last term contain only even powers of $p \sim  \sqrt{\sigma^2-1}$  (see \eqref{delta2rr} below), while the second term has only odd powers. This means that those two terms  represent half-integer-PN corrections which are  known to be a consequence of dissipative processes (so-called radiation reaction). Instead,
 the second term in \eqref{CR8} corresponds to the more conventional integer-PN expansion due to conservative dynamics. 
 
 Looking now  at the second form of \eqref{CR15} we note that its  IR divergent term (the one appearing on the first line) is simply related to the above mentioned half-integer PN terms by a trivial factor $-\frac{1}{\epsilon \pi}$. Furthermore, the same half-integer PN terms also multiply, up to a factor $\frac{1}{\pi}$, a $\log v^2 \sim  \log (4(\sigma^2-1))$ coefficient. 
 More specifically one can immediately see that the divergent part and the part with $\log (\sigma^2-1)$ of ${\operatorname{Im}} \,2 \delta_2$ are related to the two half-integer PN terms that we call  ${\operatorname{Re} } \,2 \delta_2^{(rr)}$ by the following relation:
 \begin{equation}
 2 \delta_2^{(rr)} = \left[ 1+ \frac{i}{\pi} \left(-\frac{1}{\epsilon}+ \log (\sigma^2-1)\right) \right]  \operatorname{Re}\, 2 \delta_2^{(rr)}  
\label{conne}
\end{equation}
where
\begin{equation}
\operatorname{Re}\, 2 \delta_2^{(rr)}  = \frac{16 G^3 m_1^2 m_2^2 \sigma^4}{b^2 (\sigma^2-1)^2}
\left[ \sigma^2 + \frac{\sigma (\sigma^2-2)}{(\sigma^2-1)^{\frac{1}{2}}} \cosh^{-1} (\sigma) \right]\, .
\label{delta2rr}
\end{equation}

In the next Section, using general analyticity and unitarity arguments, we will argue that these relations, rather than mere coincidences,  should be regarded as  generic relations between IR divergent terms in $\operatorname{Im} 2\delta_2$ and radiation-reaction corrections to $\operatorname{Re} 2\delta_2$.
Since analyticity and crossing are spoiled when we go from the amplitude to its partial waves (the Fourier/Legendre transform introduces spurious kinematical singularities and refers to one specific channel), we have to go back to the amplitude itself.

\section{Real-Analytic, Crossing-Symmetric Formulation}
\label{analytic}

In the previous subsection we have collected the contribution of the different diagrams both at the one and at the two-loop level, noticing that they combine into much simpler expressions than those of individual diagrams. Those expressions are typically in the form of an expansion in powers of $q^2$ and, for the two-loop calculation, of $\epsilon$.

Furthermore, as a computer output, they simply collect independently real and imaginary terms and express the result in terms of a choice of the two independent Mandelstam variables, $q^2 = -t$ and $s$ (or of quantities like $\sigma$ or $z$ which are themselves functions of $s$ and the masses).

On the other hand we expect the full amplitude to satisfy two exact properties:
\begin{itemize}
\item Real analyticity i.e. $\mathcal{A} (s^*, q^2) = (\mathcal{A} (s, q^2))^*$;
\item Crossing symmetry  i.e. $\mathcal{A} (s, q^2) = \mathcal{A} (u \equiv -s +q^2 + 2m_1^2 + 2m_2^2, q^2)$.
\end{itemize}
Those properties are not at all apparent in the formulae of the previous subsection but they must be hidden somewhere, of course up to the order in $q^2$ at which we stop our expansion. Here we will explicitly show how to rewrite  the tree, one-loop and two-loop amplitudes, given in the previous section
in a real analytic crossing-symmetric form. Besides its usefulness for the interpretation of the result this also serves as a rather stringent test of the calculations themselves.

To this purpose it is useful to introduce variables which are related by $s \leftrightarrow u$ exchange to those introduced earlier in Section~\ref{ssec:conventions}:
   \begin{eqnarray}
&& \bar{\sigma} =  \frac{u - m_1^2-m_2^2}{2m_1m_2}  = -(\sigma - \frac{q^2}{2 m_1m_2})  \nonumber  \\
&& \bar{z} = \bar{\sigma} - \sqrt{\bar{\sigma}^2-1} = - \frac{1}{z}\left( 1 - \frac{q^2/(2m_1m_2)}{\sqrt{\sigma^2-1}}\right) + O(q^4) \nonumber \\
&& 2 \bar{\sigma}  = (\bar{z} + \frac{1}{\bar{z}}) ~;~  2 \sqrt{\bar{\sigma}^2-1} = (\frac{1}{\bar{z}}- \bar{z})
\label{defs}
\end{eqnarray}

\subsection{Tree and one loop}
\label{treeLoop}
The tree-level amplitude itself  can be rewritten trivially as:
    \begin{equation}
\mathcal{A}_0 (s, q^2) = \frac{8\pi G}{q^2} \frac{1}{2} \Big[ 4m_1^2 m_2^2 \left(\sigma^2 + \bar{\sigma}^2\right) - q^4\Big]
\label{T1sym}
    \end{equation}
At one-loop, equation \eqref{A10a} suggests trying the following analytic, crossing-symmetric ansatz for the super classical term:
\begin{eqnarray}
&& \mathcal{A}_1^{scl.} (s , q^2) = - \frac{(8\pi G)^2}{(4\pi)^2} m_1^3 m_2^3 \left( \frac{4\pi}{q^2}\right)^{\epsilon} 
\frac{4^{\epsilon} \pi^{\frac{3}{2}} }{q^2 \sin (\pi \epsilon) \Gamma (\frac{1}{2} -\epsilon)} \nonumber \\
  &&\left[8  (\sigma^4 + \bar{\sigma}^4 ) \left(  \frac{\log(-z)}{\sqrt{\sigma^2-1}} + \frac{\log(-\bar{z})}{\sqrt{\bar{\sigma}^2-1}}\right) \right]
  \label{Ansatz1scl}
  \end{eqnarray} 
 With the help  of  the following relation
 \begin{eqnarray}
\frac{ 4^{\epsilon} \pi^{\frac{3}{2}}}{\sin \pi \epsilon \Gamma (\frac{1}{2} - \epsilon)} 
= - \frac{\Gamma^2 (-\epsilon) \Gamma (1+\epsilon)}{2\Gamma (-2\epsilon)}
\label{}
\end{eqnarray}
it is easy to check that, when expanded in $q^2$ using \eqref{defs},  \eqref{Ansatz1scl} reproduces both the super-classical (first line in \eqref{A10a}) and the $O(\epsilon^0 m_1 m_2)$ terms in the square bracket of \eqref{A10a}.
  
  The remaining terms  can be described as follows:
  \begin{itemize}
  \item The classical contribution proportional to $(q^2)^{-1/2}$ (second line in \eqref{A10a}). It can be trivially symmetrized in $s-u$ since the error in doing so is of order $(q^2)^{1/2}$;
  \item A quantum contribution of $O(\epsilon)$ which can be written in the crossing-symmetric form:
 \begin{eqnarray}
&& \mathcal{A}_1^{qu1} (s , q^2) =  \frac{(8\pi G)^2}{(4\pi)^2} m_1^2 m_2^2 \left( \frac{4\pi}{q^2}\right)^{\epsilon} 
\frac{4^{\epsilon} \pi^{\frac{3}{2}} }{\sin (\pi \epsilon) \Gamma (\frac{1}{2} -\epsilon)} \nonumber \\
  && \times 8 \epsilon \left[ \left(  \frac{\sigma^5 \log(-z)}{(\sigma^2-1)^{3/2}} + \frac{\bar{\sigma}^5 \log(-\bar{z})}{(\bar{\sigma}^2-1)^{3/2}}\right) +\left(\frac{\sigma^4 }{\sigma^2-1} + \frac{\bar{\sigma}^4}{\bar{\sigma}^2-1}   \right) \right]
  \label{A1q1}
  \end{eqnarray} 
  \item  A purely imaginary quantum term proportional to $m_1 m_2(m_1^2 + m_2^2) \epsilon$
  \begin{eqnarray}
&& \mathcal{A}_1^{qu2} (s , q^2) =  \frac{(8\pi G)^2}{(4\pi)^2} m_1 m_2 (m_1^2 + m_2^2) \left( \frac{4\pi}{q^2}\right)^{\epsilon} 
\frac{4^{\epsilon} \pi^{\frac{3}{2}} }{\sin (\pi \epsilon) \Gamma (\frac{1}{2} -\epsilon)} \nonumber \\
  && \times 4 \epsilon \left( \frac{\sigma^4 \log(-z)}{(\sigma^2-1)^{3/2}} + \frac{\bar{\sigma}^4\log(-\bar{z})}{(\bar{\sigma}^2-1)^{3/2}}   \right)
  \label{A1q2}
  \end{eqnarray}   
  \end{itemize}
Note that, in \eqref{A1q1} and \eqref{A1q2}, we can again neglect the difference between $\bar{\sigma}$ and $-\sigma$ as well as the difference between $\bar{z}$ and $- 1/z$.

In order to facilitate the discussion of the various terms in Eqs. \eqref{ReD1d0}-\eqref{ACV} we give here the Fourier transform in the impact parameter space (denoted as usual with a tilde) of the two previous quantum terms. We get
\begin{eqnarray}
 {\tilde{\mathcal{A}}}_1^{qu1} (s, b) =
  \frac{ 8 G^2 m_1m_2 (\pi b^2)^{2\epsilon}}{\pi b^2} \epsilon \Gamma^2 (1-\epsilon) \left[ \frac{\sigma^5}{(\sigma^2-1)^2} (i\pi +2 \log z) + \frac{2 \sigma^4}{(\sigma^2-1)^{3/2}} \right]
\label{qu1}
\end{eqnarray}
and
\begin{eqnarray}
{\tilde{\mathcal{A}}}_1^{qu2} (s,b) = i \frac{4G^2 (\pi b^2)^{2\epsilon}\Gamma^2(1-\epsilon)}{b^2}
\frac{\epsilon \sigma^4}{(\sigma^2-1)^2}(m_1^2 +m_2^2) \,\,\,.
\label{qu2}
\end{eqnarray}

\subsection{Two loops}
\label{twoloops}

Moving now to two loops we combine the different contributions in the known form of Eq.~\eqref{CR6},
which suggests to try the ansatz:
\begin{eqnarray}
&&\mathcal{A}_2 (s, q^2) =  \frac{(8\pi G)^3 }{(4\pi)^4}  \left( \frac{4\pi {\rm e}^{-\gamma_E}}{q^2}\right)^{2\epsilon}  (\sigma^4 + \bar{\sigma}^4 ) \hat{A}_2 (s, q^2)  ~;~  \nonumber \\
&& \hat{A}_2 (s, q^2) =  \frac{\hat{A}_2^{(2)} (s, q^2)}{\epsilon^2} +  \frac{\hat{A}_2^{(1)} (s, q^2)}{\epsilon}
\label{Ansatz}
\end{eqnarray}
In order to reproduce Eqs. \eqref{A22} and \eqref{A22bis} we find:
\begin{eqnarray}
&&  \hat{A}_2^{(2)} (s, q^2) = \Bigg[- \frac{8 \pi^2 m_1^4 m_2^4}{q^2} \left( \frac{\sigma^2}{\sigma^2-1} +  \frac{\bar{\sigma}^2}{\bar{\sigma}^2-1}\right) \nonumber \\
&& -  8  m_1^3 m_2^3  \Bigg(  \frac{\sigma}{(\sigma^2-1)^2}  \log^2(-z) +   \frac{\bar{\sigma}}{(\bar{\sigma}^2-1)^2}  \log^2(-\bar{z})   \nonumber \\
&& + 2\left[ \frac{\sigma^2}{(\sigma^2-1)^{3/2}}  \log(-z) +   \frac{\bar{\sigma}^2}{(\bar{\sigma}^2-1)^{3/2}}  \log(-\bar{z})  \right] \Bigg) \Bigg] 
\label{Ansatz(2)}
\end{eqnarray}
 and for Eqs. \eqref{RA10} and   \eqref{IA10}:
 \begin{subequations}
  \label{Ansatz(1bis)}
  \begin{eqnarray}
  &&  \hat{A}_2^{(1)} (s, q^2) = 4 \pi^2 m_1^2 m_2^2 (m_1^2+ m_2^2) \frac{\sigma^2}{(\sigma^2-1)^2} \label{ImD1d0}\\
  && - 32  m_1^3 m_2^3  \frac{\sigma^2}{(\sigma^2-1)^{3/2}} \left( \log(-z) - \log(z)  \right) \label{ReD1d0}\\
  && + 4 m_1^3 m_2^3 \frac{\sigma (\sigma^2-2)}{(\sigma^2-1)^2}\left( \log(-z) - \log(z)  \right) \left(\operatorname{Li}_2(z^2) - \operatorname{Li}_2\left(\frac{1}{z^2}\right)\right) \label{newwlog} \\
  && - \frac{8}{3}  m_1^3 m_2^3 \frac{1}{(\sigma^2-1)^{1/2}} \left[\left( \log^3(-z) - \log^3(z) \right) + \pi^2 \left( \log(-z) - \log(z) \right) \right] \label{P-MRZ} \\
  && - 8 m_1^3 m_2^3 \frac{\sigma^3}{(\sigma^2-1)^2}\left( \log(-z) - \log(z)  \right)\left( \log(-z) + \log(z)  \right) \label{D1d0}\\
  &&+ 8  m_1^3 m_2^3  \frac{\sigma^2}{(\sigma^2-1)^{3/2}} \left( \log(-z) - \log(z)  \right) \left( \log(1-z^2) + \log\left(1-\frac{1}{z^2}\right) \label{ACV} \right) 
\end{eqnarray}
\end{subequations}
The two previous equations have been derived using    $\log (-z) = i\pi + \log z$ and the following identities and branch choices:
 \begin{eqnarray}
&& - \operatorname{Li}_2\left(\frac{1}{z^2}\right) =  \operatorname{Li}_2(z^2)+\frac{\pi^2}{6} + \frac12 \log^2(-z^2) = \operatorname{Li}_2(z^2) - \frac{\pi^2}{3} + 2 \log^2 z+2 i \pi \log z ; \nonumber\\
&& \log\left(1-\frac{1}{z^2}\right) =  \log(1-z^2)  - 2 \log z - i \pi.
\label{ident}
\end{eqnarray}

Combining these with \eqref{enha3} we can also write the combination appearing in  \eqref{newwlog} in a  form:
 \begin{equation}
 \operatorname{Li}_2(z^2) - \operatorname{Li}_2\left(\frac{1}{z^2}\right) = - 2 \left( \operatorname{Li}_2(1- z^2) +  \log z^2 \log (1-z^2)  - \log^2 z -  i \pi \log z\right)\; ,
\label{ident1}
\end{equation}
that is useful to discuss the non relativistic ($\sigma , z \to 1$) limit.

We can check that the expressions \eqref{Ansatz(2)} and \eqref{Ansatz(1bis)}  satisfy   crossing symmetry and real-analyticity.
The first property follows simply from the fact that, at this order in $q^2$, we can use the identification ${\bar{z}} = - \frac{1}{z}$. 

 Checking real analyticity is a bit more subtle.   For \eqref{Ansatz(2)} one needs to take into account the subleading (in $q^2$) term origination from the first line. This (purely real) term exactly cancels a similar term coming from the second line. Then one is left with a purely imaginary term from the second line:
\begin{eqnarray}
 - \frac{16 m_1^3 m_2^3 \pi \sigma}{(\sigma^2-1)^2} i \log z \,\,,
\label{replace1}
\end{eqnarray}
 that can be written in real-analytic crossing-symmetric form as
 \begin{equation}
 \label{replace}
 -  8  m_1^3 m_2^3  \frac{\sigma}{(\sigma^2-1)^2} \frac12 ( \log(-z) +   \log(-\bar{z}))( \log (z^2) -   \log(\bar{z}^2)) 
 \end{equation}
 
 At this point checking that the amplitude is real in the unphysical region $\sigma^2 < 1$ is straightforward once one realizes that, in that region, $|z|^2 = 1$, i.e. $z$ is a pure phase and thus $\log z$ is purely imaginary.  This implies that the term in Eq. \eqref{replace1} is real
 and the same is true for the last term in \eqref{Ansatz(2)} because of an extra factor $i$ coming from $\sqrt{\sigma^2-1}$.
 
 Coming now to \eqref{Ansatz(1bis)}, real analyticity is easily checked along the same lines for \eqref{ImD1d0}, \eqref{ReD1d0}, \eqref{P-MRZ} and \eqref{D1d0}.
 Concerning instead \eqref{newwlog} and \eqref{ACV} one has to remember
 that, in the unphysical region, $z^{-1} = {\bar{z}}$. Since both $\operatorname{Li}_2(z^2)$ and $\log(1-z^2)$ are real-analytic functions, the combinations involving them, appearing in \eqref{newwlog} and \eqref{ACV} give a purely imaginary and purely real factor, respectively. Because of the different powers of $(\sigma^2-1)$ appearing in the two contributions this is exactly as needed for real analyticity. In conclusion, we have shown that Eq. \eqref{Ansatz(2)} and
 Eqs. \eqref{ImD1d0} $\dots$ \eqref{ACV} are real analytic functions. Note that this would not have been the case for the eikonal itself.

Before going further, in Subsection 5.3,   into considerations of analyticity let us discuss, with the help of the Fourier transform  of the pre-factor in Eq. \eqref{Ansatz}
\begin{equation}
\frac{ G^3  (\sigma^4 + {\bar{\sigma}}^4)}{\pi^2 b^2 m_1m_2 \sqrt{\sigma^2-1}} \, \, ,
\label{prefactor}
\end{equation}
 how the different terms in
 \eqref{Ansatz(1bis)}  have precise correspondences in the eikonal\footnote{Comparison among the various quantities is performed up to terms that vanish when $\epsilon \rightarrow 0$}:
\begin{itemize}
\item Eq.~\eqref{ImD1d0} is a purely real term. It is easy to check that its Fourier transform matches exactly the combination  $- 4  \delta_0 \operatorname{Im} \Delta_1$ if we insert for $\operatorname{Im} \Delta_1$ the corresponding term given by the Fourier transform of \eqref{A1q2} in  \eqref{qu2}. Therefore \eqref{ImD1d0} does not contribute to $\operatorname{Re} \delta_2$;
\item Eq.~\eqref{ReD1d0} matches only half of the corresponding contribution from $4 i \operatorname{Re} \Delta_1 \delta_0$ if we insert for $\operatorname{Re} \Delta_1$ the second term in  \eqref{qu1}.
This leaves an identical contribution in $\operatorname{Im} \delta_2$.
\item Eq.~\eqref{newwlog} provides a new   $log s$-enhanced contribution to be discussed below. It corresponds to the last term in Eq.~\eqref{CR8};
\item Eq.~\eqref{P-MRZ} gives a  $\log(s)$-enhanced contribution  whose real part can be checked to match the result for $\operatorname{Re} \delta_2$ given  in ref. \cite{Parra-Martinez:2020dzs}.
It corresponds to the second term in Eq.~\eqref{CR8};
\item Eq.~\eqref{D1d0} This contribution has both a real and an imaginary part. The real part, like with  \eqref{ImD1d0}, matches exactly  a corresponding term in $- 4  \delta_0 \operatorname{Im} \Delta_1$ (the imaginary term in \eqref{qu1}) leaving behind no contribution to $\operatorname{Re} \delta_2$. Similarly, the imaginary part matches  $4 i \delta_0 \operatorname{Re} \Delta_1 $ (the term with $\log z$ in \eqref{qu1}) leaving no residual contribution to $\operatorname{Im} \delta_2$;
\item Eq.~\eqref{ACV} is a new  non $log $-enhanced contribution reproducing at high energy  the ACV90 result. It corresponds to the first term in Eq.~\eqref{CR8}. 
\end{itemize}

In summary, for what concerns $\operatorname{Re} \delta_2$ we are left with the result given in \cite{Parra-Martinez:2020dzs} accompanied by two new contributions given by Eq.~\eqref{newwlog} and Eq.~\eqref{ACV}. It is thanks to these new contributions, given already in Eq.~\eqref{CR8}, that the high-energy limit of $\operatorname{Re} \delta_2$ is finite, universal, and in agreement with \cite{Amati:1990xe}.
The two above-mentioned  new contributions to $\operatorname{Re} \delta_2$ have some distinctive features, in particular with respect to their PN expansion discussed in the Appendix \ref{PN}. 

\subsection{Connecting real and imaginary parts via analyticity}
\label{connection}

Before turning  to the discussion of the analytic properties of our results  \eqref{Ansatz(2)} and \eqref{Ansatz(1bis)} let us look at the imaginary parts they contain picking up, in particular, those containing, in \eqref{Ansatz(1bis)}, a $\log (1-z) \sim \frac12 \log(\sigma^2-1)$ enhancement for $\sigma \to 1$.
The result is:
\begin{eqnarray}
&&  \operatorname{Im} \hat{A}_2^{(2)}  = -  16 \pi  m_1^3 m_2^3  \Bigg(  \frac{\sigma^2}{(\sigma^2-1)^{3/2}} + \frac{\sigma}{(\sigma^2-1)^2}  \log(z) \Bigg) \nonumber \\
&& \operatorname{Im} \hat{A}_2^{(1)}= 16 \pi  m_1^3 m_2^3  \Bigg(  \frac{\sigma^2}{(\sigma^2-1)^{3/2}} - \frac{\sigma (\sigma^2-2)}{(\sigma^2-1)^2}  \log(z) \Bigg) \log(1-z^2)
\label{ImhatA2}
\end{eqnarray}
After multiplying both equations by the appropriate power of $\epsilon$ we note that they fail to reproduce the combination
\begin{equation}
\label{combin}
- \frac{1}{\epsilon} + 2 \log(1-z^2)
\end{equation}
 we had noticed at the end of Subsection \ref{ssec:deltachi3PM}  for $\operatorname{Im} \delta_2$. More specifically, the contribution of \eqref{ACV} is accompanied by twice as large a contribution from  \eqref{Ansatz(2)}.  At the same time the contribution of \eqref{newwlog} has the correct counterpart in  \eqref{Ansatz(2)},   but only for its $- 2 \sigma$ piece in the last term of \eqref{ImhatA2} while the $\sigma^3$ piece is missing.

Fortunately, this apparent contradiction is completely resolved once we subtract from $\operatorname{Im} A_2$ those pieces that correspond to the elastic cut of the amplitude itself (note that this subtraction only affects Eq. \eqref{Ansatz(2)}) given by the Fourier transform from impact parameter to momentum space of the quantity $2\delta_0 2 \operatorname{Re} \Delta_1 4m_1m_2 \sqrt{\sigma^2-1}$. In other words, the full analyticity argument has to be carried out on the (analytic and crossing-symmetric) piece of the two-loop amplitude whose imaginary part (entering suitably subtracted dispersion relations as discussed below) is restricted to  inelastic contributions. Therefore, indicating by $\hat{A}_2^{(3pc)}$ that piece of the two-loop amplitude, we find:
\begin{equation}
\label{ImA23pc}
 \operatorname{Im} \hat{A}_2^{(3pc)}= 16 \pi  m_1^3 m_2^3  \Bigg(  \frac{\sigma^2}{(\sigma^2-1)^{3/2}} - \frac{\sigma (\sigma^2-2)}{(\sigma^2-1)^2}  \log(z) \Bigg) \left(- \frac{1}{2 \epsilon} + \log(1-z^2)\right)
\end{equation}
in full agreement with the structure \eqref{combin}.

The appearance of that combination is by no means an accident and can be understood \cite{DiVecchia:2021ndb} as following from unitarity and three-particle phase space considerations. In $D=4-2 \epsilon$ one finds that the imaginary part of the three-particle cut behaves, for $\sigma \to 1$ as
\begin{equation}
\label{3pphspace}
 \operatorname{Im} \hat{A}_2^{(3pc)}= \int_0^{\bar {\omega}} d \omega (\omega b)^{-2 \epsilon} \sim - \frac{(\bar{\omega} b)^{-2 \epsilon}}{2 \epsilon} \sim \left(- \frac{1}{2 \epsilon} + \log(\bar {\omega} b) \right) \sim  - \frac{(\sigma^2-1)^{- \epsilon}}{2 \epsilon}
 \end{equation}
 where $\hbar \omega$ is the energy of the massless particle and we have used the fact that the relevant upper limit in $\omega$ is ${\cal O}(v/b)$. We refer to \cite{DiVecchia:2021ndb} for further details.

Let us now look at  \eqref{Ansatz(1bis)} which corresponds to  ${\cal O}(\epsilon^0)$ contributions to $\delta_2$.
Let us look, in particular, at the new contributions appearing in \eqref{newwlog} and \eqref{ACV}. Leaving the discussion of their PN expansion to Appendix~\ref{PN} we note that, in both, the real parts we are interested in are accompanied by an imaginary piece that contains an overall $\log(\sigma^2-1)$ factor. More precisely, in the $s$-channel  physical region real and imaginary parts appear in the combination:
\begin{equation}
\label{comb1}
 \pi + i \log((\sigma^2-1)) = \pi + i  \log \left[\frac {\left( s - m^2  \right)~\left(  m^2 - u \right)}{ 4 m_1^2 m_2^2} \right]
\end{equation}
where we have used the relation:
\begin{equation}
  (\sigma^2-1) \sim \frac {\left( s - m^2  \right)~\left(  m^2 - u \right)}{4 m_1^2 m_2^2}~;~ m \equiv (m_1+m_2)
\end{equation}
showing that the position of the branch points correspond to  thresholds for the $s$ and $u$-channels.

On the other hand we have seen that, combining real and imaginary parts, the combination  \eqref{3pphspace} appears  giving, in total, the structure:
\begin{equation}
\label{comb2}
 \pi + i \left(- \frac{1}{\epsilon} + \log(\sigma^2-1)\right) = \pi + i \left( - \frac{1}{\epsilon} + \log \left[\frac {\left( s - m^2  \right)~\left(  m^2 - u \right)}{ 4 m_1^2 m_2^2} \right] \right)
\end{equation}
in full agreement with (and in addition to) what we have observed at the $\delta_2$ level.

The combination appearing in  \eqref{comb2} looks to be forced upon by general considerations of unitarity, analyticity  and crossing. Unitarity determines the behaviour of the imaginary parts of the amplitude near the branch points at $\sigma = \pm 1$. for the elastic  and  inelastic (i.e. three-particle) channels. 
The elastic channel leads to a square-root type branch point, while for the inelastic channel integration over the soft-particle momentum induces the logarithmic enhancement $\sim \log(\sigma^2-1)$ of \eqref{3pphspace}. 

Concerning the implications of analyticity it is convenient to strip out of the amplitude a harmless $(\sigma^4+\bar{\sigma}^4)$ as in~\eqref{Ansatz} and meromorphic factors such as $\sigma^2/(\sigma^2-1)$. We can then write an unsubtracted dispersion relation for the resulting ``reduced" $\hat{A}_2^{(1)}$ amplitude. Also, because of   $s \leftrightarrow u$ symmetry, the left and right hand cut contributions  combine to give the familiar form:
\begin{equation}
\label{disprel}
 \operatorname{Re} \hat{A}_2^{(1)}(\sigma) = \frac{2}{\pi}\ \mathcal P\!\! \int_1^{\infty} \frac{  \operatorname{Im} \hat{A}_2^{(1)}(\sigma')  \sigma' d \sigma' }{\sigma'^2- \sigma^2} =  \operatorname{Re} \hat{A}_2^{(1)}(-\sigma)\; ,
\end{equation}
where $\mathcal P$ denotes the principal value of the integral.

Finally, inserting in  \eqref{disprel} each contribution to the imaginary part generates automatically the corresponding real part given in $\eqref{Ansatz(1bis)}$. In other words, the full real-analytic amplitude is uniquely reconstructed from its discontinuities across the physical cuts via a Cauchy integral going around the branch cuts and a circle at infinity. In particular a $\log(\sigma^2-1)$ factor in the imaginary part  induces, via \eqref{disprel}, a corresponding factor $+\pi$ in the real part, in agreement with \eqref{comb2}.

The above general considerations are neatly illustrated by considering, as an example,  two different imaginary parts, one with and one without the $\log(\sigma^2-1)$ factor, see~\eqref{ReD1d0} and~\eqref{ACV}. The above mentioned Cauchy integrals can be computed explicitly and give, at $\sigma^2 > 1$ and above the cut,
\begin{eqnarray}
\label{2ex}
&&  \frac{2}{\pi} \int_1^{\infty}  d \sigma'  \frac{ \sigma' (\sigma'^2- 1)^{-\frac12} \log(\sigma'^2- 1)  }{\sigma'^2- (\sigma+i0)^2} = (\sigma^2- 1)^{-\frac12}(\pi + i \log(\sigma^2- 1))\nonumber \\
&&  \frac{2}{\pi} \int_1^{\infty}  d \sigma'  \frac{ \sigma' (\sigma'^2- 1)^{-\frac12}}{\sigma'^2- (\sigma+i0)^2} = i (\sigma^2- 1)^{-\frac12} , 
 \end{eqnarray}
showing the appearance of the extra $\pi$-term in the real part for the former case.

The above considerations lead us to conjecture that the combination \eqref{comb2} is a generic feature for gravitational theories containing, besides a graviton, other massless degrees of freedom. This conjecture is indeed fully confirmed \cite{DiVecchia:2021ndb} by the GR example, for which we recover the recent result of \cite{Damour:2020tta}  as well as by the ${\cal{N}}=8$ supergravity example at different values of the $\cos \phi$ parameter. It is also confirmed by the explicit calculations of the three-particle cuts given in Section~\ref{sec:directunitarity}.

\section{Direct-Unitarity Calculations}
\label{sec:directunitarity}

So far we focused on  massive ${\cal N}=8$ supergravity where the 4-point amplitudes take a particularly simple form, see~\eqref{T6}, \eqref{1L1} and~\eqref{CR1}. We then considered the soft-region expansion, which is the one relevant for the classical limit, and used the results of Section~\ref{sec:toolkit} to evaluate the resulting integrals. This allowed us to extract the complete eikonal $\delta$ up to 3PM order by applying~\eqref{eikonalgen}. When considering the contributions of the full soft region, one obtains both a real and an imaginary part for $\delta_2$, the 3PM term of the eikonal. The complete ${\cal N}=8$ results display an intriguing relation between the IR-divergent $\mathcal O(\epsilon^{-1})$ term of $\operatorname{Im}2\delta_2$ and the radiation-reaction terms $\operatorname{Re} 2 \delta_2^{(rr)}$ of the real part, see~\eqref{conne}. In the previous section, we argued that this relation is in fact a general feature of gravitational field theories in the two derivative approximation, including GR. The main goal of this section is to provide further evidence for this by deriving both the $\mathcal O(\epsilon^{-1})$ and $\mathcal O(\epsilon^0)$ terms of $\operatorname{Im} 2 \delta_2$ for GR and showing that they indeed display the structure appearing in~\eqref{comb2} with a coefficient that agrees with the radiation-reaction result of~\cite{Damour:2020tta,DiVecchia:2021ndb}.

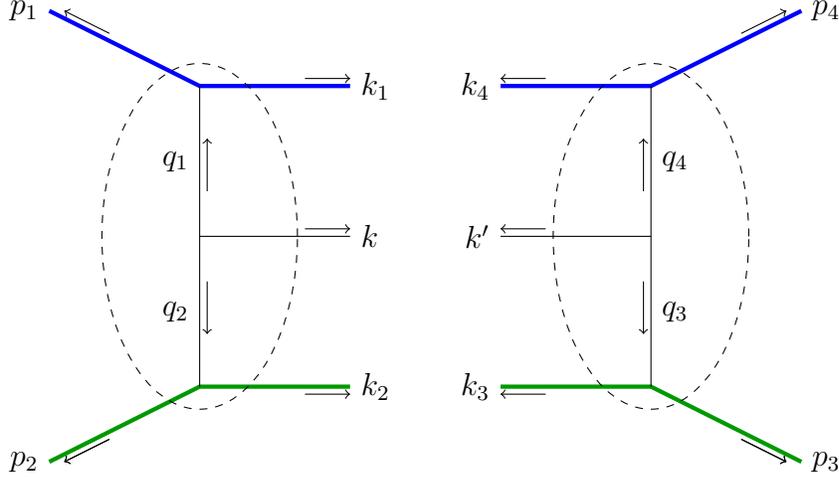
\begin{figure}
	\begin{center}
		\begin{tikzpicture}
			\draw[<-] (-4.8,6)--(-4.2,5.7);
			\draw[<-] (-4.8,0)--(-4.2,.3);
			\draw[<-] (-1,5.1)--(-1.6,5.1);
			\draw[<-] (-1,3.1)--(-1.6,3.1);
			\draw[<-] (-1,.9)--(-1.6,.9);
			\draw[<-] (-4.8,0)--(-4.2,.3);
			\draw[<-] (-2.9,1.7)--(-2.9,2.4);
			\draw[<-] (-2.9,4.3)--(-2.9,3.6);
			\draw[<-] (4.8,6)--(4.2,5.7);
			\draw[<-] (4.8,0)--(4.2,.3);
			\draw[<-] (1,5.1)--(1.6,5.1);
			\draw[<-] (1,3.1)--(1.6,3.1);
			\draw[<-] (1,.9)--(1.6,.9);
			\draw[<-] (4.8,0)--(4.2,.3);
			\draw[<-] (2.9,1.7)--(2.9,2.4);
			\draw[<-] (2.9,4.3)--(2.9,3.6);
			\path [draw, ultra thick, blue] (-5,6)--(-3,5)--(-1,5);
			\path [draw, ultra thick, color=green!60!black,ultra thick] (-5,0)--(-3,1)--(-1,1);
			\path [draw] (-3,3)--(-1,3);
			\path [draw] (-3,1)--(-3,5);
			\path [draw, ultra thick, blue] (5,6)--(3,5)--(1,5);
			\path [draw, ultra thick, color=green!60!black,ultra thick] (5,0)--(3,1)--(1,1);
			\path [draw] (3,3)--(1,3);
			\path [draw] (3,1)--(3,5);
			\draw[dashed] (-3,3) ellipse (1.3 and 2.3);
			\draw[dashed] (3,3) ellipse (1.3 and 2.3);
			\node at (-1,5)[right]{$k_1$};
			\node at (-1,3)[right]{$k$};
			\node at (-1,1)[right]{$k_2$};
			\node at (1,5)[left]{$k_4$};
			\node at (1,3)[left]{$k'$};
			\node at (1,1)[left]{$k_3$};
			\node at (-5,6)[left]{$p_1$};
			\node at (5,6)[right]{$p_4$};
			\node at (-5,0)[left]{$p_2$};
			\node at (5,0)[right]{$p_3$};
			\node at (-3,4)[left]{$q_1$};
			\node at (3,4)[right]{$q_4$};
			\node at (-3,2)[left]{$q_2$};
			\node at (3,2)[right]{$q_3$};
		\end{tikzpicture}
	\end{center}
	\caption{\label{fig:A5cl} 
		Kinematic conventions for the three-particle cut.
		The lines in bold represent classical states. The drawing inside the dashed bubbles does not represent a specific topology, but just provides a visual help to recall the definition of the $q_i$ variables.}
\end{figure}

We will take a direct approach based on the 3-particle unitarity cut, which can be evaluated starting from the tree-level $2 \to 3$ inelastic amplitude $A_5$ and taking its ``square'' to obtain the corresponding contribution to the imaginary part of the elastic two-loop amplitude, schematically
\begin{equation}
	\begin{split}
		\left[\operatorname{Im} 2 A_2\right]_{3pc} =& \int \frac{d^{D}k}{(2\pi)^{D}} \frac{d^{D}k_1}{(2\pi)^{D}} \frac{d^{D}k_2}{(2\pi)^{D}}\, (2\pi)^D\delta(p_1+p_2+k_1+k_2+k)\\
		&  2\pi\theta(k^0)\,\delta(k^2)  \ 2\pi\theta(k_1^0)\,\delta(k_1^2+m_1^2)  \ 2\pi\theta(k_2^0)\,\delta(k^2_2+m^2_2)\   | {A}_{5}|^2\,,
		\label{inelasticA}
	\end{split}
\end{equation}
where $k$ and $k_{1,2}$ are the momenta of the massless and the two massive particles involved in the cut, while $p_{1,2}$ denote the massive external momenta, as in Figure~\ref{fig:A5cl}.
We will specify more precisely how $|A_5|^2$ must be understood in the next subsection.
Starting from \eqref{inelasticA}, after taking the classical limit, a Fourier transform to impact-parameter space directly yields $\operatorname{Im}2\delta_2$, which can be interpreted as the total number of gravitons emitted in collisions at given energy and impact parameter. Moreover, multiplying the integrand in~\eqref{inelasticA} by  $k^\mu$, the very same steps lead instead to the total emitted energy-momentum for such processes.

The approach we adopt is thus complementary to the one of~\cite{Herrmann:2021lqe} as far as the integrand construction is concerned.  In our approach we determine the integrand of \eqref{inelasticA} using the 5-point amplitude in \eqref{ampliS9y}, while, in the latter reference, the 3-particle cut is extracted from the $2\to2$ elastic amplitude.
However, once the integrand is obtained, we then evaluate the phase-space integrals by trading the mass-shell delta functions for cut propagators and reducing the integrations to the cut master integrals discussed in Section~\ref{sec:cutto}, borrowing ideas and techniques from the reverse unitarity approach \cite{Anastasiou:2002yz,Anastasiou:2002qz,Anastasiou:2003yy,Anastasiou:2015yha,Herrmann:2021lqe}.

Our strategy allows us to treat both the ${\cal N}=8$ case and pure GR essentially on the same ground, bypassing the explicit form of the full 2-loop integrand in GR.\footnote{The integrand used as a starting point in~\cite{Bern:2019crd} is tailored to the derivation of the potential contribution and is not sufficient for extracting the radiation reaction effects and the imaginary part of the 3PM eikonal. Rather, the full soft-region expression in \cite{Herrmann:2021toa} should be employed.} 
In the end, we can compare the result for $\operatorname{Im} 2 \delta_2$ in ${\cal N}=8$ supergravity with~\eqref{CR15} finding perfect agreement, while the GR result is new, provides a confirmation of the general considerations of Section~\ref{connection}, and can further serve as a crosscheck for a full derivation of the 3PM eikonal in that case. 
Performing the analogous calculations for the radiated energy and momentum, we also reproduce the results of~\cite{Herrmann:2021lqe}.

Of course it is often interesting to consider differential (as opposed to inclusive) quantities such as the radiated energy per unit of frequency and/or per solid angle. In this case the direct approach taken here is probably the only one available since
it allows one to obtain the full transition rate $|A_5|^2$ without dropping any of its pieces at intermediate steps, not even scaleless terms that cancel out in the full integral \eqref{inelasticA}. Armed with the complete integrand one can then integrate (numerically if needed) on any of the remaining independent variables, obtaining for instance the graviton emission spectrum or the emitted power spectrum. Also, the remaining quantities can be expressed by using different variables depending on what is most convenient. For instance, one can take a $(D-2)$-dimensional Fourier transform to impact parameter space of $A_5$ and obtain the waveforms in terms of the orbital angular momentum and the frequency. This is what was done in~\cite{DiVecchia:2021ndb}, albeit only for the leading soft contribution, {\rm i.e.}~the leading term of $A_5$ as $k\to 0$. If one starts from the full classical amplitude $A_5$, as for instance discussed in~\cite{Kovacs:1978eu} for the non-relativistic limit and recently done exactly in~\cite{Mougiakakos:2021ckm}, then instead of the simple rational dependence on $b$, the waveforms involve Bessel functions, which provides an exponential suppression for very large frequencies. Alternatively, it is possible to perform a $(D-1)$-dimensional Fourier transform and write the waveforms in terms of the retarded time and recently refs.~\cite{Mogull:2020sak,Mougiakakos:2021ckm} showed how to derive the results of~\cite{Kovacs:1978eu} starting directly from a Feynman diagram approach.

\subsection{Direct unitarity in momentum space}
\label{pspace}

In this subsection we focus on inclusive quantities, such as the imaginary part of the 3PM eikonal or the total radiated energy-momentum to 3PM order, as discussed above.

As a starting point to calculate the five-point amplitude $A_5^{\mu\nu}(p_1,p_2;k_1,k_2,k)$
on the left in Figure~\ref{fig:A5cl}, we use the $2\to 3$ tree-level amplitude in Eq.~(3.1) of~\cite{DiVecchia:2020ymx}, which had been obtained from the low energy limit of a string amplitude in a toroidal compactification. Thus, in particular, the theory naturally includes the contribution of the dilaton, together with vectors and scalars arising in the Kaluza--Klein compactification. We then enforce the classical limit: using the conventions summarised in Fig.~\ref{fig:A5cl}, so that
\begin{equation}\label{q1q2k}
	q_1=p_1+k_1\,,\qquad
	q_2=p_2+k_2\,,\qquad
	q_1+q_2+k=0\,,
\end{equation} 
we take the momenta $q_{1,2}$ and $k$ to be simultaneously small, of the order of the elastic momentum transfer $q$ (see \eqref{hierarchy}),
\begin{equation}\label{q1q2simq}
	q_1\sim q_2 \sim k \sim \mathcal O(q)\,.
\end{equation}
Equivalently, reinstating momentarily $\hbar$, this can be regarded as a formal $\hbar\to0$ limit in which the wavenumbers $q_{1,2}/\hbar$ and $k/\hbar$ are held fixed. The particles corresponding to the bold lines are classical and so their energy and momentum are instead independent of $q$, to leading order,
\begin{equation}\label{leadingp1p2k1k2}
	p_{1} \sim -m_1 u_1^\mu + \mathcal O(q) \sim -k_1\,,\qquad 
	p_{2} \sim -m_2 u_2^\mu + \mathcal O(q)\sim -k_2\,.
\end{equation}

To include contributions stemming from the graviton and the dilaton, together with Kaluza--Klein vectors and scalars, simultaneously, we find it convenient to once again promote all vectors $p_{1,2}^\mu$, $k_{1,2}^\mu$, $q_{1,2}^\mu$, and $k^\mu$ to $10$-dimensional vectors $P_{1,2}^M$, $K_{1,2}^M$, $q_{1,2}^M$, and $k^M$. Their precise relation to the effective four-dimensional kinematics depends on the theory under consideration and will be specified below. 

The leading contribution to $A_5$ in the classical limit is of order $\mathcal O(q^{-2})$ and reproduces the result of~\cite{Goldberger:2016iau,Luna:2017dtq,Mogull:2020sak}. It can be written in the following convenient form 
\begin{equation}
	\label{ampliS9y}
	\begin{split}
		&A_{5} ^{M N} = (8\pi G)^{\frac{3}{2}}   \Bigg\{\frac{8\left(P_1 k   P_2^M - P_2 k P_1^M\right)\left( P_1 k P_2^N - P_2 k P_1^N \right)}{q_1^2 q_2^2 } \\
		& +  8 P_1P_2 \Bigg[\frac{P_1^M P_1^N \frac{kP_2}{kP_1}- \! P_1^{(M} P_2^{N)}}{q_2^2} + \frac{P_2^M P_2^N \frac{kP_1}{kP_2}-\! P_1^{(M} P_2^{N)}}{q_1^2} 
		- 2 \frac{P_1k   P_2^{(M} q_1^{N)} \! -\! P_2 k P_1^{(M} q_1^{N)}}{q_1^2 q_2^2 } \Bigg] \\
		& + \beta \Bigg[ -\frac{P_1^M P_1^N (kq_1)}{(P_1k)^2 q_2^2}
		-
		\frac{P_2^M P_2^N (kq_2)}{(P_2k)^2 q_1^2}  +2 \left(
		\frac{P_1^{(M} q_1^{N)}}{(P_1k)q_2^2} 
		- \frac{P_2^{(M} q_1^{N)} }{(P_2k)q_1^2}+\frac{q_1^M q_1^N}{q_1^2 q_2^2} \right)\Bigg]  \Bigg\} \;,
	\end{split}
\end{equation}
where $P_1^{(M} P_2^{N)} = \frac{1}{2}(P_1^M P_2^N+P_1^N P_2^M)$, while the quantity $\beta$ is defined in~\eqref{eq:betan8gr} below depending on the theory under consideration. The main feature of~\eqref{ampliS9y} is that it satisfies $k_M   A_{5} ^{M N} = k_N   A_{5} ^{M N}=0$ for arbitrary values of the free index, which makes the calculations in general dimensions easier~\cite{KoemansCollado:2019ggb,Kosmopoulos:2020pcd}. 

When focusing on ${\cal N}=8$ supergravity, it is convenient to choose the following 10D kinematics~\cite{DiVecchia:2021ndb}\footnote{For sake of simplicity here we focus directly on the case $\phi=\frac{\pi}{2}$ in the notations of that reference.} 
\begin{equation}
	\label{eq:pn8}
	\begin{split}
		P_1 &= (p_1;0,0,0,0,0,m_1)\,,\qquad \quad P_1^2=0\,,\\
		P_2 &= (p_2;0,0,0,0,m_2,0)\,,\qquad \quad P_2^2=0\,,\\
		K_1 &= (k_1;0,0,0,0,0,-m_1)\,,\qquad K_1^2=0\,,\\
		K_2 &= (k_2;0,0,0,0,-m_2,0)\,,\qquad K_2^2=0\,,\\
	\end{split}
\end{equation}
where $p_{1,2}$ and $k_{1,2}$ are 4D momenta. In contrast, in the case of GR all momenta are 4D and so we have
\begin{equation}
	\label{eq:pgr}
	\begin{split}
		P_1 &= (p_1;0,0,0,0,0,0)\,,\qquad \,P_1^2=-m_1^2\,,\\
		P_2 &= (p_2;0,0,0,0,0,0)\,,\qquad \,P_2^2=-m_2^2\,,\\
		K_1 &= (k_1;0,0,0,0,0,0)\,,\qquad K_1^2=-m_1^2\,,\\
		K_2 &= (k_2;0,0,0,0,0,0)\,,\qquad K_2^2=-m_2^2\,.\\
	\end{split}
\end{equation}
The momenta $k$ and $q_{1,2}$ are always non-trivial only in the uncompact 4D. 
Another difference between ${\cal N}=8$ and GR is related to the contribution of the dilaton as an internal state exchanged between the massive objects. This exchange can only occur in massive ${\cal N}=8$ supergravity and then the parameter $\beta$ in~\eqref{ampliS9y} should read as the $\beta^{{\cal N}=8}$ below, while for GR the contribution of the dilaton has to be subtracted, which can be done simply by using $\beta^{GR}$ instead
\begin{equation}
	\label{eq:betan8gr}
	\beta^{{\cal N}=8} = 4 m^2_1 m^2_2 \sigma^2\,,\quad \beta^{GR} =  4 m^2_1 m^2_2 \left(\sigma^2 -\frac{1}{D-2}\right)\;.
\end{equation}
Moreover, it is useful to realize that, since the amplitude \eqref{ampliS9y} is homogeneous of order $\mathcal O(q^{-2})$ under the scaling \eqref{q1q2simq}, the massive momenta can be safely approximated to leading order in $q$ as in \eqref{leadingp1p2k1k2}.

Let us now describe how the ${\cal N}=8$ supergravity and GR amplitudes are ``squared'' in order to construct the integrand in~\eqref{inelasticA}. 
In both cases, the kinematics is given as in Figure~\ref{fig:A5cl} by
\begin{equation}\label{}
	\begin{split}
		K_1+K_4=0\,,\qquad K_2+K_3=0\,,\qquad k+k'=0\,,
	\end{split}
\end{equation}
and 
\begin{equation}\label{q1q2q3q4q}
	q_1+q_4=P_1+P_4=q\,,\qquad q_2+q_3=P_2+P_3=-q\,.
\end{equation}
In the maximally supersymmetric case, we simply work in 10D and saturate the Lorentz indices with the Minkowski metric so as to get 
\begin{equation}
	\label{eq:n8int}
	|A^{{\cal N}=8}_5|^2 \to A_5^{MN}(P_1,P_2,K_1,K_2,k)\; \eta_{MR}\eta_{NS}\;  A_5^{RS}(P_4,P_3,-K_1,-K_2,-k) \,,
\end{equation}
where we do not need to subtract the contribution of the longitudinally polarised states as the amplitude $A_5$ is transverse. As already stressed, the advantage of this strategy is that, doing so, one effectively encompasses all the contributions that were taken into account separately in~\cite{DiVecchia:2021ndb}, {\em i.e.} besides the graviton and the dilaton, also the vectors and scalars arising from the toroidal compactification. 
For GR all indices become 4D and on the top of this we need to subtract the contribution of the dilaton. Thus we have 
\begin{equation}
	\label{eq:grint}
	|A^{gr}_5|^2 \to A_5^{\mu\nu}(p_1,p_2,k_1,k_2,k)\, \left[\eta_{\mu\rho}\eta_{\nu\sigma}\!-\!\frac{1}{D-2} \eta_{\mu\nu} \eta_{\rho\sigma}\right]\,  A_5^{\rho\sigma}(p_4,p_3,-k_1,-k_2,-k)\,,
\end{equation}
where the structure within square brackets arises from the sum over the physical polarisations and is the one appearing also in the de Donder propagator, simplified by using the symmetry of the $A_5$ in its two Lorentz indices.

Once the integrand $|A_5|^2$ of \eqref{inelasticA} is constructed according to the above strategy, one then needs to perform the phase space integrals. 
This task can be greatly simplified by resorting to the reverse-unitarity strategy \cite{Anastasiou:2002yz,Anastasiou:2002qz,Anastasiou:2003yy,Anastasiou:2015yha,Herrmann:2021lqe}: trading the mass-shell delta functions appearing in the second line of \eqref{inelasticA} with the corresponding cut propagators, one can recast the integral as an effectively more tractable (cut) two-loop integral. 
After performing these steps, we need to decompose the resulting integrand into structures that match the cut integrals evaluated in Section~\ref{sec:cutto}, that is, to isolate terms having each the appropriate denominator structure for a given topology once a suitable identification of the integrated momenta is applied. This decomposition can be performed in several equivalent ways associated to different possible polynomial divisions.

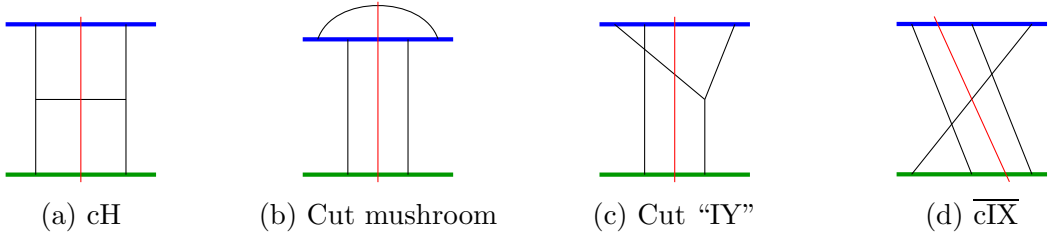
\begin{figure}
	\centering
	\begin{subfigure}[b]{.24\textwidth}
		\centering
		\begin{tikzpicture}
			\path [draw, ultra thick, blue] (-1,2)--(1,2);
			\path [draw, ultra thick, color=green!60!black,ultra thick] (-1,0)--(1,0);
			\draw (-.6,0)--(-.6,2);
			\draw (.6,0)--(.6,2);
			\draw (-.6,1)--(.6,1);
			\draw[red] (0,-.1)--(0,2.1);
		\end{tikzpicture}
		\caption{$\cHt$}
		\label{fig:cutH}
	\end{subfigure}
	\begin{subfigure}[b]{.24\textwidth}
		\centering
		\begin{tikzpicture}
			\path [draw, ultra thick, blue] (-1,1.8)--(1,1.8);
			\path [draw, ultra thick, color=green!60!black,ultra thick] (-1,0)--(1,0);
			\draw (-.4,0)--(-.4,1.8);
			\draw (.4,0)--(.4,1.8);
			\draw (-.8,1.8) .. controls (-.6,2.4) and (.6,2.4) .. (.8,1.8);
			\draw[red] (0,-.1)--(0,2.3);
		\end{tikzpicture}
		\caption{Cut mushroom}
		\label{fig:cutmush}
	\end{subfigure}
	\begin{subfigure}[b]{.24\textwidth}
		\centering
		\begin{tikzpicture}
			\path [draw, ultra thick, blue] (-1,2)--(1,2);
			\path [draw, ultra thick, color=green!60!black,ultra thick] (-1,0)--(1,0);
			\draw (-.4,0)--(-.4,2);
			\draw (.4,0)--(.4,1);
			\draw (.4,1)--(.8,2);
			\draw (.4,1)--(-.8,2);
			\draw[red] (0,-.1)--(0,2.1);
		\end{tikzpicture}
		\caption{Cut ``$\mathrm{IY}$"}
		\label{fig:cutIY}
	\end{subfigure}
	\begin{subfigure}[b]{.24\textwidth}
		\centering
		\begin{tikzpicture}
			\path [draw, ultra thick, blue] (-1,2)--(1,2);
			\path [draw, ultra thick, color=green!60!black,ultra thick] (-1,0)--(1,0);
			\draw (-.8,0)--(.8,2);
			\draw (0,0)--(-.8,2);
			\draw (.8,0)--(0,2);
			\draw[red] (-.5,2.1)--(.5,-.1);
		\end{tikzpicture}
		\caption{$\cRNbar$}
		\label{fig:cutIXbar}
	\end{subfigure}
	\caption{Cut topologies arising in the calculation of Eq.~\eqref{inelasticA}.}
	\label{fig:cuts}
\end{figure}

A simple-minded approach is to just expand out all terms in the product $|A_5|^2$ and use their  denominators written in terms of $p_1 k$, $p_2 k$ and $q_1^2$, $q_2^2$, $q_3^2$, $q_4^2$ to group them into  the following classes:
\begin{equation}\label{Rtopologies}
	\begin{split}
		&\frac{1}{q_1^2 q_2^2 q_3^2 q_4^2}\,,\quad
		\frac{1}{(p_2k)^2 q_1^2 q_4^2}\,,\quad
		\frac{1}{(p_1k)^2 q_2^2 q_3^2}\,,\quad
		\frac{1}{(p_1k)(p_2k)q_1^2 q_3^2}\,,\quad \frac{1}{(p_1k)(p_2k)q_2^2 q_4^2}\,,\\
		&\frac{1}{(p_1k) q_2^2 q_3^2 q_4^2}\,,\quad
		\frac{1}{q_1^2 (p_2k) q_3^2 q_4^2}\,,\quad
		\frac{1}{q_1^2 q_2^2 (p_3 k) q_4^2}\,,\quad
		\frac{1}{q_1^2 q_2^2 q_3^2 (p_4k)}\,,
	\end{split}
\end{equation} 
where the massive propagators are in general raised to non-negative integer powers.
These structures can be associated to the topologies of Figure~\ref{fig:cuts}.
With proper identifications of $q_1$ and $q_2$ in terms of $\ell_1$, $\ell_2$ and $q$, all such structures can be embedded in the $\cHt$ or $\cRNbar$
families of Section~\ref{sec:cutto}, possibly after use of crossing symmetry and of the $u_1\leftrightarrow u_2$ symmetry. For example, letting 
\begin{equation}\label{}
	q_1=-\ell_2+q\,,\qquad q_2=-\ell_1\,,\qquad k=\ell_1+\ell_2-q\,,
\end{equation}
the first structure of Eq.~\eqref{Rtopologies} can be recognized as belonging to the $\cHt$ family: inserting the cut propagators, denoted with an underscore,
\begin{equation} 
	\frac{1}{q_1^2 q_2^2 q_3^2 q_4^2}
	\frac{m_1 m_2}{\underline{(p_1q_1)}\,\underline{(p_2q_2)}\,\underline k^2}
	\simeq  \frac{1}{\ell_1^2\ell_2^2(\ell_1-q)^2(\ell_2-q)^2}
	\frac{1}{\underline{(u_1\ell_2)}\,\underline{(u_2\ell_1)}\,\underline{(\ell_1+\ell_2-q)}^2}
	\,.
\end{equation}
Moreover, letting
\begin{equation}\label{}
	q_1=-\ell_1-\ell_2\,,\qquad q_2 = \ell_2 \,,\qquad k=\ell_1\,,
\end{equation}
we find, for the first structure in the second line of \eqref{Rtopologies},
\begin{equation}\label{}
	\frac{1}{(p_1k) q_2^2 q_3^2 q_4^2}
	\frac{m_1 m_2}{\underline{(p_1q_1)}\,\underline{(p_2q_2)}\,\underline k^2}
	\simeq \frac{1}{(u_1\ell_1)\ell_2^2(\ell_2+q)^2(\ell_1+\ell_2+q)^2}\frac{1}{\underline{(u_1(\ell_1+\ell_2))}\,\underline{(u_2\ell_2)}\,\underline{\ell_1}^2}\,,
\end{equation}
which we recognize as belonging to the $\cRNbar$ family up to an (immaterial) interchange of $u_1$ and $u_2$ and the exploitation of crossing symmetry.
After the appropriate family is identified, the integral under scrutiny can be reduced to the corresponding cut master integrals discussed in Section~\ref{sec:cutto} using \texttt{LiteRed}.
When expanding the mushroom \ref{fig:cutmush} and ``$\mathrm{IY}$'' \ref{fig:cutIXbar}  topologies in terms of the $\cRNbar$ family, however, one should note that, again via crossing symmetry,
\begin{equation}\label{contamination}
	\tilde G_{0,0,\underline{k_1},\underline{k_2},\underline1,1,0,0,1} \leftrightarrow  G_{0,\underline{k_2},\underline{k_1},0,0,1,\underline1,0,1}
\end{equation}  
which relates integrals of the type $\tilde G_{0,0,\underline{k_1},\underline{k_2},\underline1,1,0,0,1}$ written in terms of the $\cRNbar$ family to  integrals $G_{0,\underline{k_2},\underline{k_1},0,0,1,\underline1,0,1}$ appearing in the definition of $\cHt$ masters, $f_{\cHt,8}$ for the even sector and $f_{\cHt,12}$ for the odd sector.

A slightly more systematic approach is to start out by fixing a particular choice of independent momenta, for instance
\begin{equation}\label{}
	q_1=-\ell_1\,,
	\qquad
	q_2=\ell_1+\ell_2\,,
	\qquad
	k=-\ell_2\,,
\end{equation} 
thus rewriting the integrand in the form
\begin{equation}\label{numden}
	\frac{N(u_1\ell_2,u_2\ell_2,\ell_1q,\ell_2q,u_1\ell_1,u_2\ell_1,\ell_1^2,\ell_1\ell_2,\ell_2^2)}{D_1^{\alpha_1} D_2^{\alpha_2}D_3 D_4 D_5 D_6 \underline{D_7}\, \underline{D_8}\, \underline{D_9}}\,,
\end{equation} 
where $N$ is a Lorentz-invariant numerator and the denominators are
\begin{equation}\label{}
	\begin{split}
		D_1&=u_1\ell_2\,,\qquad
		D_2=u_2\ell_2\,,\qquad
		D_3=\ell_1^2\,,\\
		D_4&=(\ell_1+\ell_2)^2\,,\qquad
		D_5=(\ell_1+q)^2\,,\qquad
		D_6=(\ell_1+\ell_2+q)^2\,\\
		D_7&=\ell_2^2\,,\qquad
		D_8=u_1\ell_1\,,\qquad
		D_9=u_2(\ell_1+\ell_2)\,.
	\end{split}
\end{equation}
In particular, the powers $\alpha_1$ and $\alpha_2$   in \eqref{numden} take the value $2$ in $\mathcal N=8$ supergravity and $4$ in GR.
One can then apply simple integrand-reduction techniques using basic tools from algebraic geometry such as Groebner basis and syzygy relations (see e.g. \cite{Zhang:2016kfo}). 
To simplify matters, one can first trade $N$ with the remainder $R$ of the polynomial division between $N$ itself and the Groebner basis of $D_7$, $D_8$ and $D_9$, thus eliminating many terms that vanish on-shell. Denoting by $\tilde g_7$, $\tilde g_8$, $\tilde g_9$ the Groebner basis of $D_7$, $D_8$ and $D_9$, this means 
\begin{equation}\label{}
	N = R + \sum_{i=7}^9\tilde q_i \tilde g_i = R + \sum_{i,j=7}^9\tilde q_i \tilde a_{ij} D_j\,, 
\end{equation}
for some quotient polynomials $\tilde q_7$, $\tilde q_8$, $\tilde q_9$ and a suitable $3\times3$ conversion matrix $\tilde a_{ij}$.
Then, one can perform the polynomial division between $R$ and the Groebner basis of the full set of denominators, thus generating a number of integrands with simpler sets of denominators, determined by the corresponding quotients. One can also check that the remainder of this division is zero. Explicitly, letting $g_i$ with $i=1,\ldots,9$ denote the Groebner basis of $D_1,\ldots, D_9$, with conversion matrix $a_{ij}$ such that
\begin{equation}\label{}
	g_i = \sum_{j=1}^9 a_{ij} D_j\,,	
\end{equation}
one then finds quotient polynomials $q_i$ such that
\begin{equation}\label{}
	R = \sum_{i=1}^{9} q_i g_i=\sum_{ij=1}^9 q_i a_{ij} D_j\,,
\end{equation}
and therefore
\begin{equation}\label{}
	\begin{split}
		\frac{R}{D_1^{k_1}D_2^{k_2}D_3D_4D_5 D_6\underline{D_7}\, \underline{D_8}\, \underline{D_9}}
		&=
		\sum_i \frac{q_i a_{i1}}{D_1^{k_1-1}D_2^{k_2}D_3D_4D_5 D_6\underline{D_7}\, \underline{D_8}\, \underline{D_9}}\\
		&+\cdots+
		\sum_i \frac{q_i a_{i6}}{D_1^{k_1}D_2^{k_2}D_3D_4D_5 \underline{D_7}\, \underline{D_8}\, \underline{D_9}}\,,
	\end{split}
\end{equation}
up to terms where the cut propagators would again cancel out. One can automatise this procedure and repeat it for each of the newly generated ratios, collecting along the way those terms for which the remainder of the polynomial division is nonzero, which constitute the irreducible integrands. Moreover, the form of the numerators generated at each step can be further simplified by using the syzygy identities obeyed by the relevant denominators.
To implement these steps we used the very convenient software \texttt{Singular} \cite{DGPS} in combination with its Mathematica interface. Proceeding this way, we were able to regroup the integrand into three contributions associated to the $\cHt$ and $\cRNbar$ topologies of figures \ref{fig:cutH} and $\ref{fig:cutIXbar}$, in the $\mathcal N=8$ case, and into seven contributions associated to the $\cHt$, ``$\mathrm{IY}$'' and $\cRNbar$ of figures \ref{fig:cutH}, \ref{fig:cutIY} and \ref{fig:cutIXbar} in the GR case. All integrals can be then mapped to the $\cRT$ and $\cRNbar$ families via appropriate changes of variables and judicious use of the $u_1\leftrightarrow u_2$ symmetry. For instance, for the denominator
\begin{equation}\label{}
	\frac1{(u_1\ell_2)\ell_1^2(\ell_1+\ell_2)^2(\ell_1+\ell_2+q)^2 \underline{\ell_2}^2\,
		\underline{u_1\ell_1}\,\underline{u_2(\ell_1+\ell_2)}}\,,
\end{equation} 
we can perform the change of variables $\ell_1\mapsto -\ell_1-\ell_2-q$, followed by $\ell_1\leftrightarrow\ell_2$ and by the harmless replacement $u_1\leftrightarrow u_2$ to obtain the denominator
\begin{equation}\label{}
	\frac1{(u_2\ell_1)\ell_2^2(\ell_1+\ell_2+q)^2(\ell_2+q)^2 \underline{\ell_1}^2\,
		\underline{u_1\ell_2}\,\underline{u_2(\ell_1+\ell_2)}}\,,
\end{equation}
which is appropriate for the $\cRNbar$ family, up to crossing symmetry. The observation \eqref{contamination} again comes in handy when performing the reduction of integrals associated with figure \ref{fig:cutIY}.
We include in an ancillary file the result of this decomposition of the  ${\cal N}=8$ and the GR integrands.

In the calculation of the three-particle cut contribution to the imaginary part of $A_2$, Eq.~\eqref{inelasticA}, only the even sectors of master integrals are actually involved. As already mentioned, the 3-particle cut is the only and full contribution to $\operatorname{Im} 2 \delta_2$, so, in the case of ${\cal N}=8$, evaluating \eqref{inelasticA} and taking the $(D-2)$-dimensional Fourier transform to impact parameter space according to~\eqref{impparsp}, we reproduce Eq.~\eqref{CR15}. For the case of GR we obtain instead the new result
\begin{equation}
	\label{GR15}
	\begin{aligned}
		\operatorname{Im} 2 \delta^{(gr)}_2 = & ~\frac{2 m_1^2 m_2^2 G^3}{\pi b^2} \frac{(2\sigma^2-1)^2}{(\sigma^2-1)^2}\Bigg\{\!-\!\frac{1}{\epsilon} \left[\frac{8-5\sigma^2}{3}-\frac{\sigma(3-2\sigma^2)}{(\sigma^2-1)^{\frac{1}{2}}} \cosh^{-1} (\sigma) \right] \\ & + \left(\log(4(\sigma^2-1))- 3 \log(\pi b^2 {\rm e}^{\gamma_E})\right) \left[\frac{8-5\sigma^2}{3}-\frac{\sigma(3-2\sigma^2)}{(\sigma^2-1)^{\frac{1}{2}}} \cosh^{-1} (\sigma) \right] \\ & + (\cosh^{-1}(\sigma))^2 \left[\frac{ \sigma \left(3 -2 \sigma^2\right) }{(\sigma^2-1)^{\frac{1}{2}}}-2\frac{4 \sigma^6-16 \sigma ^4+9 \sigma ^2+3}{\left(2 \sigma ^2-1\right)^2}\right] \\ &  +\cosh^{-1}(\sigma)\left[\frac{\sigma  \left(88 \sigma^6-240 \sigma^4+240 \sigma^2-97\right)}{3  \left(2 \sigma ^2-1\right)^2 (\sigma ^2-1)^{\frac{1}{2}}} \right] \\ & +\frac{\sigma(3-2\sigma^2)}{(\sigma^2-1)^\frac{1}{2}}  \operatorname{Li}_2(1-z^2) +  \frac{-140 \sigma^6+220 \sigma^4-127 \sigma^2+56}{9  \left(2 \sigma ^2-1\right)^2} \Bigg\}\;.
	\end{aligned}
\end{equation}
The first line agrees with the result obtained in~\cite{DiVecchia:2021ndb} by using the soft approximation which is sufficient to capture the IR-divergent term. The second line matches the analytic behaviour as $\sigma\to 1$ derived on general grounds in Section~\ref{connection}. The ultrarelativistic limit $\sigma \to \infty$ reproduces the universal behaviour derived in~\cite{DiVecchia:2020ymx} by using an argument based on analyticity/crossing and an explicit calculation in the double Regge limit. In particular the terms proportional to $(\sigma \log(2\sigma))^2$ in the second and third line cancel and the leading term is actually $(2\sigma)^2 \log(2\sigma)$ in agreement with~\cite{DiVecchia:2020ymx}. 
Combining the conservative eikonal phase that can be extracted from the scattering angle given in \cite{Bern:2019nnu,Bern:2019crd} with the radiation-reaction contribution derived in \cite{DiVecchia:2021ndb} and confirmed by \eqref{GR15}, let us complement \eqref{GR15} with the corresponding (IR-finite) real part: 
\begin{equation}
	\label{ReGR15}
	\begin{aligned}
		&\operatorname{Re} 2 \delta^{(gr)}_2 = 
		\frac{4G^3m_1^2m_2^2}{b^2}\Bigg\{
		{{\color{blue}
			\frac{(2 \sigma^2-1)^2(8-5\sigma^2)}{6(\sigma^2-1)^2}
		}} 
		-\frac{\sigma(14\sigma^2+25)}{3\sqrt{\sigma^2-1}}
		\\ &
		+\frac{s (12\sigma^4-10\sigma^2+1)}{2m_1m_2(\sigma^2-1)^\frac{3}{2}}
		 +\cosh^{-1}\sigma \left[
		 {{\color{blue}
		 \frac{\sigma(2\sigma^2-1)^2(2\sigma^2-3)}{ 2(\sigma^2-1)^\frac{5}{2}}
		}}
		+\frac{-4\sigma^4+12\sigma^2+3}{\sigma^2-1}
		\right]
		\Bigg\}.
	\end{aligned}
\end{equation}
The first term in the curly brackets and the first term in the square brackets (highlighted in blue) are the radiation-reaction contributions $\operatorname{Re}2\delta_2^{(rr)}$~\cite{Damour:2020tta,DiVecchia:2021ndb}, while the remaining ones are those arising from the potential region~\cite{Bern:2019nnu,Bern:2019crd}. One can easily check that \eqref{ReGR15}  reproduces the universal result of \cite{Amati:1990xe} in the ultra relativistic limit. In the non-relativistic limit the new radiation reaction terms start at 2.5 PN order, while the potential region ones contain a  probe-limit contribution (the one proportional to $s m_1 m_2$) starting at 1 PN, with the remaining ones (with an $m_1^2 m_2^2$ coefficient)  starting at 2 PN order.

We conclude this section by briefly describing how the same approach can be used to derive the total radiated energy and momentum $R^\mu$. It is sufficient to multiply the integrand of eq.~\eqref{inelasticA} by an extra factor of $k^\mu$. The calculation and decompositions of $|A_5|^2$ described above can be imported verbatim, as this factor of $k^\mu$ just goes along for the ride, and the only new step is to project the resulting vector integrals along the vectors $u_1^\mu$, $u_2^\mu$ and $q^\mu$, after having identified the appropriate integral topologies. Each contribution now can be calculated again by reducing it to scalar integrals by using \texttt{LiteRed} and then by using again the results of Section~\ref{sec:cutto}. Of course in this case the integrals of the odd sectors appear and we obtain Eq.~(9) of~\cite{Herrmann:2021lqe}: after all terms are taken into account with the appropriate sign dictated by the flow of the cut $k^\mu$ line, the $q^\mu$ component vanishes, while the $u_1^\mu$ and $u_2^\mu$ components are identical.
Explicitly, we obtain
\begin{equation}\label{Erad}
	R^\mu
	=
	\frac{\pi G^3m_1^2m_2^2}{b^3}\,\frac{u_1^\mu+u_2^\mu}{\sigma+1} 
	\left[
	f_1(\sigma) + f_2(\sigma)\,\log\frac{\sigma+1}{2} + f_3(\sigma)\,\frac{\sigma\cosh^{-1}\sigma}{2\sqrt{\sigma^2-1}}
	\right]\,,
\end{equation}
where in $\mathcal{N}=8$
\begin{equation}\label{f1f2f3}
	f_1=\frac{8\sigma^6}{(\sigma^2-1)^{\frac{3}{2}}}\,,\qquad
	f_2=-\frac{8\sigma^4}{\sqrt{\sigma^2-1}}\,,
	\qquad
	f_3=\frac{16\sigma^4(\sigma^2-2)}{(\sigma^2-1)^{\frac{3}{2}}}\,,
\end{equation}
while in GR
\begin{equation}\label{f1f2f3gr}
	\begin{split}
		f_1&=\frac{210\sigma^6-552\sigma^5+339\sigma^4-912\sigma^3+3148\sigma^2-3336\sigma+1151}{48(\sigma^2-1)^{\frac{3}{2}}}\,,\\
		f_2&=-\frac{35\sigma^4+60\sigma^3-150\sigma^2+76\sigma-5}{8\sqrt{\sigma^2-1}}\,,\\
		f_3&=\frac{(2\sigma^2-3)(35\sigma^4-30\sigma^2+11)}{8(\sigma^2-1)^{\frac{3}{2}}}\,,
	\end{split}
\end{equation}
finding perfect agreement with the results of \cite{Herrmann:2021lqe}.
It is interesting to compare the previous exact calculations of the total radiated energy with the ones that can be obtained
by inserting only the leading Weinberg soft term in the  five-point amplitudes and by choosing the  upper limit of integration to be
\begin{eqnarray}
 \omega_{max} = \frac{\sqrt{\sigma^2-1}}{8b} \frac{(m_1+m_2) \pi^2 }{\sqrt{m_1^2 +m_2^2 +2m_1m_2 \sigma}} 
\label{5}
\end{eqnarray}
in order to match as much as possible the two results. For massive ${\cal{N}}=8$ supergravity  we reproduce precisely $f_1$ and $f_3$ of Eq. \eqref{f1f2f3} but we get $f_2=0$. In the case of GR we get:
\begin{eqnarray}
f_1=\frac{(2\sigma^2-1)^2(8-5\sigma^2)}{3(\sigma^2-1)^{\frac{3}{2}}}~~;~~f_2 =0~~;~~f_3  =\frac{2(2\sigma^2-1)^2 (2\sigma^2-3)}{(\sigma^2-1)^{\frac{3}{2}}}
\label{f1f3z}
\end{eqnarray}
 matching the results for $f_1$ and $f_3$ in Eqs. \eqref{f1f2f3gr} only for $\sigma \to 1^+$. The vanishing of $f_2$ follows, in both cases, from  the fact that this contribution comes from diagrams where the massless particle is emitted from an internal leg and thus do not contribute to Weinberg's soft limit.

Let us also mention that, while the approach developed here directly captures the classical result \eqref{Erad}, the analogous calculation based on taking suitable cuts of the elastic two-loop amplitude relies on the cancellation of spurious superclassical contributions $\mathcal O(b^{-1})$, $\mathcal O(b^{-2})$ which appear at intermediate steps.

\subsection{Direct unitarity in impact-parameter space}
\label{bspace}

We now consider an alternative approach  to the one of the previous subsection. This approach was adopted in~\cite{DiVecchia:2021ndb}, even if in that paper the analysis was restricted to the Weinberg limit~\cite{Weinberg:1965nx} of the inelastic amplitude $A_5$, i.e. to the leading term for $k\ll q_i$. The basic idea is to introduce the Fourier transform to impact parameter space at an earlier stage in order to rewrite Eq.~\eqref{inelasticA} as the product of two inelastic amplitudes $\tilde{A}_5$ expressed in terms of the impact parameter $b$, which is then integrated over the momentum $k$ of the massless particle in the cut. Most importantly, the integrand obtained by this procedure represents the  number of produced quanta per unit of phase space for each massless particle species and polarisation. 

As a first step we write Eq.~\eqref{inelasticA} for the three-particle cut in the explicit form
\begin{eqnarray}
	\begin{split}
		[\operatorname{Im} 2A_2]_{3pc} &= \int \frac{d^{D-1} k_1}{(2\pi)^{D-1} 2k_1^0} \frac{d^{D-1} k_2}{(2\pi)^{D-1}2k_2^0}  \frac{d^{D-1} k}{(2\pi)^{D-1}2k^0}  \\
		& \times   {\it A}_5^{M N}(P_1, P_2 , K_1, K_2 ,k)  \left[ \sum_i   \epsilon_{M N}^{(i)} \epsilon_{R S}^{(i)}\right ] {\it A}_5^{RS} (P_4, P_3 , -K_1,-K_2, -k) \\
		& \times  (2\pi)^{D} \delta^{(D)} (p_1+p_2 + k_1 +k_2 +k) \,\, ,
	\end{split}
	\label{3CT1}
\end{eqnarray}
where $k_1^0$, $k_2^0$, $k^0$ are the energies of the three intermediate particles and we are using the kinematics of Eqs.~\eqref{eq:pn8} or~\eqref{eq:pgr} depending on whether we consider ${\cal N}=8$ or GR. The sum over physical polarisations in the square brackets reproduces the structures in~\eqref{eq:n8int} or~\eqref{eq:grint} again depending on the case under consideration. 

To fix ideas, it is useful to further specialize the kinematics \eqref{covBreit} of the elastic process, separating transverse and longitudinal components with respect to the classical direction of propagation:
\begin{equation}
	\begin{split}
		p_1 &= \left(-E_1,\frac{\mathbf{q}}{2}, -{\bar{p}}\right),\qquad
		p_4 = \left(E_1,  \frac{\mathbf{q}}{2}, {\bar{p}}\right),
		\\
		p_2&= \left(-E_2,  -\frac{\mathbf{q}}{2}, {\bar{p}}\right),\qquad
		p_3 = \left(E_2,  -\frac{\mathbf{q}}{2}, -{\bar{p}}\right),
	\end{split}
	\label{TC1a}
\end{equation}
Here and in the following, where needed, transverse $(D-2)$-vectors shall be denoted by boldface letters.
Conversely, components along the $(D-1)$th direction, referred to as longitudinal, shall be labelled by a superscript $L$. 
On the other hand, spatial $(D-1)$-vectors shall be denoted with an arrow.
For instance, for the $q_{1,2}$ defined in \eqref{q1q2k} and the massless momentum $k$, 
\begin{equation}\label{}
	q_{1,2}=(q_{1,2}^0,\vec q_{1,2})=(q_1^0,\mathbf q_{1,2},q_{1,2}^L)\,,\qquad
	k=(k^0,\vec k)=(k^0,\mathbf k,k^L)\,.
\end{equation}

In Eq.~\eqref{3CT1}, the integral over the longitudinal components $k_{1,2}^L$ of $k_{1,2}$ can be performed using the longitudinal and energy  components of the $\delta$-function of momentum conservation, explicitly
\begin{equation}\label{}
	\begin{split}
		&\delta(k_1^0+k_2^0+k^0+p_1^0+p_2^0)\delta(k_1^L+k_2^L+k^L+p_1^L+p_2^L)\\
		=&
		\delta\left(
		\sqrt{\mathbf k_1^2+(k_1^L)^2+m_1^2}
		+
		\sqrt{\mathbf k_2^2+(k_2^L)^2+m_2^2}
		+
		k^0
		-E_1-E_2
		\right)
		\delta(k_1^L+k_2^L+k^L)\,.
	\end{split}
\end{equation} 
The resulting Jacobian cancels the factors of $k_{1,2}^0$ in the measure and produces an extra factor of $| k_1^0 k_2^L - k_2^0 k_1^L|^{-1}$. We are then left with only the  integrals over the remaining $D-2$ transverse directions: 
\begin{equation}
	\begin{split}
		[\operatorname{Im} 2 A_2]_{3pc} &= \int \frac{d^{D-2} q_1}{(2\pi)^{D-2}} \int \frac{d^{D-2} q_2}{(2\pi)^{D-2}} \int  \frac{d^{D-1} k}{(2\pi)^{D-1}2  k^0}   \\ &\times \frac{1}{4| k_1^0 k_2^L - k_2^0 k_1^L| } \, (2\pi)^{D-2} \delta^{(D-2)} (k+q_1 +q_2) \label{3CT2b} \\  
		&\times   {\it A}_5^{M N}(P_1, P_2 , K_1, K_2 ,k)  \left[ \sum_i   \epsilon_{M N}^{(i)} \epsilon_{R S}^{(i)}\right ] {\it A}_5^{RS} (P_4, P_3 , -K_1,-K_2, -k) \;,
	\end{split}
\end{equation}
where we changed variables in the integration to the $q_{1,2}$ introduced in \eqref{q1q2k} (see also Fig.~\ref{fig:A5cl} for a visual help).
In the classical limit we can safely approximate $k_1^L \simeq \bar p \simeq p$ and $k_2^L\simeq-\bar p \simeq -p$, where $p$ is the absolute value of the momentum in the centre  of mass frame. Similarly $k_1^0\simeq E_1$ and $k_2^0\simeq E_2$, so that, in terms of the total energy $E= \sqrt{s} =E_1+E_2$ in the centre of mass frame, we have $4| k_1^0 k_2^L - k_2^0 k_1^L|\simeq 4E p$. This is precisely the factor appearing in the Fourier transform to impact parameter space \eqref{impparsp}.

In order to treat the two $5$-point amplitudes in Eq.~\eqref{3CT2b} more symmetrically, let us impose the relations in~\eqref{q1q2q3q4q} by introducing  integrals over  $q_3$ and $q_4$ and two extra $\delta$-functions.
This can be done inserting the factor
\begin{equation}
	1=  
	\int d^{D-2} {q}_4 \,\delta^{(D-2)} ({q}_1 +{q}_4-{q})
	\int d^{D-2} {q}_3 \,\delta^{(D-2)} ({q}_3 +{q}_2+q)   
	\label{3CT6}
\end{equation} 
in Eq.~\eqref{3CT2b}. Going to impact parameter space then gives
\begin{align}
	\nonumber
		& 2 \operatorname{Im} 2 \delta_2 (b, s) =  \int \frac{d^{D-2} {q}}{(2\pi)^{D-2}} \, e^{-ib \cdot {q}} \, \frac{[\operatorname{Im} 2 A_2]_{3pc}}{4Ep} =   \int \frac{d^{D-1} k}{(2\pi)^{D-1}2k^0}  \sum_i    \\
		&   \Bigg[  \int \frac{d^{D-2} {q}_1d^{D-2} q_2}{(2\pi)^{D-2}}  \delta^{(D-2)} (q_1 + q_2 + k) \frac{ e^{-i \frac{b}{2}
				 ({q}_1-{q}_2)}}{4Ep}  A_5^{M N}(P_1, P_2 , K_1, K_2 ,k)  \epsilon_{M N}^{(i)}  \Bigg]   \label{3CT8}  \\
		&  \Bigg[  \int \frac{d^{D-2} {q}_3 d^{D-2} {q}_4}{(2\pi)^{D-2}}  \delta^{(D-2)} (q_3 +q_4-k) \frac{e^{-i \frac{b} {2} (q_4-q_3)} }{4Ep}
		{\it A}_5^{RS} (P_4, P_3 , -K_1,-K_2, -k)  \epsilon_{R S}^{(i)}   \Bigg],
	\nonumber
\end{align}
where we used $q= \frac{1}{2}(q_1-q_2 + q_4-q_3)$ in the exponent.
As reviewed in Section~\ref{sec:preliminaries}, $\operatorname{Re} 2 \delta_2$ is directly related to the classical deflection angle, while, as will become clear shortly, $2 \operatorname{Im} 2 \delta_2$ is equal to
the total number of massless states that are produced in a collision of the two massive particles, for fixed energy and impact parameter, to order $G^3$.
The form obtained in Eq.~\eqref{3CT8} suggests also to focus on each one of the square parenthesis separately, which represents, for each $i$, the corresponding wave-form\footnote{In particular, for $i = +, \times, $ it will give the two linear polarizations of the gravitational wave.}  in terms of its frequency and angular direction. Defining $A_{5,i} =  A_5^{M N} \epsilon_{M N}^{(i)}$ we are thus lead to consider:
\begin{equation}
	\label{eq:A5bk}
	\tilde{A}_{5,i}(b,\vec{k}) = \int \frac{d^{D-2} \Delta}{(2\pi)^{D-2}} \frac{e^{-i b  \Delta}}{4Ep}  A_{5,i}(P_1, P_2 , K_1, K_2 ,k) \;,
\end{equation}
where $\mathbf{\Delta} \equiv \frac12 (\mathbf{q}_1 - \mathbf{q}_2)$, and we used the delta function to carry out the integration over  $\mathbf{q}_1 + \mathbf{q}_2$, which sets
\begin{equation}\label{bfq1bfq2bfk}
\mathbf q_1 + \mathbf q_2=-\mathbf k\,.	
\end{equation}
We display only the dependence on $b$ and $k$ after the Fourier transform, even if, of course, $\tilde{A}$ depends also on the masses and  the centre of mass energy. At leading order in the Weinberg limit $k\ll q_{1,2}$ only the last line of~\eqref{ampliS9y} survives and, since in this limit $q_2^2 \simeq q_1^2 \simeq {\mathbf{q}}_1^{\;2}$, the Fourier transform can be easily performed and one obtains the results summarised in Section~4 of~\cite{DiVecchia:2021ndb}. As discussed in that reference, this limit is sufficient to obtain the IR divergent part of $\operatorname{Im} 2 \delta_2$,  the zero-frequency limit of the radiated energy, and the radiation reaction on $\operatorname{Re} 2 \delta_2$.

One can go beyond this approximation by keeping $k$ to be generically of the same order as $q_1, q_2$ as in~\eqref{q1q2simq} and then obtain the full spectrum for each massless particle in the classical limit. In the rest of this section we will illustrate this procedure for the case of the dilaton spectrum in ${\cal N} = 8$ supergravity.

Let us start from the classical amplitude $A_{5, \mathrm{dil}}$  for  dilaton emission that is obtained from Eq. \eqref{ampliS9y} by taking $\beta=4(p_1p_2)^2$ and saturating it  with the dilaton projector
\begin{equation}\label{}
 \epsilon^{(\mathrm{dil})}_{\mu\nu}
 =	
 \frac{\eta_{\mu\nu}}{\sqrt{D-2}}\,.
\end{equation}
Using in particular the exact relations
\begin{equation}\label{piqiqi2}
	2p_1 q_1 = q_1^2\,,\qquad 2p_2q_2=q_2^2\,,\qquad 
	kq_1=-k q_2\,,
\end{equation}
which follow from the mass-shell constraints, and neglecting irrelevant terms in the classical limit,
we obtain
\begin{equation} 
	\begin{split}
		A_{5, \mathrm{dil}} &= - \frac{2(8\pi G)^{\frac{3}{2}}}{\sqrt{D-2}}\Bigg\{(p_1p_2)^2  \left( \frac{m_1^2}{(p_1 k)^2} \frac{2 k q_2}{q_2^2} + \frac{m_2^2}{(p_2 k)^2} \frac{2 k q_1}{q_1^2} \right)      \\
		&+4  (p_1p_2) \left(  \frac{m_1^2 (p_2 k)  }{q_2^2 (p_1 k) }+ \frac{m_2^2 (p_1 k)  }{q_1^2 (p_2 k) } \right) +4 \left(\frac{m_1^2 (p_2 k)^2 + m_2^2 (p_1 k)^2 }{q_1^2 q_2^2}   \right) \Bigg\}\,.
		\label{traceampliB4bis}
	\end{split}
\end{equation} 

Let us now introduce rapidity variables $y_1$, $y_2$ and $y$ according to
\begin{equation}
\begin{split}
	p_1 &= \left(-\bar{m}_1 \cosh y_1,  \frac{\mathbf{q}}{2} ,-\bar{m}_1 \sinh y_1\right),  \\
	p_2 &= \left(-\bar{m}_2 \cosh y_2,  - \frac{\mathbf{q}}{2} ,-\bar{m}_2 \sinh y_2\right), \\
	k &= \left(|{\mathbf{k}}|\cosh y, {\mathbf k},  |{\mathbf{k}}|\sinh y\right),      
	\label{rapid}
\end{split}
\end{equation}
where 
\begin{equation}\label{}
	 y_1 >0\,, \qquad  y_2 <0\,,
\end{equation}
and $\bar m_{1,2}$ were introduced in \eqref{eq:mbar}.
Using the conservation conditions
\begin{equation}\label{}
		k^0+q_1^0+q_2^0=0\,,\quad
		k^L+q_1^L+q_2^L=0\,,
\end{equation}
and solving first two constraints \eqref{piqiqi2} to leading order in the classical limit, so that effectively
\begin{equation}\label{}
	-p_1^0 q_1^0+p_1^L q_1^L\simeq 0\,,\qquad -p_2^0q_2^0+p_2^Lq_2^L\simeq0\,,
\end{equation}
we obtain
\begin{equation}\label{explq1q2L0}
\begin{split}
	q_1^L &= q_1^0 \coth y_1 = |\mathbf k| \, \frac{\sinh y \tanh y_2-\cosh y}{\tanh y_1 - \tanh y_2} \,,\\	
	q_2^L &= q_2^0 \coth y_2 = |\mathbf k| \, \frac{\cosh y-\sinh y \tanh y_1}{\tanh y_1 - \tanh y_2} \,,
\end{split}
\end{equation}
so that
\begin{equation}\label{c1c2}
	\begin{split}
			q_1^2 &= (\mathbf{q}_1)^2 + c_1^2 {{\mathbf{k}}^2}\,,\qquad  c_1 = \frac{\cosh (y-y_2)}{\sinh(y_1-y_2)}\,, \\
		q_2^2 &= (\mathbf{q}_2)^2 + c_2^2 {{\mathbf{k}}^2}\,,\qquad   c_2 = \frac{\cosh (y_1-y)}{\sinh(y_1-y_2)} \,.
	\end{split}
\end{equation}
We note that $c_{1,2}$ are positive. In fact, in the following, they will play the role of effective masses with respect to the traverse dynamics and will regulate some infrared divergences.
We also find:
 \begin{equation}\label{}
 	\begin{split}
k q_1 &= {\mathbf{k}}{\mathbf{q}_1} + {\mathbf{k}}^2 d_1\,,\qquad
d_1 =\frac{\cosh (y-y_2) \sinh(y_1-y)}{\sinh(y_1-y_2)}\,,  \\
k q_2 &= {\mathbf{k}}{\mathbf{q}}_2 + {\mathbf{k}}^2 d_2\,,\qquad
d_2 = \frac{\cosh (y_1-y) \sinh(y-y_2)}{\sinh(y_1-y_2)} \,,
\label{d1d2}
 	\end{split}
 \end{equation}
where 
\begin{equation}\label{}
	d_1 + d_2 = 1\,,
\end{equation}
which ensures $k(q_1+q_2)=0$.

In Eq. \eqref{rapid} we have parametrised $k$ in terms of its rapidity and its two components in transverse space. On the other hand, depending on the problem at hand, it can also be convenient to parametrise $k$ in terms of its frequency $\omega = |{\vec k}|$ and two angles $\theta$ and $\phi$, all measured in the centre of mass frame we defined earlier:
\begin{equation}
	k = \omega (1, \sin \theta \cos \phi, \sin \theta \sin \phi, \cos \theta) \,.
	\label{komega}
\end{equation}
We can then rewrite the previous quantities $c_{1,2}$ and $d_{1,2}$ in terms of $\sigma$, $\cos \theta$ and the two masses $m_1$ and $m_2$ 
using
\begin{equation}
  \label{eq:yitv}
  \sinh y_1 = \frac{m_2}{\sqrt s}\,\sqrt{\sigma^2-1}\,,\qquad
  \sinh y_2 = -\frac{m_1}{\sqrt s} \sqrt{\sigma^2-1}\,,\qquad
  \sinh y = \frac{\cos\theta}{\sin\theta}\;,
\end{equation}
which are obtained using~\eqref{rapid} and noting that
\begin{equation}
	\begin{split}
		\sigma &\simeq \frac{p_1^0 p_2^0 - p_1^L p_2^L}{\bar m_1 \bar m_2} =\cosh(y_1-y_2)\,, \\
		0 &=p_1^L+p_2^L \simeq m_1 \sinh y_1+m_2 \sinh y_2 \,,\\
		\cos \theta &= \frac{k^L}{k^0}=\tanh y\\
		\omega &=k^0= |\mathbf k|\,\cosh y  
		\label{variousre}
	\end{split}
\end{equation}
The resulting expressions are
\begin{equation}
\begin{split}
	 |{\mathbf{k}}| c_1 &= \frac\omega{\sqrt s} \frac{m_2+m_1 \sigma + m_1 \sqrt{\sigma^2-1} \cos \theta}{\sqrt{\sigma^2-1}}\,,  \\
 |{\mathbf{k}}| c_2 &=  \frac\omega{\sqrt s} \frac{m_1+m_2 \sigma - m_2 \sqrt{\sigma^2-1} \cos \theta}{\sqrt{\sigma^2-1}}  
	\label{M1M2}
\end{split}
\end{equation}
and
\begin{align}
	\begin{split}
		 {\mathbf{k}}^2 d_1 &= \omega^2 \frac{ \left( m_2+m_1 \sigma+m_1 \cos \theta  \sqrt{\sigma^2-1} \right)\left( m_2 \sqrt{\sigma^2-1} - (m_1+m_2 \sigma) \cos \theta \right)}{s\sqrt{\sigma^2-1}} \,,  \\
		{\mathbf{k}}^2  d_2 &= \omega^2  \frac{ ( m_1+m_2 \sigma- m_2 \cos \theta \sqrt{\sigma^2-1})(m_1 \sqrt{\sigma^2-1}+ (m_2+m_1 \sigma)\cos \theta)}{  s \sqrt{\sigma^2-1}} \,.
		\label{N1N2}
	\end{split}
\end{align}
Note, however, that the expressions in terms of rapidities hold in a general frame boosted in the longitudinal direction (corresponding to an overall shift of all the rapidities).

Consider now, for fixed $\vec{k}$,  the Fourier transform of \eqref{traceampliB4bis} rewritten in the form:
\begin{equation}
	A_{5,\mathrm{dil}}  =  \frac{2(8\pi G)^{\frac{3}{2}}}{\sqrt{D-2}}\left(B^{(1)} + B^{(2)} + B^{(3)} \right)  \equiv \frac{2(8\pi G)^{\frac{3}{2}}}{\sqrt{D-2}} B \, ,
	\label{Bidef}
\end{equation}
where each $B^{(a)}$ corresponds to a different scaling with $k$ at fixed $q_i$.
The basic objects  to be computed are then:
\begin{equation}
	\tilde{B}^{(i)} = \frac{1}{4 Ep}\int \frac{d^{D-2}\Delta}{(2 \pi)^{D-2}} ~ e^{-i \Delta b} ~B^{(i)}(\Delta)
	\label{FTdil}
\end{equation}
Let us remark once again that, at the classical level, one can neglect corrections of order $q_i, k$ approximating for instance $-p_1\simeq \bar p$.

The simplest case is $\tilde{B}^{(2)}$, which reduces to the standard Bessel integral
\begin{equation}
	\int \frac{d^{D-2}\Delta}{(2 \pi)^{D-2}} ~ e^{-i \Delta b} \frac{1}{{\mathbf \Delta}^2 + m^2} = \frac{m^{D-4}}{(2\pi)^{\frac{D-2}{2}}}  \frac{K_{\frac{D}{2}-2} (mb)}{(mb)^{\frac{D}{2}-2}} \,,
	\label{BesselInt1}
\end{equation}
using equations \eqref{bfq1bfq2bfk} and \eqref{c1c2} to rewrite
\begin{equation}\label{}
	q_1^2 = \left(\tfrac12(\mathbf q_1-\mathbf q_2)+\tfrac12(\mathbf q_1+\mathbf q_2)\right)^2+c_1^2\mathbf k^2= \left(\mathbf \Delta-\tfrac12 \mathbf k \right)^2+c_1^2\mathbf k^2
\end{equation}
(and similarly for $q_2^2$), and shifting $\mathbf \Delta$ appropriately.
We thus find, restricting for simplicity to $D=4$,\footnote{To evaluate the finite part of $2\operatorname{Im}2\delta_2$, one ought to retain the first $\epsilon$-correction as well.} 
\begin{equation}
	(8 \pi p E)~\tilde{B}^{(2)} = - 4 (p_1 p_2) \left[ \frac{ m_1^2~ p_2 k  }{p_1 k}e^{i \frac{k b}{2} } K_0 (c_2 |{\mathbf{k}}|b) +\frac{ m_2^2~ p_1 k  }{p_2 k}e^{-i \frac{k b}{2}} K_0 (c_1 |{\mathbf{k}}| b)\right].
	\label{B2}
\end{equation}

The case of $\tilde{B}^{(1)}$ is slightly more involved because of the $k q_i$ appearing in the numerators. Using \eqref{d1d2} we can separate the longitudinal part, for which we can again apply \eqref{BesselInt1} while for the transverse part we need to replace $\mathbf{q}_i$ by the appropriate derivative with respect to $\mathbf{b}$.
Using $K'_0 = - K_1$, we  get:
\begin{equation}
	\begin{split}	\label{B1}
		(8 \pi p E)~ \tilde{B}^{(1)} 
		&=  -2 (p_1\cdot p_2)^2 {\mathbf{k}}^2  
		\Big[ \frac{ m_1^2 e^{i \frac{k b}{2} }}{(p_1 k)^2} \left(d_2 K_0 (c_2 |{\mathbf{k}}| b) + i \frac{k \cdot b}{|\mathbf k| b} c_2 K_1(c_2 |{\mathbf{k}}| b)\right) 
		\\
		&+  \frac{ m_2^2 e^{-i \frac{k \cdot b}{2} }}{(p_2 k)^2} \left(d_1 K_0 (c_1 |{\mathbf{k}}| b) - i \frac{k \cdot b}{|{\mathbf{k}}| b } c_1 K_1(c_1 |{\mathbf{k}}| b)\right) \Big]   \;.
	\end{split}
\end{equation}

Let us finally turn to $\tilde{B}^{(3)}$. The most convenient way to compute it is by introducing two Schwinger parameters $t_1,t_2$ in order to write the two propagators in exponential form. One then performs the Gaussian integral over $\mathbf{\Delta}$ followed by the integral over $T=t_1+t_2$ that can be done in terms of a $K_1$ Bessel function.
One is  left with an integral over $x \equiv t_1/T$. The final result is:
\begin{equation}
\begin{split}
	(8 \pi p E)~ \tilde{B}^{(3)} &= - 2b^2  \left(m_1^2~ (p_2 k)^2 + m_2^2~ (p_1 k)^2 \right) \int_0^1 dx\, e^{i \frac{k  b}{2} (1-2x)} \frac{K_1(b |{\mathbf{k}}| \sqrt{f})}{b |{\mathbf{k}}| \sqrt{f}}\,,  \\
f &\equiv x(1-x) + c_1^2 x + c_2^2 (1-x)
\label{B3}
\end{split}
\end{equation}

We may further simplify \eqref{B2}, \eqref{B1}, \eqref{B3} by replacing the remaining scalar products in terms of masses and rapidities. The outcome is that each $\tilde{B}$ depends only upon the products $m_1m_2$ and $|{\mathbf{k}}| b$, the rapidities $y, y_1,y_2$, and the azimuthal angle $\phi$ between $\mathbf{b}$ and  $\mathbf{k}$.

Assembling together the various terms and prefactors we arrive at the following result:
\begin{equation}\label{}
	{\tilde{A}}_{5,\mathrm{dil}}(b,\vec{k}) =   \frac{4 m_1 m_2 G \sqrt{8\pi G}}{\sqrt2 \sinh (y_1-y_2)}  \, {\cal M}_{\mathrm{dil}}\,,\qquad
	{\cal M}_{\mathrm{dil}} \equiv \frac{4 \pi p E}{m_1^2 m_2^2}\,\tilde{B}\,,
\end{equation}
with
\begin{equation}
	\begin{split}
		\mathcal M_\mathrm{dil}&= 
			2  \cosh (y_1-y_2)  
			\frac{ \cosh(y-y_2)  }{\cosh(y_1-y)}
			e^{i \frac{k b}{2} } 
			K_0 (c_2 |{\mathbf{k}}| b) \\
			& +
			2  \cosh (y_1-y_2) 
			\frac{ \cosh(y_1-y)  }{\cosh(y-y_2)}
			e^{-i \frac{k b}{2}} 
			K_0 (c_1|{\mathbf{k}}| b)
		\\  &-  \frac{\cosh^2(y_1-y_2)}{\cosh^2(y_1-y)}  
		e^{i \frac{k b}{2}} \left[d_2 K_0 
		(c_2 |{\mathbf{k}}| b) 
		+ i \frac{k \cdot b}{|{\mathbf{k}}|b}\, 
		c_2 K_1(c_2 |{\mathbf{k}}| b) \right] \\
		&
		-\frac{\cosh^2(y_1-y_2)}{\cosh^2(y-y_2)}
		e^{-i \frac{k \cdot b}{2} }  \left[d_1 K_0 (c_1 |{\mathbf{k}}| b) - i \frac{k \cdot b}{|{\mathbf{k}}|b } \,c_1 K_1(c_1 |{\mathbf{k}}| b) \right]
		 \\
		&  - b |{\mathbf{k}}|  \left(\cosh^2(y-y_2) +  \cosh^2(y_1-y)\right) \int_0^1 dx \, e^{i \frac{k  b}{2} (1-2x)} \frac{K_1(b |{\mathbf{k}}| \sqrt{f})}{\sqrt{f}}.
		\label{AtildeGV}
	\end{split}
\end{equation}
Here, $c_i$, $d_i$ and $f$ are defined, respectively, in \eqref{c1c2}, \eqref{d1d2} and \eqref{B3}. 
The above result uses the transverse momentum ${\mathbf k}$ and  rapidity $y$ to represent the dilation's momentum. Note that $ {\cal M}_{\mathrm{dil}}$ is dimensionless and, when written in these variables, it is also independent of the masses.

Depending on circumstances it can be more convenient to express\eqref{AtildeGV} in terms of the centre of mass energy $\omega$ and angles $\theta, \phi$ of the produced dilaton. One finds:
\begin{equation}
\begin{split}
	& {\tilde{A}}_{5,\mathrm{dil}}(b, \vec{k})  =
	- \frac{ 4   m_1m_2  G  \sigma^2\sqrt{8\pi G}}{ \sqrt{2(\sigma^2-1)}}    \\
\times \Bigg\{&  \frac{m_1^2\,\mathbf{k}^2\,
	{ e}^{\frac{i}{2} k b}  }{(p_1k)^2} \left[ \left(  {d_2  + \frac{2(p_1k)(p_2k)}{\mathbf{k}^2 (p_1p_2)}}\right) K_0 ( c_2 |{\mathbf{k}}| b)  + i \frac{ k \cdot b}{|{\mathbf{k}}| b} c_2
K_{1} ( c_2 |{\mathbf{k}}| b) \right]  \\
& + \frac{m_2^2\,\mathbf{k}^2\, {e}^{ -\frac{i}{2} k b} }{(p_2k)^2} \left[ \left(  {d_1   + \frac{2(p_1k)(p_2k)}{\mathbf{k}^2(p_1p_2)}}\right) K_0 (c_1 |{\mathbf{k}}|b)-i \frac{ k \cdot b}{b |{\mathbf{k}}|} c_1
K_{1} ( c_1 |{\mathbf{k}}| b)\right]  \\
& + \frac{(m_1^2 (p_2k)^2 + m_2^2 (p_1k)^2)}{(p_1p_2)^2}\int_0^1 dx\, e^{ \frac{i}{2} b  k (1-2x) }  \frac{b K_{1} (\sqrt{f} |{\mathbf{k}}| b)}{\sqrt{f} |{\mathbf{k}}|}    \Bigg\}
\label{DDD29a}
\end{split}
\end{equation}
where one should  insert the relations given in  equations \eqref{M1M2}, \eqref{N1N2}, \eqref{variousre} and write explicitly the different scalar products.

The amplitude in impact parameter space  \eqref{DDD29a}, appropriately squared as in \eqref{3CT8}, determines the differential spectrum of the number of emitted dilatons according to
\begin{equation}
	d N_{\mathrm{dil}}= \left|{\tilde{A}}_{5,\mathrm{dil}}(b,\vec{k})\right|^2\frac{d^3 k}{(2 \pi)^32\omega} 
	\label{dilrate}\,.
\end{equation}
Equivalently, using the relation  $d^3 k = d^2 k d ( |{\mathbf{k}}| \sinh y)= d^2 k |{\mathbf{k}}| \cosh y dy$ and the fact that $\omega= |{\mathbf{k}}| \cosh y$, one rewrite $dN_\mathrm{dil}$ in the form
\begin{equation}
    d N_{\mathrm{dil}}=
    \frac{(8\pi G)^3 m_1^2 m_2^2}{4 (2 \pi)^5 \sinh^2 (y_1-y_2)}\left|{\cal M}_{\mathrm{dil}} \right|^2 d^2 k dy\,,
\end{equation}
with $\mathcal M_\mathrm{dil}$ given in \eqref{AtildeGV}.
Other useful relations that allow one to recast $dN_\mathrm{dil}$ in different forms are \eqref{variousre}, \eqref{M1M2}, \eqref{N1N2}, together with
\begin{equation}
	d^2 k dy =  \frac{1}{2} d {\mathbf{k}}^2 d \phi dy  =  d\omega \omega d^2\Omega\,, \quad d^3 k = \omega^2 d \omega d^2 \Omega\,,\quad
	d^2\Omega = \sin \theta d\theta d \phi\,.
	\label{dilrate1}
\end{equation}

It is easy to check that, if we interpret $\vec{k}$ and $\omega$ as a classical wave-vector and a frequency, respectively, \eqref{dilrate} must have an extra factor $\hbar^{-1}$ on its right-hand side. Indeed the number of emitted dilatons per unit of phase space diverges in the $\hbar \to 0$ limit. On the other hand the dilaton's energy emission spectrum is simply obtained multiplying $dN_\mathrm{dil}$ by $\hbar \omega$,
\begin{equation}
		dE_{\mathrm{dil}} =  \hbar \omega\, dN_\mathrm{dil}
		=  |{\mathbf{k}}| \cosh y\, \frac{(8\pi G)^3 m_1^2 m_2^2}{4 (2 \pi)^5 \sinh^2 (y_1-y_2)}|{\cal M}_{\mathrm{dil}} |^2d^2 k d y\,,
	\label{Edil}
\end{equation} 
and  has therefore a smooth classical limit.

One can also compute less differential quantities, such as the frequency spectrum $\frac{d N_{\mathrm{dil}}}{d\omega}$ obtained after angular integration or the angular distribution $\frac{d N_{\mathrm{dil}}}{ d^2\Omega}$ integrated over the frequency.
It should be clear how the above considerations can be applied to the other massless particles produced in the collision, in particular to the graviton in GR, where we can also disentangle the two polarizations of the gravitational wave.
Furthermore, in principle, evaluating the integrals of our spectra corresponding to the total number or total energy of the emitted particle one ought to reproduce the results derived in the previous section.

So far we have been discussing  waveforms, as well as number and energy differential rates in the frequency (i.e.~$\omega$) domain.
Performing, at the level of the amplitude, one more Fourier transform from $\omega$ to the retarded time $u$, one can obtain the corresponding waveforms and rates in the time domain and make contact with other results in the literature~\cite{Kovacs:1978eu,Jakobsen:2021smu,Mougiakakos:2021ckm}.

\section{Discussion and Outlook}
\label{sec:outlook}

In this paper we provided a detailed derivation of the full 3PM eikonal in  massive ${\cal N}=8$ supergravity from the two-loop $2 \to 2$ elastic  amplitude. The real part of the eikonal captures the classical deflection angle while its imaginary part is related to the total number of massless particles emitted in the scattering process. This analysis  explicitly shows the necessity of taking into account contributions from the full soft region of the loop integrals in order to obtain the correct  results for the various physical observables, with a smooth high-energy limit. This is done be using the appropriate boundary conditions of the key scalar integrals entering the relevant differential equations  for the classical limit.

We also showed  how to rewrite the ${\cal N}=8$ result so as to make the analyticity properties and  crossing symmetry manifest. These properties can be used to obtain a direct link between certain terms of the imaginary part and the radiation-reaction contribution to the real part which, in turn, determines the corresponding terms in the deflection angle. The dispersion relation ensuring this relation makes it manifest that the radiation-reaction contributions have a characteristic PN expansion which is shifted by a single power of the velocity $v$ with respect to the contribution arising from the potential region. However, in the ultrarelativistic limit, the two structures scale in the same way and cannot be separated. In fact, as already discussed in~\cite{DiVecchia:2020ymx}, there are important cancellations of terms enhanced an by extra-factor of $\log(s)$ and this ensures that the deflection angle for the massless case~\cite{Amati:1990xe} is smoothly reproduced. This 3PM result is a universal property of all gravitational field theories in the two derivative approximation.

In Section~\ref{sec:directunitarity}, the imaginary part of the 3PM eikonal was derived by following an alternative approach where the $2\to 3$ tree-level classical amplitude is used to calculate explicitly the 3-particle cut. One can again use the same technical tools developed to evaluate the 2-loop integrals relevant to the elastic two-loop process. In the ${\cal N}=8$ case we reproduce the same result for $\operatorname{Im} 2 \delta_2$ as the one obtained from the evaluation of the full two-loop amplitude, but this ``direct unitarity'' approach can be easily extended to obtain $\operatorname{Im} 2 \delta_2$, waveforms and fully differential spectra even for the case of pure GR. By following this route we derived the radiation-reaction part of the 3PM eikonal for GR which, when combined with the previous results~\cite{Bern:2019nnu,Bern:2019crd} for the potential region contribution, provides the full 3PM eikonal for massive scattering in GR given in~\eqref{GR15} and~\eqref{ReGR15}. This fully amplitude-based approach reproduces the physical deflection angle derived  in~\cite{Damour:2020tta} by using the linear response formula of~\cite{Bini:2012ji}.

Of course, a first natural generalisation of our analysis is to use again ${\cal N}=8$ massive supergravity as a laboratory for studying the full soft-region contribution at 4PM. Very recently, the potential contribution in GR was derived~\cite{Bern:2021dqo} and it would be certainly interesting to complete that result by including the radiation-reaction and the tail effects so as to extract a physical, IR-finite result for the deflection angle both in ${\cal N}=8$ and in GR. More in general, the lesson of our 3PM analysis is that analyticity properties and crossing symmetry can be very useful to organise the result and find connections between different terms. On the practical side, this may simplify the derivation of (some parts of) the full 4PM eikonal. On the conceptual side it would be important to extend systematically the discussion of the eikonal exponentiation to the inelastic terms: this presumably requires to promote the eikonal phase to an operator containing  creation and annihilation modes for the massless particles that can be emitted in the process, see e.g.~\cite{Ciafaloni:2018uwe}. Another interesting development would be to use the direct unitarity approach to continue the analysis of gravitational waveforms, energy spectra and  angular momentum emission at 3PM as a step towards an extension to higher orders. 

The generalisations mentioned above will also help to provide a more clear physical interpretation of the radiation-reaction contributions to the deflection angle also beyond the 3PM order. Again, working in a maximally supersymmetric setup will provide the perfect arena to uncover the physical principles and  technical shortcuts that can then be applied  in more general situations. Of course, an ambitious goal would be to provide a complete post-Minkowskian, amplitude-based approach to the analysis of the inspiral phase of gravitational binaries. In the spirit of~\cite{Kalin:2019rwq,Kalin:2019inp}, it would be useful not to rely on gauge-dependent intermediate results and obtain the relevant physical observables directly from on-shell scattering amplitudes. This approach has been successful when restricted to the conservative part of the problem, but, as this work hopefully illustrates, radiation effects should be included in order to complete the programme.

\subsection*{Acknowledgements} 

We thank Enrico Herrmann, Julio Parra-Martinez, Michael Ruf and Mao Zeng for useful discussions and comparisons of intermediate results, as well as for sharing with us a first draft of their paper \cite{Herrmann:2021toa}.
We also thank Thibault Damour and Stephen Naculich for helpful comments on a preliminary version of this paper.
We are grateful to Claude Duhr and Vladimir Smirnov for helping us check with independent methods some of our integrals and to Gudrun Heinrich for some numerical checks.
\newline\noindent
The research of RR is partially supported by the UK Science and Technology Facilities Council (STFC) Consolidated Grant ST/P000754/1 ``String theory, gauge theory and duality''. The research of CH (PDV) is fully (partially) supported by the Knut and Alice Wallenberg Foundation under grant KAW 2018.0116.

\appendix

\section{Static Boundary Conditions}
\label{app:SoftBC}

This appendix is devoted to the calculation of the (near-)static boundary conditions that are employed in the body of the text to obtain the solutions of the differential equations in the soft region. Our goal is thus to evaluate the leading asymptotics of the master integrals as $x\to1$. We shall first focus on the $\RT$ family, from which most of the integrals appearing in the $\RN$ family can be deduced, to then move to the $\Ht$ family.

Let us begin by discussing those master integrals that can be actually evaluated for generic $x$ because they effectively reduce to iterated one-loop integrals, as is the case for $f_{\RT,1}$, $f_{\RT,5}$ and $f_{\RT,8}$. 
The first master integral can be written in the form
\begin{equation}\label{f1}
	f_{\RT,1} = -\epsilon^2 q^2 \int_{\ell_1}\int_{\ell_2} \frac{1}{((\ell_1+\ell_2+q)^2-i0)(\ell_1^2-i0)^2(\ell_2^2-i0)^2}\,,
\end{equation}
after sending $\ell_1\to\ell_1+q$ and $\ell_2\to\ell_2+q$. Feynman's $-i0$ prescription has been explicitly spelled out in Eq.~\eqref{f1}, but it will be mostly left implicit from here on.
Using Schwinger or Feynman parameters one obtains (see e.g. \cite[Eq.~(A.7)]{Smirnov:2002pj})
\begin{equation}\label{alphabetabubble}
	\int_{\ell} \frac{1}{
		(\ell^2)^\alpha((\ell+q)^2)^\beta}
	=
	e^{\gamma_E\epsilon}\frac{\Gamma\left(2-\epsilon-\alpha\right)\Gamma\left(2-\epsilon-\beta\right)\Gamma\left(\alpha+\beta+\epsilon-2\right)}{\Gamma\left(\alpha\right)\Gamma\left(\beta\right)\Gamma\left(4-\alpha-\beta-2\epsilon\right)(q^2)^{\alpha+\beta+\epsilon-2}}\,,
\end{equation}
and applying Eq.~\eqref{alphabetabubble} twice, once for each loop integral, the integral \eqref{f1} yields the first relation \eqref{fIII1}. 
For $f_{\RT,5}$, sending $\ell_1\to\ell_1+q$ and recalling that $u_1$, $u_2$ are orthogonal to $q$, we obtain instead
\begin{equation}\label{}
	f_{\RT,5} = \int_{\ell_1}\int_{\ell_2} \frac{ \epsilon^3 q^2 \tau }{(2u_1\cdot \ell_1)(-2u_2\cdot  \ell_1)(\ell_1+q)^2\ell_2^2((\ell_1+\ell_2)^2)^2}\,.
\end{equation}
The integral over $\ell_2$ can be evaluated by means of Eq.~\eqref{alphabetabubble}, obtaining
\begin{equation}\label{}
	f_{\RT,5} = 
	e^{\gamma_E \epsilon}\, \frac{\Gamma\left(1+\epsilon\right)\Gamma\left(1-\epsilon\right)\Gamma\left(-\epsilon\right)}{\Gamma\left(1-2\epsilon\right)}
	\int_{\ell}\frac{ \epsilon^3 q^2 \tau}{(2u_1\cdot \ell)(-2u_2\cdot  \ell)(\ell+q)^2(\ell^2)^{1+\epsilon}}\,.
\end{equation}
Going to Schwinger parameters, the remaining integral takes the form
\begin{equation}\label{f5almost}
	\begin{aligned}
		f_{\RT,5} &= 
		e^{2\gamma_E \epsilon}\, \frac{\Gamma\left(1-\epsilon\right)\Gamma\left(-\epsilon\right)}{\Gamma\left(1-2\epsilon\right)}
		\,\epsilon^3 q^2 \tau\\
		&\times \int_{\mathbb R_+^2} \frac{ds_1 ds_2}{(s_1+s_2)^{1-\epsilon}}\,s_2^\epsilon\, e^{-\frac{s_1 s_2 q^2}{s_1+s_2}}
		\int_{\mathbb R_+^2} dt_1 dt_2\, e^{-[t_1^2+2(-y-i0)t_1 t_2+t_2^2]}\,.
	\end{aligned}
\end{equation}
One is therefore left with the elementary integral
\begin{equation}\label{inttworoots}
	\int_{\mathbb R_+^2} dt_1 dt_2\, e^{-[t_1^2+2(-y-i0)t_1 t_2+t_2^2]}
	=
	\frac{1}{2} \int_{0}^\infty \frac{dt}{t^2+2(-y-i0)t+1}
	=
	\frac{\log x+i\pi}{2\tau }\,,
\end{equation}
and with a standard integral over Feynman parameters
\begin{equation}\label{ints1s2}
	\int_{\mathbb R_+^2} \frac{ds_1 ds_2\,s_2^{\epsilon}\, e^{-\frac{s_1 s_2 q^2}{s_1+s_2}}}{(s_1+s_2)^{1-\epsilon}} =
	\frac{\Gamma\left(1+2\epsilon\right)}{(q^2)^{1+2\epsilon}} \int_0^1 \frac{dx}{x^{1+2\epsilon}(1-x)^{1+\epsilon}}\, =\frac{\Gamma\left(1+2\epsilon\right)\Gamma\left(-\epsilon\right)\Gamma\left(-2\epsilon\right)}{(q^2)^{1+2\epsilon}\Gamma\left(-3\epsilon\right)}\,.
\end{equation}
Substituting these two integrals into Eq.~\eqref{f5almost} then gives the expression \eqref{fIII5}.
As far as $f_{\RT,8}$ is concerned 
\begin{equation}\label{}
	f_{\RT,8}=- \int_{\ell_1} \int_{\ell_2} \frac{\epsilon^3 q}{(2u_1\cdot \ell_1) \ell_1^2 \ell_2^2 ((\ell_1+\ell_2-q)^2)^2}\,,
\end{equation}
the integral over $\ell_2$ can be again evaluated thanks to Eq.~\eqref{alphabetabubble}, so that
\begin{equation}\label{}
	f_{\RT,8}= - e^{\gamma_E\epsilon}\,
	\frac{\Gamma\left(1+\epsilon\right)\Gamma\left(1-\epsilon\right)\Gamma\left(-\epsilon\right)}{\Gamma\left(1-2\epsilon\right)}
	\int_{\ell} \frac{\epsilon^3 q}{(2u_1\cdot \ell) \ell^2 ((\ell-q)^2)^{1+\epsilon}}\,.
\end{equation}
This integral can be tackled with Schwinger parameters, obtaining 
\begin{equation}\label{}
	f_{\RT,8}= - e^{2\gamma_E\epsilon}\, q\,
	\frac{\epsilon^3\Gamma\left(1-\epsilon\right)\Gamma\left(-\epsilon\right)}{\Gamma\left(1-2\epsilon\right)}
	\int_{\mathbb R_+^2} \frac{ds_1 ds_2\,s_2^{\epsilon}\, e^{-\frac{s_1 s_2 q^2}{s_1+s_2}}}{(s_1+s_2)^{\frac{3}{2}-\epsilon}}
	\int_{0}^{\infty} e^{-t^2} dt\,.
\end{equation}
The integral over $t$ yields $\frac{\sqrt \pi}{2}$ while the remaining integration over $s_1$ and $s_2$ can be handled in a manner similar to the one in Eq.~\eqref{ints1s2}, eventually retrieving
\eqref{fIII8}.

The simplest master integral with a genuine two-loop structure is  $f_{\RT,6}$, which can be put in the form
\begin{equation}\label{}
	f_{\RT,6}
	= 
	- \int_{\ell_1} \int_{\ell_2} \frac{\epsilon^3 (1-6\epsilon)}{(2u_1\cdot \ell_1) (-2u_1\cdot \ell_2)\ell_1^2 (\ell_2+q)^2 (\ell_1+\ell_2)^2}
\end{equation} 
sending $\ell_2\to\ell_2+q$.
Going to Schwinger parameters affords
\begin{equation}\label{}
	f_{\RT,6}
	= 
	\epsilon^3 (6\epsilon-1) e^{2\gamma_E\epsilon}
	\int_{\mathbb R_+^3} dt_1\, dt_2\, dt_3  \,\frac{e^{-\frac{t_1 t_2 t_3}{T}\,q^2}}{T^{1-\epsilon}} \int_{\mathbb R_+^2} dt_4\, dt_5 e^{-(t_{23}t_4^2+2 t_3 t_4 t_5+t_{13} t_5^2)}\,,
\end{equation}
where
\begin{equation}\label{Tt13t23}
	T(t_1,t_2,t_3) = t_1 t_2 + t_2 t_3 + t_3 t_1\,,\qquad
	t_{13}=t_1+t_3\,,\qquad t_{23}=t_2+t_3\,.
\end{equation}
The integral over $t_4$, $t_5$ can be evaluated by  imitating the steps followed in Eq.~\eqref{inttworoots}, although the discriminant of the relevant quadratic polynomial is negative in this case. We thus find
\begin{equation}\label{}
	f_{\RT,6}
	= 
	\epsilon^3 (6\epsilon-1) e^{2\gamma_E\epsilon}
	\int_{\mathbb R_+^3} dt_1\, dt_2\, dt_3  \,\frac{e^{-\frac{t_1 t_2 t_3}{T}\,q^2}}{2T^{\frac{3}{2}-\epsilon}} \left(
	\frac{\pi}{2}-\arctan\frac{t_3}{\sqrt T}
	\right)\,.
\end{equation}
Performing the integral with respect to the overall scale according to
\begin{equation}\label{intscale}
	1=\int_0^\infty d\lambda^2 \delta(\lambda^2-T)\,,\qquad 
	(t_1,t_2,t_3)=\lambda(x,y,z)
\end{equation}
then leads
to
\begin{equation}\label{}
	f_{\RT,6}
	= 
	\epsilon^3 (6\epsilon-1) e^{2\gamma_E\epsilon} \frac{\Gamma(2\epsilon)}{(q^2)^{2\epsilon}}
	\int_{\mathbb R_+^3} \frac{dx\, dy\, dz}{(xyz)^{2\epsilon}} \delta(1-T(x,y,z))  \left[
	\frac{\pi}{2}-\arctan z
	\right]\,.
\end{equation}
At this point it should be noted that the integration is completely symmetric under permutations of $x$, $y$ and $z$, so that the factor within square brackets can be replaced by its permutation average
\begin{equation}\label{}
	\left[\frac{\pi}{2}-\arctan z\right] \longmapsto \left[\frac{\pi}{2}-\frac{1}{3}\left( \arctan x + \arctan y + \arctan z \right)\right]\,.
\end{equation}
Furthermore, since\footnote{This can be seen by first noting that  $\log(1+ix)+\log(1+iy)+\log(1+iz)=\log(i(x+y+z-xyz))$ with $x+y+z-xyz>0$, if $T(x,y,z)=1$, and then taking the imaginary part.}
\begin{equation}\label{permidxyz}
	\arctan x + \arctan y + \arctan z=\frac{\pi}{2} \qquad \text{for }T(x,y,z)=1\,,
\end{equation}
we have 
\begin{equation}\label{}
	f_{\RT,6}
	= 
	\epsilon^3 (6\epsilon-1) e^{2\gamma_E\epsilon} \frac{\Gamma(2\epsilon)}{(q^2)^{2\epsilon}}
	\frac{\pi}{3} \int_{\mathbb R_+^3} \frac{dx\, dy\, dz}{(xyz)^{2\epsilon}} \delta(1-T(x,y,z))  \,.
\end{equation}
Letting 
\begin{equation}\label{hat}
	\hat x=yz\,,\qquad\hat y=zx\,,\qquad \hat y=zx
\end{equation}
finally reduces the integral to a multivariate Beta function
\begin{equation}\label{}
	f_{\RT,6}
	= 
	\epsilon^3 (6\epsilon-1) e^{2\gamma_E\epsilon} \frac{\Gamma(2\epsilon)}{(q^2)^{2\epsilon}}
	\frac{\pi}{6} \int_{\mathbb R_+^3} \frac{d\hat x\, d\hat y\, d\hat z}{(xyz)^{\frac{1}{2}+\epsilon}} \delta(1-\hat x-\hat y-\hat z) \,, 
\end{equation}
which allows one to read off Eq.~\eqref{fIII6}\,.

The master integrals $f_{\RT,2}$, $f_{\RT,3}$ and $f_{\RT,4}$ are instead more challenging as they depend on $x$ in a nontrivial manner. Their static limit must be discussed with care, since setting $x=1$ in the integrand in an ordinary fashion fails to capture certain non-analytic contributions that behave as $(1-x)^{-2\epsilon}$ in the static limit. These singular static terms are actually crucial to retrieve the imaginary parts of the boundary conditions. We shall therefore distinguish these two types of terms as \emph{ordinary} and \emph{singular} static contributions.

To discuss these integrals, let us first introduce the shorthand notation
\begin{equation}\label{Gjk}
	G_{j,k}=G_{0,j,j,0,0,0,k,1,1}\,,
\end{equation}
where the labels
$(j,k)$ can take values $(1,1)$, $(1,2)$ or $(2,1)$,
so that explicitly
\begin{equation}\label{}
	G_{j,k}
	=
	\int_{\ell_1}\int_{\ell_2}
	\frac{1}{((-2u_2\cdot \ell_1)(-2u_1\cdot \ell_2))^j((\ell_1+\ell_2)^2)^k\ell_1^2(\ell_2-q)^2}
\end{equation}
after shifting $\ell_1\to \ell_1+q$.
Going to Schwinger parameters then affords
\begin{equation}\label{GjkGj}
	G_{j,k}=
	e^{2\gamma_E\epsilon}
	\int_{\mathbb R_+^3} dt_1\,dt_2\,dt_3\, t_3^{k-1} \frac{e^{-\frac{t_1t_2t_3}{T}\,q^2}}{T^{2-j-\epsilon}} \, G_j\,,
\end{equation}
where we defined
\begin{equation}
	G_j
	=
	\int_{\mathbb R_+^2} dt_4\,dt_5 (t_4 t_5)^{j-1}
	e^{-[t_{13}t_4^2+2t_3(-y-i0)t_4t_5+t_{23}t_5^2]}\,.
\end{equation}
Here $T$, $t_{13}$ and $t_{23}$ are as in Eq.~\eqref{Tt13t23}. We can deal with integrals of this type paralleling the steps in Eq.~\eqref{inttworoots}, which leads to
\begin{align}
	\label{G1}
	G_1 &= \frac{\log\big(-t_3 y + \sqrt{\tau^2t_3^2-T+i0}\big)-\log\big(-t_3 y - \sqrt{\tau^2t_3^2-T+i0}\big)}{4\sqrt{\tau^2t_3^2-T+i0}}\,,\\
	\label{G2}
	G_2 &= -\frac{1
		+2yt_3\, G_1}{4(\tau^2t_3^2-T+i0)}\,.
\end{align}

Now we consider the static limit $\tau\to0^+$. A first option in this respect is to simply interchange the limit and the integral sign in \eqref{GjkGj}, which corresponds to enforcing the following scaling limit, 
\begin{equation}\label{ordinary}
	t_1,\, t_2,\,t_3\sim \mathcal O(\tau^0)\qquad
	(\tau\to0^+)
\end{equation} 
in \eqref{G1} and \eqref{G2}. This gives rise to the \emph{ordinary} contributions
\begin{equation}
	\label{Gordinary}
	G_1^{(o)} = \frac{1}{2\sqrt{T}}\left[\frac{\pi}{2}+\arctan\frac{t_3}{\sqrt{T}}\right]\,,\qquad
	G_2^{(o)} = \frac{1
		+2t_3\, G_1^{(o)}}{4T}\,.
\end{equation}
Plugging these values back into Eq.~\eqref{GjkGj} and integrating out an overall scale according to Eq.~\eqref{intscale}, we find
\begin{equation}\label{}
	G_{j,k}^{(o)}=e^{2\gamma_E\epsilon}\,\frac{\Gamma(-2+j+k+2\epsilon)}{2^{j-1}(q^2)^{-2+j+k+2\epsilon}}\int_{\mathbb R_+^3} \frac{z^{k-1}g_j(z)}{(xyz)^{-2+j+k+2\epsilon}}\, \delta(1-T(x,y,z))\, dx\, dy\, dz\,,
\end{equation}
where
\begin{equation}\label{}
	g_j(z)=\begin{cases}
		\frac{\pi}{2}+\arctan z & (j=1)\\
		1+z\left(\frac{\pi}{2}+\arctan z\right) & (j=2)\,.
	\end{cases}
\end{equation}
The simplest case is $j=k=1$, for which we find
\begin{equation}\label{}
	G_{1,1}^{(o)}=e^{2\gamma_E\epsilon}\frac{\Gamma(2\epsilon)}{(q^2)^{2\epsilon}}
	\int \frac{\frac{2\pi}{3}\delta(1- xy -yz - zx)}{( x  y z)^{2\epsilon}}\, d x\, d y\, d z\,,
\end{equation}
after symmetrizing under permutations of the three variables and making use of the identity \eqref{permidxyz}. Employing the change of variables \eqref{hat} then gives
\begin{equation}\label{G11}
	G_{1,1}^{(o)}=\frac{\pi}{3} \frac{\Gamma(2\epsilon)}{(q^2)^{2\epsilon}} \frac{\Gamma\left(\tfrac{1}{2}-\epsilon\right)^3}{\Gamma\left(\tfrac{3}{2}-3\epsilon\right)}\,e^{2\gamma_E \epsilon}
\end{equation}
and hence, in view of the prefactor $\tau$  appearing in the definition of $f_{\RT,2}$, we see that the boundary condition for $f_{\RT,2}$ receives no contribution from the ordinary static region
\begin{equation}\label{}
	f^{(o)}_{\RT,2}\big|_{x=1}=0\,.
\end{equation} 

For the evaluation of the remaining two integrals, we refer to Appendix~\ref{app:Xepsilon} and quote here the final results
\begin{align}\label{G12}
	G_{1,2}^{(o)}&=e^{2\gamma_E \epsilon}\frac{\Gamma(1+2\epsilon)}{(q^2)^{1+2\epsilon}}
	\left[
	-\frac{1}{6(1+2\epsilon)}\frac{\Gamma\left(-\epsilon\right)^3}{\Gamma\left(-3\epsilon\right)}
	+
	\frac{\pi}{3}\frac{\Gamma\left(-\tfrac{1}{2}-\epsilon\right)\Gamma\left(\tfrac{1}{2}-\epsilon\right)^2}{\Gamma\left(\frac{1}{2}-3\epsilon\right)}
	\right]\,,
	\\ 
	\label{G21}
	G_{2,1}^{(o)}&=e^{2\gamma_E \epsilon}\frac{\Gamma(1+2\epsilon)}{(q^2)^{1+2\epsilon}}
	\left[
	\frac{1+3\epsilon}{6(1+2\epsilon)}\frac{\Gamma\left(-\epsilon\right)^3}{\Gamma\left(-3\epsilon\right)}
	+
	\frac{\pi}{6}\frac{\Gamma\left(-\tfrac{1}{2}-\epsilon\right)\Gamma\left(\tfrac{1}{2}-\epsilon\right)^2}{\Gamma\left(\frac{1}{2}-3\epsilon\right)}
	\right]\,.
\end{align}
In view of \eqref{G12} and of the prefactor $\tau$ appearing in the definition of $f_{\RT,3}$ we thus obtain
\begin{equation}\label{}
	f_{\RT,3}^{(o)}\big|_{x=1}=0\,.
\end{equation}
Furthermore, 
\begin{equation}\label{}
	f^{(o)}_{\RT,4}\big|_{x=1}
	=
	\epsilon^{2} q^2 
	\left(
	G^{(o)}_{2,1}
	+\epsilon\,  G_{1,2}^{(o)}
	\right)\,,
\end{equation}
so that \eqref{G12} and \eqref{G21} give
\begin{equation}\label{f4ordinary}
	f^{(o)}_{\RT,4}\big|_{x=1}
	=
	\frac{e^{2 \gamma_E  \epsilon }}{(q^2)^{2\epsilon}}
	\,\Gamma(1+2\epsilon)\left(
	\frac{\Gamma(1-\epsilon)^3}{2\Gamma(1-3\epsilon)}
	-
	\frac{\pi\epsilon^2}{3}
	\frac{\Gamma\left(\frac{1}{2}-\epsilon\right)^3}{\Gamma\left(\frac{1}{2}-3\epsilon\right)}
	\right)\,.
\end{equation}

However, a second option is available when performing the static limit.
Indeed, note that setting $\tau=0$ in the expressions \eqref{G1}, \eqref{G2} one neglects a nontrivial region corresponding to large $t_3$ that give rise to an imaginary part for any nonzero $\tau$. To capture these contributions, we consider the scaling limit, 
\begin{equation}\label{singular}
	t_1,\, t_2\sim \mathcal O(\tau^0)\,,\qquad t_3\sim\mathcal O(\tau^{-2})\qquad 
	(\tau\to0^+)
\end{equation} 
in \eqref{G1} and \eqref{G2}. To leading order in this limit, we thus obtain the \emph{singular} static contributions
\begin{equation}\label{G1sG2s}
	G_1^{(s)} \sim \frac{i\pi}{2\tau \sqrt{t_3} \sqrt{t_3+\frac{-t_{12}+i0}{\tau^2}}}\,,
	\qquad
	G_2^{(s)} \sim -\frac{i\pi }{4\tau^3\sqrt{t_3}\left(t_3+\frac{-t_{12}+i0}{\tau^2}\right)^{\frac{3}{2}}}\,,
\end{equation}
where $t_{12}=t_1+t_2$.
We now employ these asymptotics to evaluate Eq.~\eqref{GjkGj}, where we also need to expand the integrand to leading order according to the scaling relations \eqref{singular}. Changing variable
\begin{equation}\label{}
	t_3 = \left(\frac{-t_{12}+i0}{\tau^2}\right) z
\end{equation} 
in the $t_3$ integration, we then find
\begin{equation}\label{}
		G_{j,k}^{(s)} \sim \frac{i\pi(-1)^{j-1}}{2^j \tau^{2j-1}} e^{2\gamma_E\epsilon} 
	\int_{\mathbb R_+^2}dt_1\, dt_2\, \frac{e^{-\frac{t_1 t_2}{t_{12}}\,q^2}}{t_{12}^{2-j-\epsilon}} \left(\frac{-t_{12}+i0}{\tau^2}\right)^{k-2+\epsilon}
	\!\!\! 
	\int_0^{\infty} \frac{dz}{z} \frac{z^{k+j-\tfrac52+\epsilon}}{(z+1)^{j-\tfrac12}}\,.
\end{equation}
The integration thus factorizes and the resulting integrals can be reduced to Euler Beta functions. To leading order in $\tau$ we therefore have
\begin{equation}\label{}
	\begin{split}
G^{(s)}_{j,k}
&\sim
(-1)^{k+j-1}
\frac{i\pi\, e^{i\pi\epsilon}}{\tau^{2j+2k-5+2\epsilon}}
\,
e^{2\gamma_E\epsilon}\\
&\times
\frac{\Gamma(k+j-2+2\epsilon)\Gamma(3-k-j-2\epsilon)^2\Gamma(2-k-\epsilon)\Gamma(k+j+\epsilon-\tfrac52)}{2^j(q^2)^{k+j-2+2\epsilon}\Gamma(6-2k-2j-4\epsilon)\Gamma(j-\tfrac12)}\,,
	\end{split}
\end{equation}
or more explicitly
\begin{align}
	\label{G11s}
	G^{(s)}_{1,1} &\sim -i\pi \tau^{1-2\epsilon} e^{i\pi\epsilon}\frac{e^{2\gamma_E\epsilon}}{(q^2)^{2\epsilon}}
	\frac{\Gamma\left(2\epsilon\right)\Gamma\left(1-2\epsilon\right)^2\Gamma\left(1-\epsilon\right)\Gamma\left(-\frac{1}{2}+\epsilon\right)}{2\Gamma\left(2-4\epsilon\right)\Gamma\left(\frac{1}{2}\right)}\,, \\ 
	\label{G12s}
	G^{(s)}_{1,2} &\sim i\pi \tau^{-1-2\epsilon} e^{i\pi\epsilon}\frac{e^{2\gamma_E\epsilon}}{(q^2)^{1+2\epsilon}}
	\frac{\Gamma\left(1+2\epsilon\right)\Gamma\left(-2\epsilon\right)^2\Gamma\left(-\epsilon\right)\Gamma\left(\frac{1}{2}+\epsilon\right)}{2\Gamma\left(-4\epsilon\right)\Gamma\left(\frac{1}{2}\right)} \,,\\ 
	\label{G21s}
	G^{(s)}_{2,1} &\sim i\pi \tau^{-1-2\epsilon} e^{i\pi\epsilon}\frac{e^{2\gamma_E\epsilon}}{(q^2)^{1+2\epsilon}}
	\frac{\Gamma\left(1+2\epsilon\right)\Gamma\left(-2\epsilon\right)^2(-\epsilon)\Gamma\left(-\epsilon\right)\Gamma\left(\frac{1}{2}+\epsilon\right)}{2\Gamma\left(-4\epsilon\right)\Gamma\left(\frac{1}{2}\right)} \,. 
\end{align}
In view of the additional power of $\tau$ involved in the definitions of $f_{\RT,2}$ and $f_{\RT,3}$, we see that their singular static contributions are
\begin{align}\label{}
	f_{\RT,2}^{(s)}\big|_{x=1}&=0\,, \\
	f_{\RT,3}^{(s)}\big|_{x\to1}& \sim i\pi \bigg(\frac{e^{i\pi/2}e^{\gamma_E}}{(1-x)q^2}\bigg)^{2\epsilon} \frac{\epsilon\, \Gamma(1+2 \epsilon) \Gamma(1-2 \epsilon)^{2} \Gamma(1-\epsilon)\Gamma\left(\frac{1}{2}+\epsilon\right)}{2 \Gamma(1-4 \epsilon) \Gamma\left(\frac{1}{2}\right)}\,.
\end{align}
Since the corresponding ordinary static contributions are trivial, as discussed above, we therefore retrieve eqs.~\eqref{fIII2} and \eqref{fIII3}. Conversely, for $f_{\RT,4}$, we encounter the combination
\begin{equation}\label{}
	f_{\RT,4}^{(s)}\big|_{x\to1} \sim \epsilon^2 q^2 \left(G_{2,1}^{(s)}+\epsilon\, G_{1,2}^{(s)}\right)\,,
\end{equation}
which vanishes to leading order
due to a cancellation that occurs between eqs.~\eqref{G12s} and \eqref{G21s}. The first subleading contribution, which must be evaluated retaining the next order in the expansion \eqref{G1sG2s}, is also zero because it gives rise to scaleless integrals. Therefore, the boundary condition \eqref{fIII4} is entirely dictated by the ordinary static contribution \eqref{f4ordinary}.

The boundary conditions \eqref{fIII9}, \eqref{fIII10} for $f_{\RT,9}$ and $f_{\RT,10}$ can be discussed in a similar manner: one first explicitly performs the integrals with respect to the Schwinger parameters associated to the massive lines, and then discusses the static limit of the remaining integrals distinguishing between ordinary \eqref{ordinary} and singular \eqref{singular} scaling. The leftover integrals are either elementary or can be evaluated systematically using the Mellin--Barnes representation. The final result is that Eq.~\eqref{fIII10} and the first line of Eq.~\eqref{fIII9} emerge from ordinary static contributions, while the second line of Eq.~\eqref{fIII9} emerges from singular static contributions.

The remaining boundary condition for $f_{\RT,7}$ can be instead discussed as follows. In view of the prefactor $\tau^{2}$, nontrivial contributions to the static limit can only emerge from the leading singularities of the integral $G_7\equiv G_{1,1,1,1,1,1,1,0,0}$ as $y\to1$. These can be evaluated by first writing the integral as
\begin{equation}\label{G7t1t2t3}
	G_{7}=e^{2\gamma_E\epsilon}\int_{\mathbb R^3_+}dt_1\, dt_2\, dt_3\, 
	T^{\epsilon}e^{-\frac{t_1\, t_2\, t_3}{T}\, q^2} I_4\,,
\end{equation}
where
\begin{equation}\label{I4t1t2t3}
	I_4
	=
	\int_{\mathbb R^4_+} dt_4\, dt_5\, dt_6\, dt_7\,
	e^{-
		\mathbf t^T
		M(y)\,
		\mathbf t}
	\,,\qquad
	\mathbf t=
	\left(\begin{matrix}
		t_4\\ t_5 \\t_6 \\t_7
	\end{matrix}\right)\,,
\end{equation}
and
\begin{equation}\label{}
	M(y) =
	\left(
	\begin{matrix}
		t_{23} & t_{23}(-y-i0) & t_3 & t_3 (-y-i0)\\
		t_{23}(-y-i0) & t_{23} & t_3(-y-i0) & t_3\\
		t_3 & t_3(-y-i0) & t_{13} & t_{13}(-y-i0)\\
		t_3(-y-i0) & t_3 & t_{13}(-y-i0) &  t_{13}
	\end{matrix}
	\right)\,.
\end{equation}
Singularities as $y\to1$ in Eq.~\eqref{I4t1t2t3} arise from the direction in $\mathbf t$-space where the exponent vanishes, i.e.~from the zero-modes of $M(1)$. To highlight them let us introduce
\begin{equation}\label{}
	\mathbf k = \left(\begin{matrix}
		1\\ 1\\0 \\0
	\end{matrix}\right)\,,\qquad
	\mathbf k_\perp = \left(\begin{matrix}
		1\\ -1\\0 \\0
	\end{matrix}\right)\,,\qquad
	\mathbf q = \left(\begin{matrix}
		0\\ 0\\1 \\1
	\end{matrix}\right)\,,\qquad
	\mathbf q_\perp = \left(\begin{matrix}
		0\\ 0\\1 \\ -1
	\end{matrix}\right)\,,
\end{equation}
and change variables to
\begin{equation}\label{}
	\mathbf t = 
	\frac{\alpha}{2}\, \mathbf k_\perp +  
	\frac{\beta}{2}\, \mathbf k +  
	\frac{a}{2}\, \mathbf q_\perp + 
	\frac{b}{2}\, \mathbf q \,.
\end{equation}
In this fashion, $\beta$, $b$ parametrize the directions along the two independent zero modes $\mathbf q$, $\mathbf k$ of $M(1)$ and $\alpha$, $a$ parametrize the orthogonal directions $\mathbf q_\perp$, $\mathbf k_\perp$.
Therefore, to leading order small positive $\tau$, we find
\begin{equation}\label{}
	\begin{aligned}
		I_4
		&\sim \frac{1}{4} 
		\int_{\mathbb R^2} e^{-(t_{13}a^2+2t_3 a \alpha + t_{23} \alpha^2)} d\alpha \, da
		\int_{\mathbb R_+^2} e^{-\frac{(-\tau^2-i0)}{4}(t_{13}b^2+2t_3 b \beta + t_{23} \beta^2)} d\beta  \, db\\
		&=-\frac{\pi}{2\tau^2 T}
		\left(
		\frac{\pi}{2}-\arctan\frac{t_3}{\sqrt{T}}
		\right)\,.
	\end{aligned}
\end{equation}
Note the appearance of a $\tau^{-2}$ singularity as $\tau\to0$.
Substituting into Eq.~\eqref{G7t1t2t3}, the resulting integral over $t_1$, $t_2$ and $t_3$ can be tackled in a manner analogous to  $f_{\RT,6}$ discussed above. The final result indeed reproduces \eqref{fIII7} as the overall factor of $\tau^2$ and the $\tau^{-2}$ singularity cancel out.

Let us now briefly comment on the boundary conditions for the $\RN$ family. In view of eqs.~\eqref{III-IXrelations}, all master integrals for the $\RN$ family can be deduced from the ones discussed for the $\RT$ family except $f_{\RN,10}$ and $f_{\RN,14}$. The derivation of the corresponding boundary conditions \eqref{otherbc} proceeds in a manner similar to the ones for $f_{\RT,7}$ and $f_{\RT,10}$. The crucial point is that, for these topologies, the singularities arising as $\tau \to 0$ are not sufficiently strong in order to compensate the overall factors of $\tau^2$ or $\tau$ appearing in their definitions, so that the static limit eventually gives zero.  

Moving to the $\Ht$ family, let us first recall that some integrals coincide with those appearing in the $\RT$ family by Eq.~\eqref{coinc}. The master integrals $f_{\Ht,2}$, $f_{\Ht,3}$, $f_{\Ht,4}$ can be directly evaluated via Schwinger parameters. Direct evaluation is also viable for the last three integrals appearing on the right-hand side of the definition \eqref{f10Hdef} of $f_{\Ht,10}$. 
To discuss the boundary condition for $f_{\Ht,8}$, it is again necessary to distinguish between ordinary and singular static contributions to the small-$\tau$ limit of $G_{0,1,1,0,0,1,1,0,1}$. One can verify that the ordinary static contributions cancel between
the two terms on the right-hand side of the definition \eqref{f8Hdef} of $f_{\Ht,8}$. The corresponding boundary condition \eqref{fH8} is therefore entirely due to the singular static contributions to $G_{0,1,1,0,0,1,1,0,1}$.
Finally, the (ordinary) static limit of the integral 
\begin{equation}\label{G5}
	G_5=G_{-1,1,1,-1,1,1,1,1,1}
\end{equation} 
appearing on the left-hand side of the definition \eqref{f10Hdef} of $f_{\Ht,10}$ can be discussed as in Appendix~\eqref{app:Xepsilon}, obtaining
\begin{equation}\label{G5ev}
G^{(o)}_{5}
=\frac{e^{2\gamma_E\epsilon}}{(q^2)^{1+2\epsilon}}\left[
\frac{2 \Gamma (2 \epsilon -1) \Gamma (1-\epsilon )^3}{\epsilon ^2 \Gamma (1-3 \epsilon
	)}+
\frac{\Gamma (\epsilon )^2 \Gamma (1-\epsilon )^4}{\epsilon  \Gamma (1-2 \epsilon)\Gamma (2-2 \epsilon )}
\right]\,.
\end{equation}
The discussion of the odd integrals proceeds in a similar fashion. The integral $f_{\Ht,11}$ can be evaluated directly since it is independent of $x$, while the boundary condition for $f_{\Ht,12}$ given in eq.~\eqref{fH12} emerges both from the ordinary static region (first line) and from the singular static region (second line).

\subsection{Evaluation of $G_{1,2}^{(o)}\,,$ $G_{2,1}^{(o)}$ and $G_{5}^{(o)}$}
\label{app:Xepsilon}

Let us consider a family of integrals which naturally encompasses the ordinary static limit of the general integrals $G_{i_1,\ldots, i_9}$ in Eq.~\eqref{bigG}, 
\begin{equation}\label{}
	J_{i_1,\ldots,i_7}
	=
	\int_{\ell_1} \int_{\ell_2} \frac{1}{D_1^{i_1}D_2^{i_2}D_3^{i_3}D_4^{i_4}D_5^{i_5}D_6^{i_6}D_7^{i_7}}\,,
\end{equation}
where
\begin{equation}\label{}
	D_1=-2u\cdot\ell_1\,,\qquad 
	D_2=-2u\cdot \ell_2\,,
\end{equation}
with $u=-1$ and $u\cdot q=0$, together with
\begin{equation}
	D_3=\ell_1^2\,,\quad
	D_4=\ell_2^2\,,
	\quad
	D_5=(\ell_1+\ell_2-q)^2\,,
    \quad
	D_6=(\ell_1-q)^2\,,\quad
	D_7=(\ell_2-q)^2\,.
\end{equation}
 In particular, comparing with Eq.~\eqref{Gjk}
\begin{equation}\label{}
	G^{(o)}_{j,k}=J_{j,j,0,0,0,k,1,1}.
\end{equation}

Integration by parts reduction performed with \texttt{LiteRed} allows one to express the desired integrals $G_{1,2}^{(o)}$ and $G_{2,1}^{(o)}$ in terms of $G_{1,1}^{(0)}$ and of the (elementary) integral
\begin{equation}\label{00111000}
	G_0= J_{0,0,0,0,1,1,1}= e^{2\gamma_E\epsilon} \frac{\Gamma(-1+2\epsilon)}{(q^2)^{-1+2\epsilon}}\frac{\Gamma(1-\epsilon)^3}{\Gamma(3-3\epsilon)}
\end{equation}
according to
\begin{align}
	G_{1,2}^{(o)}&=\frac{(-1+2\epsilon)(-2+3\epsilon)(1-3\epsilon)}{\epsilon(1+2\epsilon)(q^2)^2}\, G_0+\frac{2\epsilon(-1+6\epsilon)}{(1+2\epsilon)q^2} G_{1,1}^{(o)}\,,\\
	G_{2,1}^{(o)}&=\frac{(-1+2\epsilon)(-2+3\epsilon)(-1+9\epsilon^2)}{\epsilon(1+2\epsilon)(q^2)^2}\,G_0+ \frac{\epsilon(-1+6\epsilon)}{(1+2\epsilon)q^2} G_{1,1}^{(o)}\,.
\end{align}
Substituting Eqs.~\eqref{G11} and \eqref{00111000} into these expressions yields Eqs.~\eqref{G12} and \eqref{G21}.

A similar strategy can be applied to the evaluation of the ordinary static limit of $G_5$ in Eq.~\eqref{G5}. The relevant relations in this case are
\begin{equation}\label{}
	G_5^{(o)}=J_{0,0,1,1,1,1,1}
\end{equation}
and
\begin{equation}\label{0011111}
J_{0,0,1,1,1,1,1}
=\frac{2(2-3\epsilon)(1-3\epsilon)}{\epsilon^2q^4}J_{0,0,0,0,1,1,1} + \frac{1-2\epsilon}{\epsilon\, q^2}\,J_{0,0,1,1,0,1,1}\,.
\end{equation}
The first integral on the right-hand side coincides with $G_0$ given in Eq.~\eqref{00111000}, while the second one easily evaluates to
\begin{equation}\label{0011011}
	J_{0,0,1,1,0,1,1}=\frac{e^{2\gamma_E\epsilon}}{(q^2)^{2\epsilon}}
	\left[ \frac{\Gamma(1-\epsilon)^2\Gamma(\epsilon)}{\Gamma(2-2\epsilon)}\right]^2\,.
\end{equation}
Substituting Eqs.~\eqref{00111000} and \eqref{0011011} into the decomposition \eqref{0011111} leads to the result \eqref{G5ev} given for $G_5^{(o)}$ in the previous appendix.

\section{Non-Relativistic Limit and PN Expansion}
\label{PN}

In this Appendix we classify the different contributions according to whether they correspond to an integer ($\mathbb Z$) or half-integer ($\mathbb Z + 1/2$) term in the PN expansion.

Concerning the PN classification of the different contributions we use the following standard terminology:
\begin{itemize}
\item A contribution to the scattering angle $\chi \sim G^m v^p$ is $m$PM and $n$PN with $n = m + \frac{p}{2}$.
\item A contribution to the eikonal phase $\delta_{m-1} \sim G^m v^p$ is $m$PM and $n$PN with $n = m + \frac{p-1}{2}$.
\item A contribution to the amplitude $\mathcal{A}_{m-1} \sim G^m v^p$ is $m$PM and $n$PN with $n = m -1 +\frac{p}{2}$.
\end{itemize}

We will carry out the classification at each loop (PM) order using that $\sigma^2-1 \sim v^2$ and $\log z \sim v$.
\begin{itemize}
\item{{\bf Tree level (1PM)}}

This is, by definition, $0$PN and corresponds to:
\begin{equation}
\mathcal{A}_0 \sim (\sigma^2 -1)^0   \Rightarrow \delta_0 \sim (\sigma^2 -1)^{-1/2} ;~~\chi_0 \sim \frac{G}{J v} \sim \frac{G}{v^2}
\end{equation}
\item{\bf{One loop (2PM).}}

Here $0$PN is defined by
\begin{equation}
\chi_1 \sim \chi_0^2 \sim  \frac{G^2}{J^2 v^2}~~;~~ \delta_1 \sim \frac{G^2}{J v^2} \sim \frac{G^2}{ v^3}
\end{equation}
The super classical contribution, Eq.~\eqref{A11}, has $n=\frac12$ (it has an extra $J/\hbar \sim v$)

The $O(q^{-1})$ classical contribution, Eq.~\eqref{A12}, is indeed $0$PN.

The remaining quantum  terms behave as follows: 

At $O(\epsilon^0)$:
 \begin{eqnarray}
&&  \operatorname{Re} \Delta_1 \sim (\sigma^2 -1)^{-3/2}\, \Rightarrow n = 0 ,  \nonumber \\
&&  \operatorname{Im} \Delta_1 \sim (\sigma^2 -1)^{-1} \,  \Rightarrow n = \frac12
\end{eqnarray}
At $O(\epsilon)$:
 \begin{eqnarray}
&&\operatorname{Re} \Delta_1 \sim (\sigma^2 -1)^{-3/2}\, \Rightarrow n = 0 , \nonumber   \\
&& \operatorname{Im} \Delta_1 \sim (\sigma^2 -1)^{-2}\, \Rightarrow n = - \frac12
\end{eqnarray}

\item{\bf{Two loop (3PM).}}

Here the situation is richer. The $0$PN order is given by:
\begin{equation}
 \delta_2 \sim G^3 (\sigma^2 -1)^{-5/2}   \Rightarrow \mathcal{A}_2^{0\mathrm{PN}} \sim (\sigma^2 -1)^{-2}
\end{equation}

For  the $O(\epsilon^{-2})$ (IR singular) terms in \eqref{Ansatz(2)} we find, for both the superleading (real) and the classical contributions:
 \begin{eqnarray}
&& \operatorname{Re} \mathcal{A}_2^{(2)} \sim  (\sigma^2 -1)^{-1} \Rightarrow \operatorname{Re} \delta_2 \sim (\sigma^2 -1)^{-3/2} \, \Rightarrow n = 1  \, ,\nonumber \\
&& \operatorname{Im} \mathcal{A}_2^{(2)} \sim (\sigma^2 -1)^{-3/2} \Rightarrow \operatorname{Im} \delta_2 \sim (\sigma^2 -1)^{-2}\,  \Rightarrow n = \frac12 \, ,
\end{eqnarray}
meaning that $\operatorname{Re} \mathcal{A}_2^{(2)} $ is $\mathbb Z\mathrm{PN}$  while $\operatorname{Im} \mathcal{A}_2^{(2)} $ is $(\mathbb Z+1/2)\mathrm{PN}$.

For  the $O(\epsilon^{-1})$ (i.e. IR finite in $b$-space) terms in \eqref{Ansatz(1bis)} we find,

\begin{itemize}
\item {\bf \eqref{ImD1d0}}: $\mathcal{A}_2 \sim G^3 (\sigma^2-1)^{-2}  \, \Rightarrow n = 0   \sim \operatorname{Im} \Delta_1^{(\epsilon)} \delta_0$;

\item {\bf \eqref{ReD1d0}}: $\mathcal{A}_2 \sim i G^3 (\sigma^2-1)^{-3/2}  \, \Rightarrow n = 1/2   \sim \operatorname{Re} \Delta_1^{(\epsilon)} \delta_0$;

\item {\bf \eqref{newwlog}}: $\operatorname{Re} \mathcal{A}_2 \sim G^3 (\sigma^2-1)^{-3/2}  \, \Rightarrow n = 1/2$ , \\
$ \operatorname{Im} \mathcal{A}_2 \sim G^3 (\sigma^2-1)^{-3/2}  \, \Rightarrow n = 1/2$,   with a $\log v$-enhancement;

\item {\bf \eqref{P-MRZ}}: $\operatorname{Re} \mathcal{A}_2 \sim G^3   \, \Rightarrow n = 2$ , 
$ \operatorname{Im} \mathcal{A}_2  \sim G^3 (\sigma^2-1)^{1/2}  \, \Rightarrow n = 5/2$;

\item {\bf \eqref{D1d0}}: $\operatorname{Re} \mathcal{A}_2 \sim G^3 (\sigma^2-1)^{-2}  \, \Rightarrow n = 0  \sim \operatorname{Im} \Delta_1^{(\epsilon)} \delta_0$,\\
$ \operatorname{Im} \mathcal{A}_2 \sim G^3 (\sigma^2-1)^{-3/2}  \, \Rightarrow n = 1/2 \sim \operatorname{Re} \Delta_1^{(\epsilon)} \delta_0$;

\item {\bf \eqref{ACV} }: $\operatorname{Re} \mathcal{A}_2 \sim G^3 (\sigma^2-1)^{-3/2}  \, \Rightarrow n = 1/2$ , \\
$ \operatorname{Im} \mathcal{A}_2 \sim G^3 (\sigma^2-1)^{-3/2}  \, \Rightarrow n = 1/2$,  with a $\log v$-enhancement \, ,
\end{itemize}
where we have indicated by $ \Delta_1^{(\epsilon)}$ the ${\cal O}(\epsilon)$ terms in \eqref{1L7}, \eqref{1L71} (recall that $\delta_0 \sim \epsilon^{-1}$).

\end{itemize}
Note that imaginary parts are all $(\mathbb Z+1/2)\mathrm{PN}$ but the new contributions are distinguished by exhibiting an extra $\log v$ enhancement. This makes the corresponding real parts to be also $(\mathbb Z+1/2)\mathrm{PN}$. For the usual $(\mathbb Z+1/2)\mathrm{PN}$ imaginary parts the corresponding real parts are $\mathbb Z\mathrm{PN}$.
We should also mention that in \eqref{newwlog} and \eqref{ACV} the leading $0.5\mathrm{PN}$ term cancels both in the real part and in the logarithmically-enhanced imaginary part so that 
their sum is  $1.5\mathrm{PN}$ (it becomes $2.5\mathrm{PN}$ in pure Einstein gravity as discussed in \cite{Damour:2020tta} and \cite{DiVecchia:2021ndb}, and after equation \eqref{ReGR15}).

\providecommand{\href}[2]{#2}\begingroup\raggedright\endgroup

\end{document}